\PassOptionsToPackage{hypertexnames=false}{hyperref}
\documentclass[twocolumn,tighten,twocolappendix]{aastex701}

\usepackage{newtxtext,newtxmath}
\usepackage{needspace}
\usepackage{orcidlink}
\usepackage{enumitem}
\usepackage{mathtools}

\usepackage{fancyhdr}
\journalinfo{The Astrophysical Journal, 1003:102 (18pp), 2026 May 20}

\newcommand{\zacc}[1]{\ensuremath{z_\mathrm{acc}}}
\newcommand{\ftr}{\ensuremath{w_\mathrm{tr}}}

\newcommand{\subfind}{\textsc{Subfind}}
\newcommand{\sublink}{\textsc{Sublink}}

\newcommand{\Msun}{\ensuremath{\mathrm{M}_\odot}}
\newcommand{\Mdyn}{\ensuremath{M_\mathrm{dyn}}}
\newcommand{\Mstar}{\ensuremath{M_\star}}
\newcommand{\kms}{\ensuremath{\text{km}~\text{s}^{-1}}}
\newcommand{\osim}{\mathord{\sim}}

\begin{document}

\title{\vspace{-\baselineskip}\Large How to Build an Empirical Speed Distribution\\ for Dark Matter in the Solar Neighborhood}
\shorttitle{How to Build an Empirical Speed Distribution}

\author{\rm Tal Shpigel~\orcidlink{0009-0003-5629-5848}}
\affiliation{\rm Department of Physics, Princeton University, Princeton, NJ 08544, USA; \href{mailto:talshpigel@princeton.edu}{talshpigel@princeton.edu}, \href{mailto:dfolsom@princeton.edu}{dfolsom@princeton.edu}}
\email{talshpigel@princeton.edu}

\author{\rm Dylan Folsom~\orcidlink{0000-0002-1544-1381}}
\affiliation{\rm Department of Physics, Princeton University, Princeton, NJ 08544, USA; \href{mailto:talshpigel@princeton.edu}{talshpigel@princeton.edu}, \href{mailto:dfolsom@princeton.edu}{dfolsom@princeton.edu}}
\email{dfolsom@princeton.edu}

\author{\rm Mariangela Lisanti~\orcidlink{0000-0002-8495-8659}}
\affiliation{\rm Department of Physics, Princeton University, Princeton, NJ 08544, USA; \href{mailto:talshpigel@princeton.edu}{talshpigel@princeton.edu}, \href{mailto:dfolsom@princeton.edu}{dfolsom@princeton.edu}}
\affiliation{\rm Center for Computational Astrophysics, Flatiron Institute, New York, NY 10010, USA}
\email{mlisanti@princeton.edu}

\author{\rm Lina Necib~\orcidlink{0000-0003-2806-1414}}
\affiliation{\rm Physics Department and Kavli Institute for Astrophysics and Space Research, Massachusetts Institute of Technology, Cambridge, MA 02139, USA}
\email{lnecib@mit.edu}

\author{\rm Mark Vogelsberger~\orcidlink{0000-0001-8593-7692}}
\affiliation{\rm Physics Department and Kavli Institute for Astrophysics and Space Research, Massachusetts Institute of Technology, Cambridge, MA 02139, USA}
\affiliation{\rm The NSF AI Institute for Artificial Intelligence and Fundamental Interactions, Massachusetts Institute of Technology, Cambridge, MA 02139, USA}
\email{[mvogelsb@mit.edu]}

\author{\rm Lars Hernquist~\orcidlink{0000-0001-6950-1629}}
\affiliation{\rm Center for Astrophysics, Harvard \& Smithsonian, 60 Garden Street, Cambridge, MA 02138, USA\\\it Received 2025 December 17; revised 2026 April 13; accepted 2026 April 14; published 2026 May 19}
\email{lhernquist@cfa.harvard.edu}

\shortauthors{Shpigel et al.}

\begin{abstract}
    \noindent
    The dark matter flux in a direct detection experiment depends on its local speed distribution.  
    This distribution has been inferred from simulations of Milky Way--like galaxies, but such models serve only as proxies, given that no simulation directly captures the detailed evolution of our own Galaxy.  
    This motivates alternative approaches that obtain this distribution directly from observations.  
    In this work, we utilize 98 Milky Way analogs from the TNG50 simulation to develop and validate a procedure for inferring the dark matter speed distribution using the kinematics of nearby stars. 
    We find that the dark matter that originated from old mergers, plus that from recent nonluminous accretions, is well described by a Maxwell--Boltzmann speed distribution centered at the local standard-of-rest velocity.  
    Meanwhile, recently accreted dark matter from massive mergers has speeds that can be traced from the associated stellar debris of these events.  
    The stellar populations systematically underestimate the velocity dispersion of their dark matter counterparts, but a simple kinematic boost brings the two into good alignment.  
    Using the TNG50 host galaxies, we demonstrate that combining these two contributions provides an accurate reconstruction of the local dark matter speeds.  
    As  an application of the procedure to our own Galaxy, we utilize stellar kinematic data from Gaia to quantify how the dark matter remnants from the Milky Way's last major merger impact its speed distribution in the solar neighborhood.

    \vspace{0.5\baselineskip}\noindent\textit{Unified Astronomy Thesaurus concepts:} \href{https://astrothesaurus.org/uat/2178}{Galactic archaeology (2178)}; \href{https://astrothesaurus.org/uat/353}{Dark matter (353)}; \href{https://astrothesaurus.org/uat/575}{Galaxy accretion (575)}; \href{https://astrothesaurus.org/uat/1051}{Milky Way dynamics (1051)}
\end{abstract}

\section{Introduction} \label{sec:intro}
\thispagestyle{fancy}

Characterizing the speed distribution of dark matter~(DM) near the Sun is crucial to interpret the data from direct detection experiments, which search for DM interactions in terrestrial detector targets (see, e.g., \citet{1996PhR...267..195J,2013RvMP...85.1561F} for reviews). Variations in the assumed speed distribution can impact the expected number of interaction events in these experiments, affecting the interpretation of a signal as well as inferred constraints. Knowledge of the local DM speed distribution is essential to reduce this source of uncertainty~\citep{2014arXiv1404.4130D, 2017JPhG...44h4001G}. 

The DM phase-space distribution near the Sun is a product of the Milky Way's~(MW's) evolutionary history. In the current cosmological paradigm, DM halos grow hierarchically by accreting smaller satellite systems that are tidally disrupted by the host galaxy's gravitational field~\citep{1978MNRAS.183..341W,2008MNRAS.391.1685S, 2011ApJ...740..102K}. DM debris from the earliest mergers, whose time since accretion is many times its orbital timescale, should reach a quasi-equilibrium state with a speed distribution described by a Maxwell--Boltzmann distribution~\citep{1986PhRvD..33.3495D}. However, DM that is accreted more recently will not be relaxed and may retain kinematic signatures of the merger from which it originated~\citep{2008Natur.454..735D}. In the MW's stellar population, there is evidence for such unequilibrated kinematic structure; groups of stars with particular chemical abundances and orbital dynamics appear differently in this chemodynamical space from the stars thought to have formed in the MW itself---see, e.g., \cite{2020ARA&A..58..205H} for a review. These ex situ stellar populations are thought to originate from previous galactic accretion events.

One approach to model the local DM speed distribution is to simulate the formation of galaxies like the MW in a cosmological setting. This has been done in both DM-only~\citep{2005MNRAS.361L...1W,2006JCAP...01..014H,2008MNRAS.385..236V,2009MNRAS.395..797V,2010JCAP...10..034G,2010JCAP...02..030K,2013MNRAS.430.1722V,2017IJMPA..3230016B} and fully hydrodynamical~\citep{2010JCAP...02..012L,2010MNRAS.406..922T,2014ApJ...784..161P,2016JCAP...05..024B,2016MNRAS.462..663B,2016JCAP...08..071K,2016ApJ...831...93S,2017IJMPA..3230016B,  2020JCAP...07..036B,2020JHEP...07..081H,2020JCAP...11..016P, 2023MNRAS.524.2606L,2023JCAP...05..012N,2024JCAP...08..022S} contexts. These simulations can be used to characterize the halo-to-halo variance in the local DM speeds by sampling many possible formation histories. Using the largest high-resolution suite of simulated MW-like galaxies to date, \cite{2025arXiv250507924F} showed that the \emph{ensemble} of DM speed distributions is well modeled by a Maxwell--Boltzmann distribution.  However, they found that the distributions for individual galaxies do deviate from this model, especially at the high-speed tails. Where our own Galaxy falls on this spectrum of possibilities remains an open question. 

Current simulation efforts are trying to produce MW-like galaxies whose formation histories are in better alignment with our own Galaxy~\citep[e.g.,][]{2022MNRAS.512.5823M, 2023MNRAS.521..995R, 2024ApJ...971...79B}, but the challenge of this task motivates complementary efforts to obtain the DM speed distribution directly from observations.  This can be done by harnessing the correlation between the kinematics of ex situ stars and the DM that is accreted from the same merger. For instance, \citet{2018PhRvL.120d1102H}, using the \textsc{Eris} hydrodynamic simulation~\citep{2011ApJ...742...76G}, demonstrated that the oldest and most metal-poor halo stars trace the velocities of the relaxed DM, as both would have originated from the earliest mergers. Furthermore, stellar substructure, such as debris flows~\citep{2012PDU.....1..155L}, also exhibits similar kinematic features to their corresponding DM distributions, a finding supported by the work of \citet{2015ApJ...807...14L} using the Via Lactea simulation~\citep{2008Natur.454..735D} and stellar tagging catalog from~\cite{2012ApJ...745..142R}. This correspondence between stellar and DM velocities has been further confirmed by \citet{2019ApJ...883...27N}, both for the relaxed component and present-day debris flows, using two different simulated halos from the FIRE suite~\citep{2018MNRAS.480..800H}.\footnote{Using six halos from the Auriga project~\citep{2017MNRAS.467..179G}, \citet{2019JCAP...06..045B} did not find a correlation between old stars and the \emph{total} DM distribution. This is likely because they did not separate out the DM contributions from more recent mergers when making the comparison~\citep{2018arXiv181204114L}.} 

As detailed in \citet{2019ApJ...883...27N}, an empirical approach to reconstructing the local DM speed distribution requires considering different subpopulations of DM, as defined by merger origin.  Focusing on DM accreted from luminous mergers, they demonstrated that the oldest DM (with accretion redshift $\zacc{} \gtrsim 3$) is well traced by the most metal-poor stars.  DM from later  mergers ($\zacc{} \lesssim 3$) may be traced by intermediate-metallicity stars; this is especially true if the DM is in velocity substructure, such as debris flow, but less so for the most recent accretions that leave behind  streams. In this paper, we build  on the work of \citet{2019ApJ...883...27N} in several key ways.  First, we vastly increase the sample size of MW analogs studied and also use a separate galaxy formation model.  Second, we account for the contribution of DM from nonluminous mergers and diffuse accretion~\citep{2011MNRAS.413.1373W, 2015MNRAS.448.2941S, 2010MNRAS.401.1796A}, which we refer to as ``dark accretion'' throughout.

In particular, we use a sample of 98 MW-like galaxies from the TNG50 simulation~\citep{2019MNRAS.490.3196P,2019MNRAS.490.3234N} and find that the DM in the solar neighborhood can be modeled as two distinct components: a background that follows a Maxwell--Boltzmann distribution peaking at the local standard-of-rest speed and a population originating from recent, high-mass accretion events.  The former is the combination of DM accreted from the oldest luminous mergers, as well as that from dark accretion.  By combining a Maxwell--Boltzmann model with a stellar-informed model for the merger component, we demonstrate how to reconstruct the total DM speed distribution in the solar neighborhood, accounting for the contribution of dark accretion. These findings, validated on simulation data, establish a practical framework for building a complete empirical speed distribution for the local DM. 

As a first application of this technique, we update previous empirical distributions for the local DM~\citep{2019ApJ...874....3N, 2024ApJ...974..167Z}.  Our result accounts for both dark accretion, as well as the Gaia Sausage--Enceladus merger \citep[GSE;][]{2018MNRAS.478..611B,2018Natur.563...85H}, believed to be the MW's last major merger. We find that the GSE-contributed DM is modestly slower than the local standard-of-rest speed. The mode of the resulting speed distribution is 11~\kms{} slower than that of the Maxwell--Boltzmann-only model, resulting in an $\osim20\%$ suppression to the high-speed tail.

This paper is organized as follows. \autoref{sec:methods} introduces the TNG50 simulation, the criteria used to select the sample of MW analogs, the methodology for tracking the origin of DM and stellar particles in the solar neighborhood, and the selection of analyzed mergers. \autoref{sec:results} presents the core results of the study, focusing on the kinematic correlations between DM and stellar components within the selected merger sample, and the reconstruction of the local DM speed distribution. \autoref{sec:discussion} applies our formalism to the MW and discusses the implications for direct detection experiments. Finally, \autoref{sec:conclusions} summarizes the main conclusions of the study and their broader implications. The Appendices provide an expanded discussion of the DM--merger association procedure, the offset in velocity dispersion between stars and DM, the GSE observational data, and other supplementary figures.

\rhead{\footnotesize Shpigel et al.}
\section{Methodology}\label{sec:methods}
\subsection{MW Analogs}

The IllustrisTNG project\footnote{The IllustrisTNG data are available online at \url{https://tng-project.org/}.}~\citep{2019ComAC...6....2N} comprises a suite of magnetohydrodynamic cosmological simulations that utilize the moving-mesh code AREPO~\citep{2010ARA&A..48..391S}. The sample of MW analogs used in this work is taken from the highest-resolution simulation in this suite, TNG50~\citep{2019MNRAS.490.3234N, 2019MNRAS.490.3196P}, which covers a volume of $(51.7~\rm{Mpc})^3$ from early times (redshift $127$) to the present day, taking cosmological parameters as measured by the~\citet{2016A&A...594A..13P}. The mass resolution of a DM~(star) particle in TNG50 is $4.5\times10^5$~($\osim8.5\times10^4)~\Msun$.  Details regarding the IllustrisTNG Project can be found in \citet{2019ComAC...6....2N}; here, we only focus on those elements that are most relevant for this study.

At each saved time step (``snapshot'') of the simulation, structures are found within the volume using a Friends-of-Friends~(FoF) clustering algorithm~\citep{1985ApJ...292..371D} that is run only on the DM particles. Other particle types (gas, stars, and black holes) are associated with the FoF group of the nearest DM particle. The resulting FoF groups typically consist of a central halo and its neighboring substructure, but the groups can include multiple systems if the central halos are close enough. Further, the FoF algorithm does not require that the particles in the resulting groups be gravitationally bound. Therefore, the \subfind{} algorithm~\citep{2001MNRAS.328..726S, 2009MNRAS.399..497D} is used to determine the bound structures (i.e., DM halos) within each FoF group at each snapshot. Since this is done on a snapshot-by-snapshot basis, the same physical object will be given different \subfind{} IDs at each snapshot. The \sublink{} algorithm identifies persistent physical halos from the \subfind{} catalog, tracking when \subfind{} halos merge with each other and are disrupted, resulting in a ``merger tree'' that traces the assembly of the \subfind{} halos. \subfind{} provides a total mass \Mdyn{} for each halo, which includes all particles bound to the halo but excludes mass bound to satellite halos. The contribution to \Mdyn{} that comes from star particles is denoted $M_\star$.

MW analogs are selected from the \subfind{} catalog for the present-day snapshot following the procedure outlined in \citet{2025ApJ...983..119F}. To summarize, MW analogs are defined as those halos that 
\begin{enumerate}
    \item have $M_\star$ within $4\times10^{10}~\Msun$ and $7.3\times10^{10}~\Msun$,
    \item are farther than 500~kpc from any halo with \Mdyn{} larger than that of the candidate MW, and
    \item are farther than 1~Mpc from any halo with $\Mdyn>10^{13}~\Msun$.
\end{enumerate}

The first criterion selects galaxies with stellar masses that are comparable to that of the MW, based on $68\%$ confidence intervals quoted in the literature~\citep{2006MNRAS.372.1149F, 2011MNRAS.414.2446M, 2015ApJ...806...96L, 2016ARA&A..54..529B, 2016ApJ...831...71L, 2017MNRAS.465...76M, 2020MNRAS.494.4291C}. The next criterion ensures that the MW analog is not too close to a larger galaxy, excluding interacting galaxies residing in close pairs or groups. This does still allow for an M31-like partner, as it is 760~kpc away~\citep{2017MNRAS.468.3428P, 2018MNRAS.475.4043K,2021ApJ...920...84L, 2022ApJ...928L...5B, 2023PhRvD.107j3003V, 2023MNRAS.521.4863S, 2023ApJ...956...15L}. The last criterion removes galaxies that reside near galaxy clusters, as the Local Group is characterized by its relative isolation and the absence of large galaxy groups in its proximity~\citep{2005AJ....129..178K}. Applying these selection criteria yields a final sample of 98 MW-like galaxies. The MW analogs in this sample have peak dynamical masses of $1.16^{+0.51}_{-0.28} \times 10^{12}~\Msun$.\footnote{Throughout this work, we quote the median, 16th, and 84th percentiles.}

While these selection criteria permit certain features of the Local Group, such as a Large Magellanic Cloud--like satellite or an M31-like partner halo, these are not requirements of our selection. Further, the $\mathcal{O}(10~\mathrm{Mpc})$-scale environment of the TNG50 simulation is modestly dissimilar from that of the Local Group. For example, the simulation volume contains two Virgo-mass galaxy clusters, with $\Mdyn \gtrsim 10^{14}~\Msun$. The MW itself is $\osim 16$~Mpc away from the Virgo cluster, but 81\% of the MW analogs in TNG50 are within 16~Mpc of one of these large clusters, and three analogs are close enough to the largest cluster to be in its FoF group. Therefore, the results of this analysis must be interpreted with the understanding that the simulated environments do not exactly reproduce such Local Group features, which may bias the formation histories of the MW analogs.

\subsection{Tracking of DM and Star Particles}
\label{subsec:mergers}

For each host galaxy, we define the region of interest~(ROI) to be a cylindrical shell spanning galactocentric radii $r\in[6,10]$~kpc and height $|z|\leq 2$~kpc from the disk midplane, chosen to be centered around the Sun's location in our Galaxy, $R_\odot \sim 8$~kpc~\citep{2021A&A...647A..59G,2024A&A...692A.242G}.\footnote{Note that the simulated galaxies are generically of different sizes from the MW, which is itself anomalously compact; see \citet{2007ApJ...662..322H,2013ApJ...779..115B,2016ApJ...833..220L,2020MNRAS.498.4943B,2025MNRAS.540.3493T}. The resulting speed distributions at 8~kpc are therefore not indicative of the MW's speed distribution \citep{2025arXiv250507924F}.} The goal is to determine the origin of the DM and star simulation particles that are located within this volume in order to reconstruct the merger history near the solar neighborhood of the host galaxy.

The tracking procedure for star particles follows that of \citet{2025ApJ...983..119F}. To review, mergers are first extracted from the \sublink{} trees for each MW analog. Next, each star particle is tracked across all simulation snapshots, recording the subhalos to which it is bound to over time. Ex situ stars are identified as those that form outside of the MW analog and that are bound to a merger for at least one snapshot.\footnote{There is, however, ambiguity in $\osim1\%$ of cases \citep{2025ApJ...983..119F}.} These stars are associated with the mergers to which they are bound for the greatest number of snapshots. The definition of ex situ chosen by \citet{2025ApJ...983..119F} prioritizes the purity of the sample, ensuring that the ex situ stars can be identified with a contributing merger. 

Tracking the origin of DM particles in the ROI carries additional challenges relative to the stars. This is because the DM halo surrounding a satellite galaxy is more extended than its  stellar component, so it is easier to miss a satellite's more loosely bound DM particles or to misidentify DM from the host as belonging to the subhalo. The DM tracking algorithm used here is designed to minimize these effects.  For each DM particle, we find the first snapshot where \subfind{} considers it bound to the host MW.  This is labeled as the accretion redshift, \zacc{}, for that particle. If \zacc{} is within the first 2~Gyr of the simulation's start, we do not reconstruct the particle's history any further. If not, then we consider the 2~Gyr preceding the accretion redshift.\footnote{Because stars form over the course of the simulation, this 2~Gyr time window would induce a bias if the algorithm were to be na\"ively applied to the stellar component. Stars formed within 2~Gyr of a merger's infall may not appear as bound to that merger for long enough and would be incorrectly categorized as unassociated. Therefore, we do not apply this algorithm to stars and instead use the algorithm from \citet{2025ApJ...983..119F} described above.} The DM is associated with a subhalo if it is bound to it for more than 70\% of the snapshots in this window.  Those particles that do not satisfy this criterion remain unassociated. \autoref{appendix:merger_association} discusses this algorithm in more detail, including robustness tests for the choice of parameters.  Varying the lookback time from 0.5 to 3~Gyr and the snapshot fraction from 10\% to 90\% does not significantly affect the results.

\begin{figure*}
    \centering
    \includegraphics{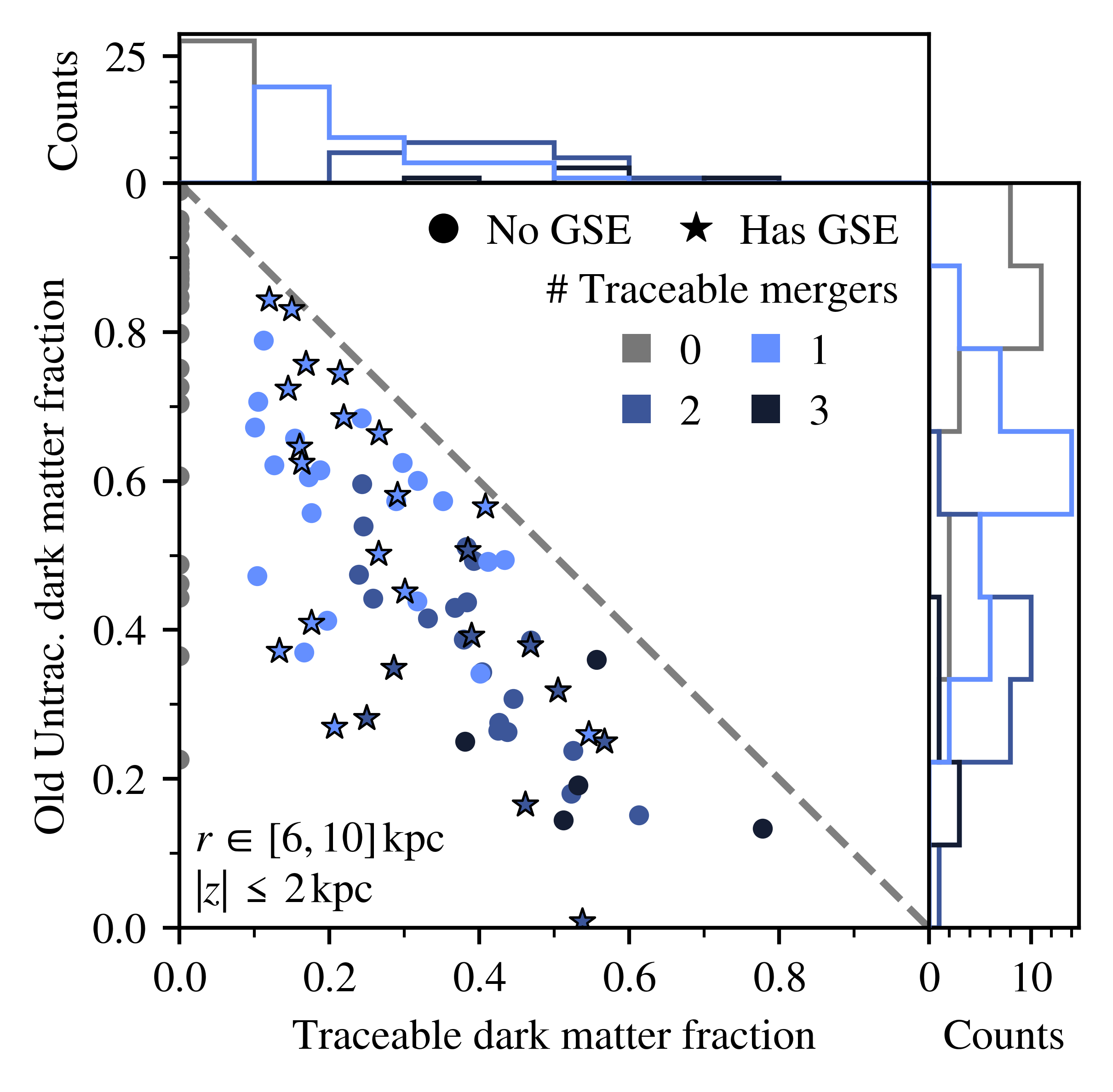}
    \includegraphics{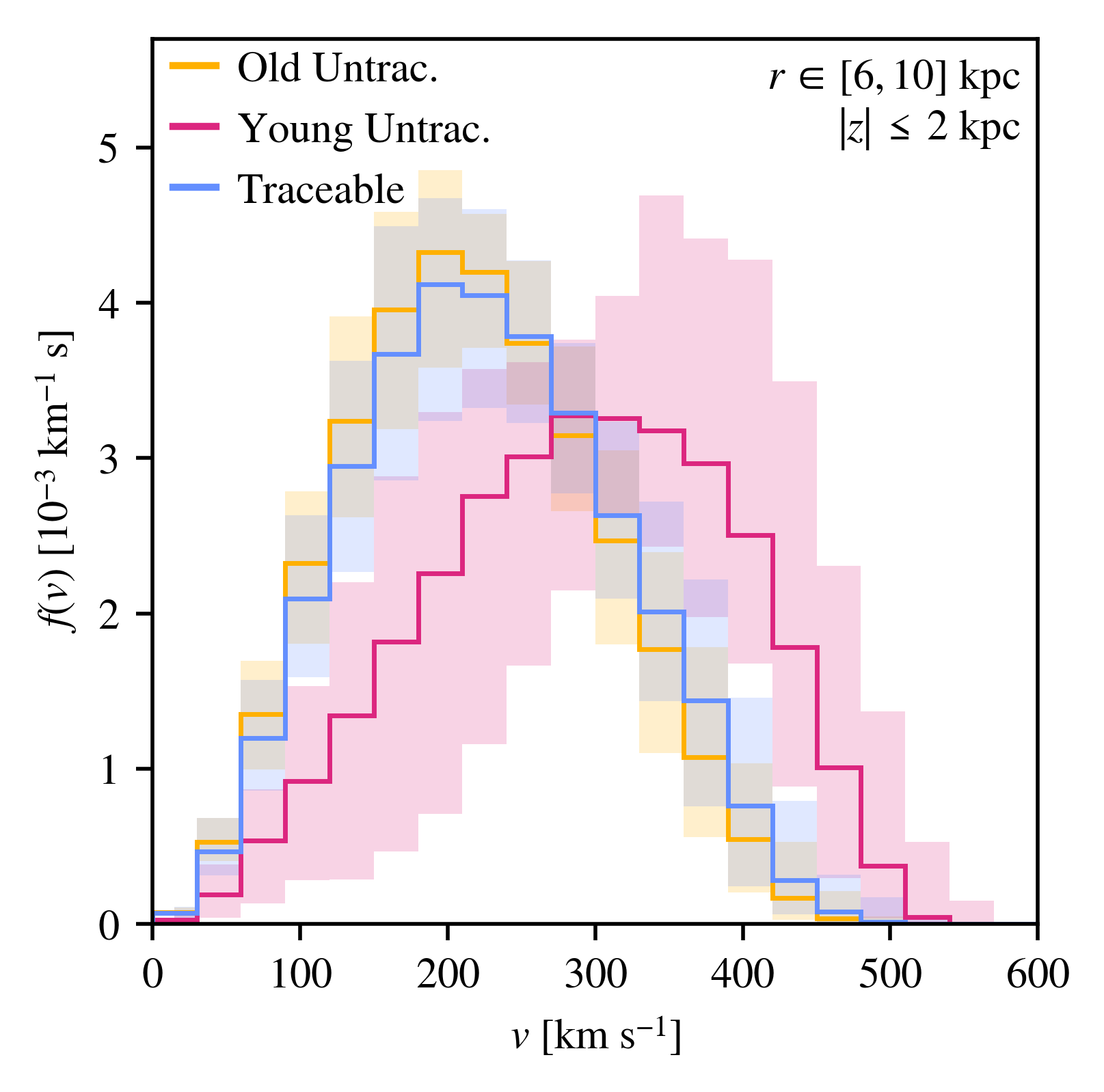}
    \caption{(Left) Fraction of DM in the solar annulus (6--10 kpc in cylindrical radius and height $|z|\leq2$ kpc) originating from Traceable mergers vs. the fraction accreted prior to redshift 3 (Old Untraceable DM), across the 98 MW analogs.  Points are colored by the number of Traceable mergers and shaped by the presence~(star markers) or absence~(circle markers) of a GSE-like event in that galaxy's history. Marginal histograms show the distribution of each quantity across the sample, given the number of Traceable mergers. For systems with one Traceable merger, $58^{+12}_{-17}\%$ of the DM in the solar annulus is Old Untraceable, while $18^{+15}_{-5}\%$ is Traceable. The remaining $21^{+11}_{-12}\%$ is contributed after redshift three but lacks a substantial population of luminous tracers (Young Untraceable). (Right) Speed distributions for the three DM components in the solar annulus: Old Untraceable~(yellow), Young Untraceable~(magenta), and Traceable~(blue). The solid line represents the median probability density, while shaded regions indicate the 16th--84th percentile range across the sample. While the Young Untraceable component exhibits greater halo-to-halo variance and is biased toward higher speeds, it typically represents a small fraction of the total Untraceable DM, such that the combined distribution is well described by a Maxwell--Boltzmann profile. The Traceable component can be modeled using the kinematics of the associated stellar debris. The DM components (Old Untraceable, Young Untraceable, and Traceable) are defined in \autoref{subsec:mergers}.}
    \label{fig:1}
\end{figure*}

Once the origin of each simulation star and DM particle in the ROI is identified, we divide the particles into three categories based on infall time and the mass of their associated merger:
\begin{enumerate}[wide=0pt, leftmargin=\parindent]
    \item \textit{Traceable}. DM particles with $\zacc{}<3$ that originated from a ``Traceable'' merger, defined as a merger with an infall halo mass of $\Mdyn \geq 10^9~\Msun$ that contributes more than 10\% of the DM in the ROI.
    \item \textit{Young Untraceable}. DM particles with $\zacc{}<3$ that are not Traceable. This population can arise from late-time dark accretion as well as from luminous mergers that do not contribute a substantial fraction of DM to the local neighborhood.
    \item Old Untraceable. DM particles with $\zacc{}>3$, consisting of early-time dark accretion and early luminous mergers, both of which have had sufficient time to virialize in the innermost regions of the MW halo. Both ``Untraceable'' components may include contributions from low-mass subhalos with insufficiently resolved stellar populations. 
\end{enumerate}
Though we do not do so here, the DM from early luminous mergers may be directly modeled using metal-poor stars, as shown by \citet{2018PhRvL.120d1102H} and \citet{2019ApJ...883...27N}.  

The criteria used to select the Traceable component identify large mergers that contribute debris to the ROI, so that this component may be traced by stars that are stripped from the same events. Indeed, every Traceable merger contributes at least 186 star particles ($\osim 1.6\times 10^7~\Msun$) to the ROI. In practice, the requirement that a merger contributes more than 10\% of the DM in the ROI already restricts the sample to relatively massive systems; even without imposing the $\Mdyn\geq10^9~\Msun$ cut, Traceable mergers have peak masses of $8^{+14}_{-5}\times10^{10}~\Msun$.

\subsection{Identifying GSE-like Mergers}

For the primary results of this work, we do not require that the TNG50 galaxies have merger histories resembling that of the MW. For illustrative purposes, however, we do identify those MW analogs that have a GSE-like merger. The GSE is thought to be the most recent major merger experienced by the MW, occurring 8--10~Gyr ago and depositing $\osim10^{8}$--$10^{9}~\Msun$ of stars primarily to the inner 30~kpc of the Galaxy~\citep{2018ApJ...862L...1D, 2018Natur.563...85H,2018ApJ...856L..26M, 2019NatAs...3..932G, 2019MNRAS.486..378L, 2019ApJ...874....3N, 2020ApJ...897L..18B,2021MNRAS.508.1489F, 2021MNRAS.502.5686I,2021NatAs...5..640M, 
 2021ApJ...923...92N}.

In our simulations, a merger is classified as GSE-like if it 
\begin{enumerate}
    \item contributes more than 50\% of all ex situ stars in the volume $|z|\in[9,15]$~kpc and
    \item exhibits a stellar velocity anisotropy $\beta>0.5$ for the stars within the volume $|z|\in[9,15]$~kpc,
\end{enumerate}
where the velocity anisotropy is defined in terms of the galactocentric spherical velocity components:
\begin{equation}
    \beta=1-\frac{\sigma_\phi^2+\sigma_\theta^2}{2\sigma_r^2} \, . 
\end{equation}
These selection criteria reflect observations of the GSE stellar debris, which accounts for more than 50\% of the ex situ stars within a height $z\in[9,15]$~kpc of the Galactic plane and exhibits significant radial velocity anisotropy~\citep{2018ApJ...862L...1D, 2018ApJ...856L..26M, 2019MNRAS.486..378L, 2019ApJ...874....3N, 2021MNRAS.502.5686I, 2021ApJ...923...92N}. We choose this spatial footprint because it is a region in the MW within which the GSE's stellar contribution is most significant~\citep{2020ApJ...901...48N}, and it was also used in previous studies by \citet{2019MNRAS.484.4471F}. Also, while our GSE-like mergers tend to be more recently accreted than the GSE itself, with infall times of $6^{+3}_{-2}$~Gyr ago, adding in a requirement that the merger be before a redshift of 1 does not significantly change the results.

\needspace{2\baselineskip}
\section{Results} \label{sec:results}

This section characterizes the local speed distribution of accreted DM in the MW analogs of TNG50. We divide the DM into the three components described in \autoref{sec:methods}: Old Untraceable, Young Untraceable, and Traceable. The analysis proceeds in three parts. First, we show that the Untraceable components together follow a Maxwell--Boltzmann distribution; second, we demonstrate a procedure for modeling the Traceable component by its associated stellar debris; and finally, we combine these results to reconstruct the total local DM speed distribution. 

\autoref{fig:1} summarizes the fractions of the three DM components and their speed distributions across the 98 MW analogs. The left panel shows the fraction of the DM in the ROI contributed by Old Untraceable mergers versus the fraction considered Traceable, with histograms on the top and right edges indicating the marginal distributions of these values, split up by the number of Traceable mergers. The markers in the central panel are colored according to this number and are shaped according to the presence~(star marker) or absence~(circle marker) of a GSE-like event in the analog's history. Across the full sample, 70 out of 98 analogs (71\%) host one or more Traceable mergers. Among these 70 systems, 37~(53\%) have exactly one Traceable merger, 28~(40\%) have two, and 5~(7\%) have three, corresponding to an average of 1.1 Traceable mergers per halo across the sample.

The MW itself is characterized by a major merger, the GSE, which occurred $\osim 8$--10~Gyr ago. In our sample, 26 MW analogs have a GSE-like Traceable merger. Of these, the GSE-like merger is usually the only Traceable merger, with only 9 analogs (35\% of GSE-hosting halos) having a second. In the MW analogs with one Traceable merger, most of the DM in the ROI ($58^{+12}_{-17}\%$) is Old Untraceable. $18^{+15}_{-5}$\% of the DM is contributed by its Traceable merger, and the remaining $21^{+11}_{-12}\%$ is Young Untraceable. Requiring that the single Traceable merger be considered GSE-like does not meaningfully impact these fractions. This is expected, since selecting GSE analogs among these halos amounts to a cut on the merger's kinematics rather than on the merger history per se, and it is the merger history that sets the DM composition.

\begin{figure*}
    \centering
    \includegraphics{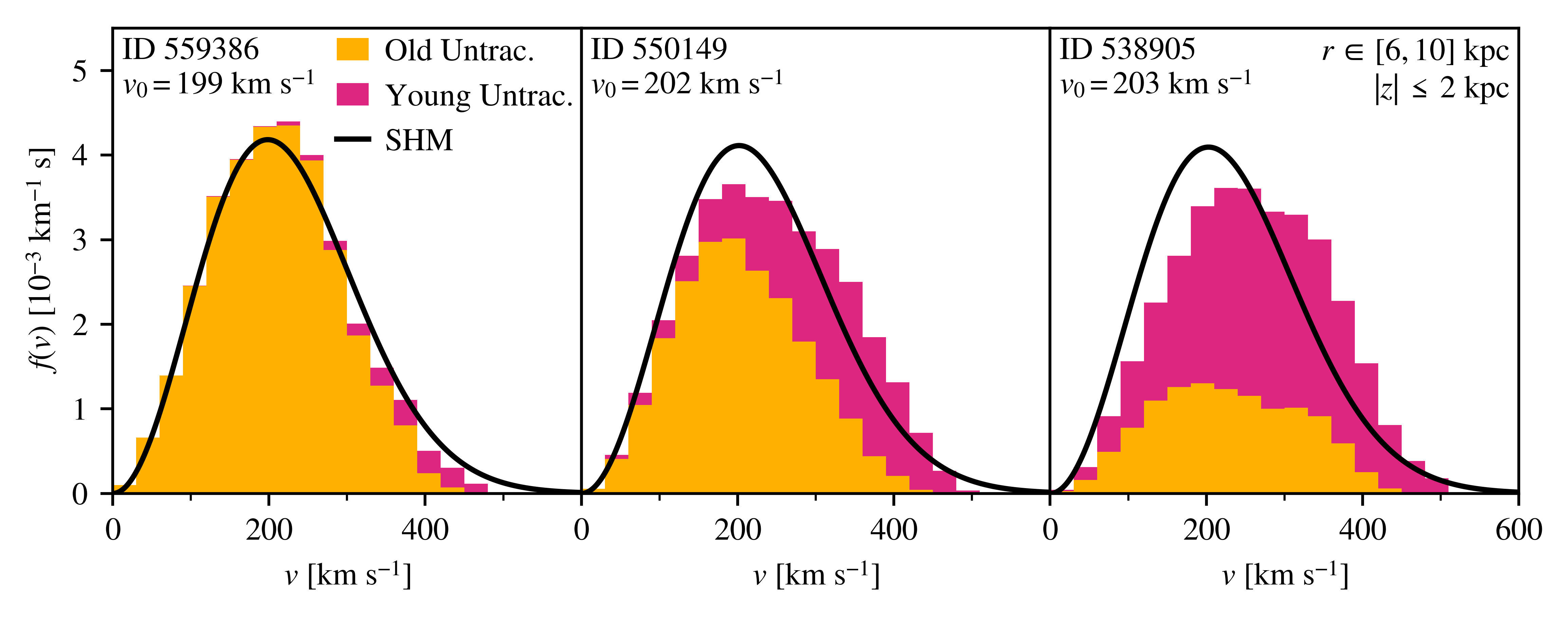}
    \caption{Stacked speed distributions of the Old Untraceable~(yellow) and Young Untraceable~(magenta) DM components in the solar annulus for three representative MW analogs (each containing one GSE-like Traceable merger) with varying Young Untraceable fractions. The SHM~(black curve) is shown for each halo as a Maxwell--Boltzmann distribution with scale velocity $v_0$ set by the mass enclosed within the solar radius. When the Young Untraceable fraction is small, as in the left panel, the combined distribution is well described by the SHM, yielding an EMD of 9~\kms{} between the SHM and the Untraceable DM. For the middle and right panels, the EMD values increase to 14~\kms{} and 28~\kms{}, respectively, reflecting the growing deviations from a Maxwell--Boltzmann shape as the Young Untraceable fraction becomes larger.} 
    \label{fig:2}
\end{figure*}
\begin{figure}
    \centering
    \includegraphics{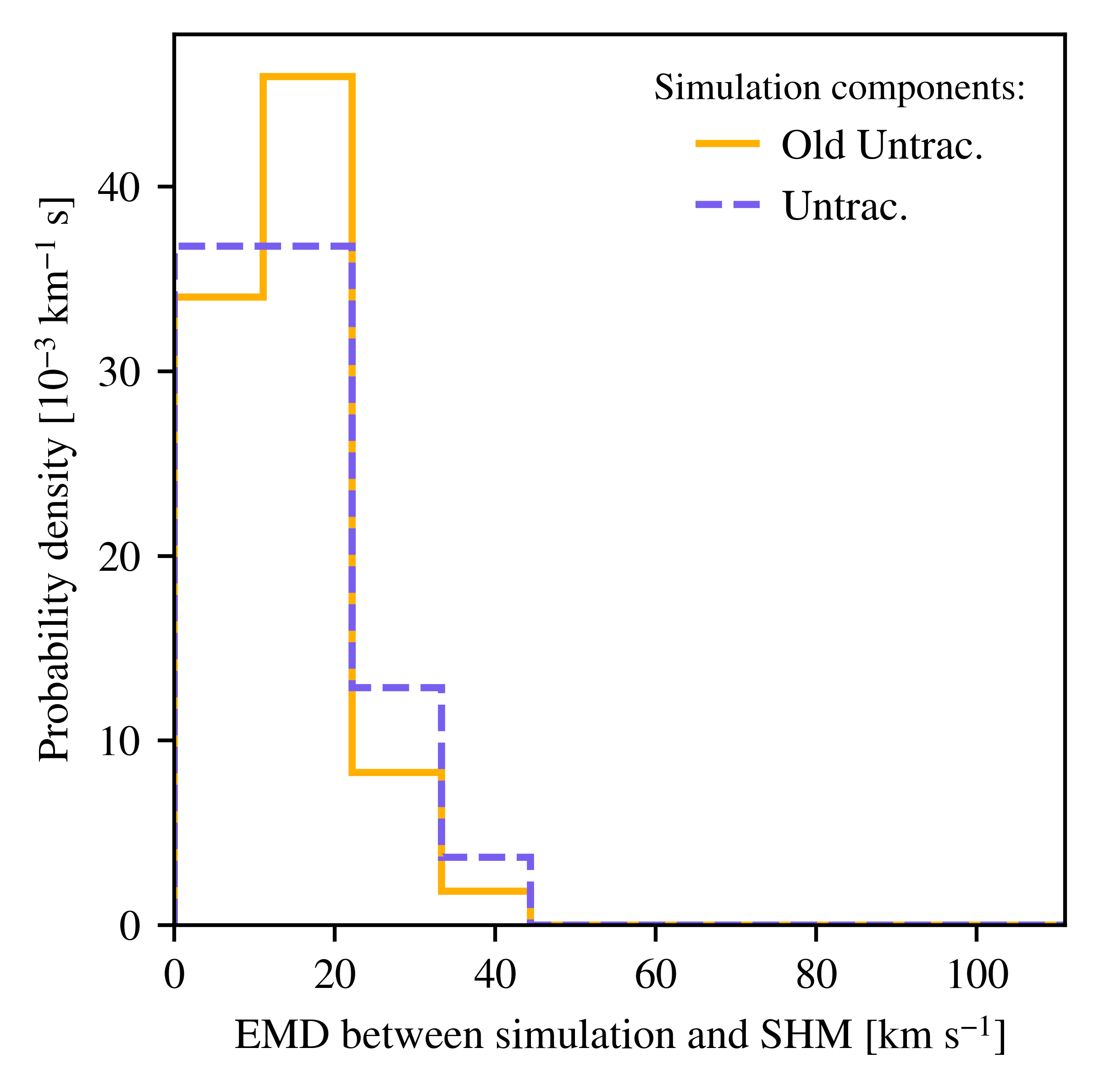}
    \caption{EMDs between the SHM and the speed distribution of the Old Untraceable DM~(solid yellow) or the Untraceable components combined~(dashed purple). The Old Untraceable component on its own is well described by a Maxwell--Boltzmann, with an EMD of $13^{+8}_{-6}~\kms{}$, while the Young Untraceable component alone (not shown in the figure) is not, lying $65^{+52}_{-32}~\kms{}$ from the SHM. However, since the Young Untraceable component is typically a small fraction of the DM in the solar annulus, combining the Old and Young Untraceable components together is still well modeled by the SHM, with an EMD of $13^{+13}_{-6}~\kms$.}
    \label{fig:3}
\end{figure}

The right panel of \autoref{fig:1} shows the speed distributions of the three components across all 98 MW analogs. In each bin, the solid line indicates the median value for the probability density $f(v)$ for each of the three components, individually normalized to an integral of unity, and the shaded region indicates the 16th--84th percentile range across the sample. The speed distributions for Traceable DM~(blue), Young Untraceable DM~(magenta), and Old Untraceable DM~(yellow) are shown. The speed  distributions of the Young Untraceable component exhibit large halo-to-halo variance compared to the other two components, and they are also biased toward higher values. However, it typically contributes a small fraction relative to the Old Untraceable contribution, so the speed distribution of the \emph{total} (Old and Young) Untraceable component does not typically exhibit these large non-Maxwellian features.

\subsection{Untraceable Components} \label{subsec:old_untraceable}
Given that the Untraceable components contain contributions from dark accretion, it is difficult to model them directly using stellar tracers. Therefore, we use a simple analytic parameterization to describe this DM distribution. We model the speed distribution of the combined Untraceable population in each halo with a Maxwell--Boltzmann profile, 
\begin{equation}
    f(v) = \operatorname{SHM}(v|v_0) = \frac{4v^2}{\sqrt{\pi}v_0^3}\exp\left(-\frac{v^2}{v_0^2}\right) \, ,
    \label{eq:maxwell}
\end{equation}
which we label as the standard halo model~(SHM).\footnote{While it is common to truncate the SHM at the galactic escape speed,  we do not implement such a truncation for simplicity.} The free parameter of the model, $v_0$, is estimated as
\begin{equation} \label{eq:v0}
    v_0 = \sqrt{\frac{GM(<R_\odot)}{R_\odot}} \, ,
\end{equation}
where $G$ is Newton's constant and $M(<R_\odot)$ is the mass enclosed within a radius $R_\odot=8$~kpc, the midpoint of the ROI. We choose this value to correspond to the predicted circular speed in the halo at the solar radius, assuming a spherically symmetric gravitational potential.

To quantify the agreement between the exact speed distribution and this parameterization, we use the Earth mover's distance~(EMD), 
\begin{equation}
    \operatorname{EMD}(f_1,f_2) = \int\!\left| F_1(v) - F_2(v) \right|\ \mathrm{d}v \,,
\end{equation}
where $f_1$ and $f_2$ are speed distributions and $F_1$ and $F_2$ are their respective cumulative distribution functions. Intuitively, the EMD represents the minimum ``work'' required to transform one probability distribution into the other, where ``work'' is the amount of probability that must be moved times the distance it is moved in speed space. The EMD, therefore, has units of speed. This metric is  indicative of shifts in the mean value of the speed distribution, where the probability density is highest, and should be understood as a rough measurement of the similarity of the distributions. In particular, the EMD is not very sensitive to changes at the high-speed tail, as the work required to shift this portion of the speed distributions is proportional to the probability density there. 

\autoref{fig:2} shows the sum of the Old and Young Untraceable speed distributions for three example MW analogs with increasing fractions of Young Untraceable DM~(3\%, 25\%, and 52\%, respectively).\footnote{Note that the first and last cases are outliers in terms of their Young Untraceable fractions; the second case is more typical for the one Traceable merger population.} The speed distributions for the Old Untraceable DM~(yellow) and Young Untraceable DM~(magenta) populations are stacked such that the top edge of the filled histogram is the distribution for the total Untraceable population. The SHM~(solid black) is shown for each halo as a Maxwell--Boltzmann distribution with characteristic speed $v_0$ set by \autoref{eq:v0}. For the leftmost panel, which has the smallest Young Untraceable fraction, the combined Old and Young Untraceable distribution closely matches the SHM, with an EMD between the two of 9~\kms{}. For the two other panels, with intermediate and high Young Untraceable fractions, the SHM does a poorer job of modeling the distribution, and the EMD increases to 14~\kms{} and 28~\kms{}, respectively.

\begin{figure*}
    \centering
    \includegraphics{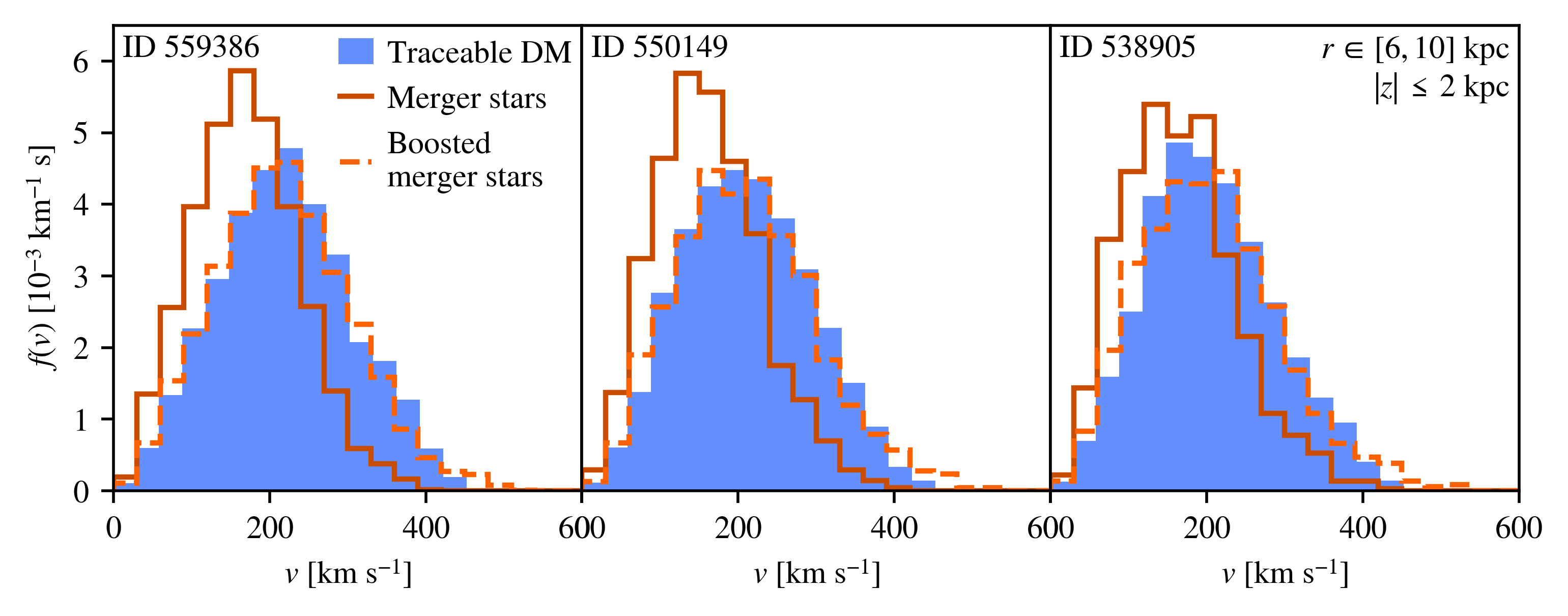}
    \caption{Speed distributions of Traceable DM~(filled blue) and stars accreted from the same merger~(solid orange) for the Traceable merger in each of the three MW analogs from \autoref{fig:2}. In each case, the stars underpredict the DM speed. To correct for this, we boost the dispersion of the stellar tracers according to \autoref{eq:boost}, giving the boosted stellar distribution (dashed orange). Before the boost, the EMDs between the DM and stellar speed distributions are 49, 48, and 40~\kms{} for the left, middle, and right panels, respectively. The boosted stellar distribution more closely traces the DM distribution, and the resulting EMDs are reduced to 4, 5, and 5~\kms{}, respectively.}
    \label{fig:4}
\end{figure*}

\autoref{fig:3} shows the EMD values between the SHM and the speed distribution of the Old Untraceable component alone~(solid yellow) and the combined distribution of Old and Young Untraceable components~(dashed purple), across all 98 MW analogs. The Old Untraceable component, which is likely virialized and well phase mixed, has an EMD of $13^{+8}_{-6}~\kms{}$ relative to the SHM, with the largest deviation being 43~\kms{}. The Young Untraceable component alone (not shown in the figure) deviates much more strongly from the SHM parameterization, with an EMD of $65^{+52}_{-32}~\kms{}$, with the most discrepant halo 230~\kms{} from the SHM. When the Untraceable components are taken together, the SHM remains a good model; the distributions are $13^{+13}_{-6}~\kms{}$ from the SHM, with a maximum discrepancy of 41~\kms{}. These distances are similar to the distance between the SHM and the Old Untraceable component alone, indicating that the typically small Young Untraceable fraction in most MW analogs does not significantly affect one's ability to model the combined Untraceable populations of DM with the SHM.

\subsection{Traceable Component} \label{subsec:traceable}

\begin{figure*}
    \centering
    \includegraphics{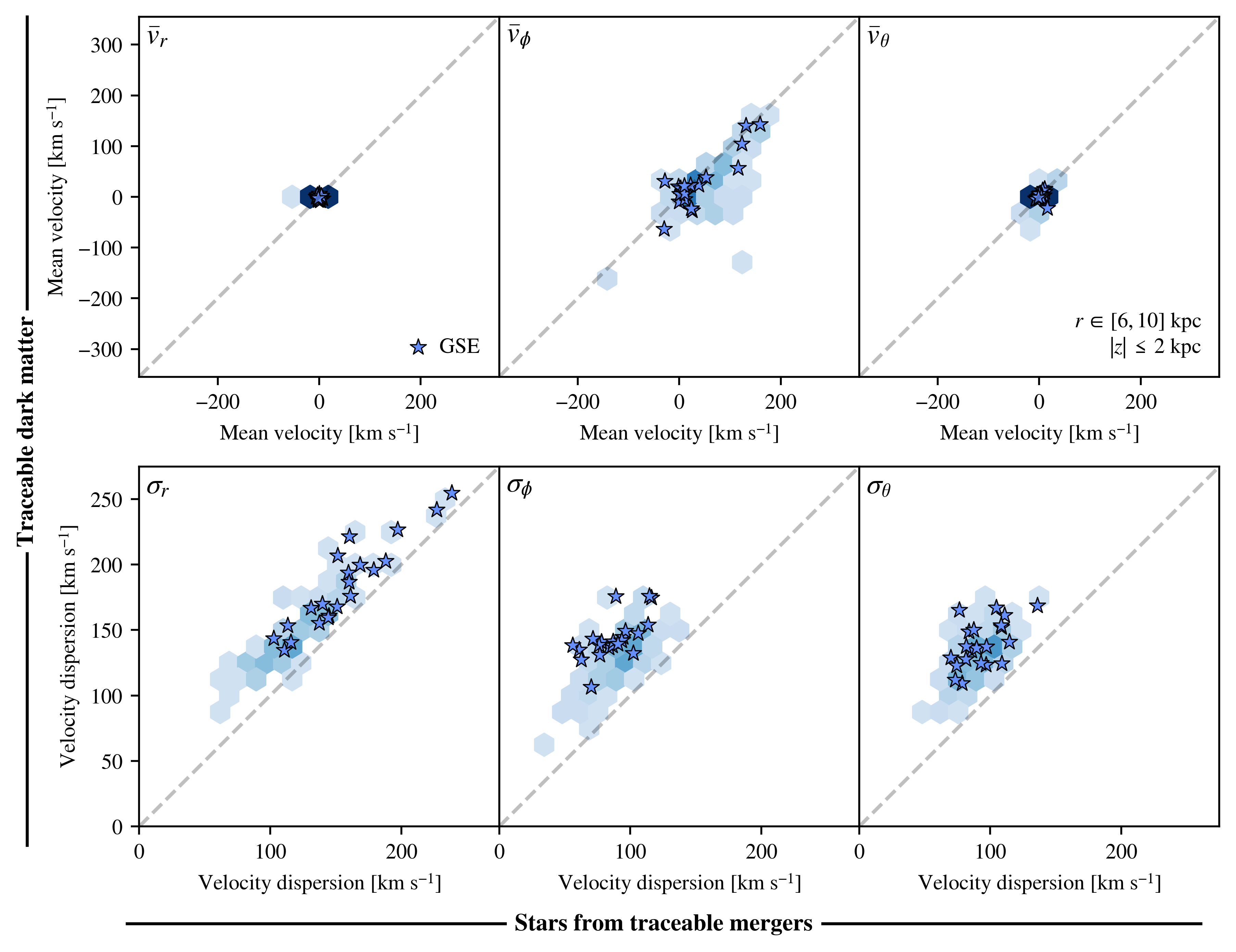}
    \caption{Correlation between stellar and DM kinematics for the 108 Traceable mergers. The top row shows the mean galactocentric velocities $\bar{v}_i$ for each of the spherical components, $i\in \{r,\phi,\theta\}$, while the bottom row shows the velocity dispersion in each of these components. In each panel, the horizontal axis shows the value for the stars contributed by each merger, while the vertical axis shows the value for the DM contributed by the same merger. Equality is indicated by the dashed gray line, such that probability density along this line indicates good agreement between the DM and the stellar tracers. The probability density across the full sample of mergers is shown in the background~(blue, with darker indicating higher probability), with the GSE-like mergers highlighted using star markers. Across the sample, the DM and stars have similar mean velocities, with a difference in $\bar{v}_i$ of $(-1^{+5}_{-3}, -15^{+16}_{-34}, 0^{+4}_{-6})~\kms{}$ for $(v_r, v_\phi, v_\theta)$. The velocity dispersions, on the other hand, are shifted: the DM exhibits an offset of $\Delta\sigma = 34_{-11}^{+10}~\kms{}$ with respect to the stars, for $\Delta\sigma$ the directionally averaged difference in dispersions (\autoref{eq:delta_sigma}). 
    The dispersions are particularly discrepant ($\Delta \sigma = 43^{+11}_{-10}~\kms{}$) for the GSE-like mergers, which are chosen based on their radially biased stars and have lower stellar tangential dispersions relative to the overall population of mergers. The offset in dispersions biases the stellar tracers to predict lower speeds for the DM, as seen in \autoref{fig:4}, which is corrected for by \autoref{eq:boost}.}
    \label{fig:5}
\end{figure*}

We next consider the DM and stellar debris deposited by Traceable mergers. \autoref{fig:4} compares the speed distributions of Traceable DM~(blue) and the corresponding stars~(solid orange) for the same three MW analogs shown in \autoref{fig:2}. In all three halos, the DM is peaked toward higher speeds than the stars, indicating that the stellar speeds provide a biased model for the Traceable DM distributions. Therefore, we develop a procedure for modifying the stellar speeds to better trace the DM. 

To assess this discrepancy within the full sample of MW-like galaxies, \autoref{fig:5} compares the mean velocities ($\bar{v}_i$, in the top row) and velocity dispersions ($\sigma_i$, in the bottom row) for each spherical component $i\in\{r, \phi, \theta\}$ of the debris deposited by the Traceable mergers. In each panel, the quantity for the Traceable DM is shown on the vertical axis, while the corresponding quantity for the stars is shown on the horizontal axis. The distribution across all 108 Traceable mergers is shown in the background~(blue), with the GSE-like mergers~(star markers) highlighted.

Generally, the mean velocity of the Traceable DM ($\bar{v}^\mathrm{DM}$) is in agreement with the mean velocity for the stars contributed by the same merger ($\bar{v}^\star$), with $\bar{v}^\star - \bar{v}^\mathrm{DM} = (-1^{+5}_{-3}, -15^{+16}_{-34}, 0^{+4}_{-6})~\kms{}$ for the components $(v_r, v_\phi, v_\theta)$. In contrast, the velocity dispersions for the DM ($\sigma^\mathrm{DM}$) are systematically higher than those of the stars ($\sigma^\star$), for which we see $(\sigma^\text{DM} - \sigma^\star) = (27^{+10}_{-11}, 38^{+17}_{-14}, 38^{+14}_{-12})~\kms{}$ for $(\sigma_r, \sigma_\phi, \sigma_\theta)$. This is especially true for the GSE-like mergers, which select for radially biased stellar distributions. For these mergers, the $\sigma_\phi^\star$ tends to be much smaller than the $\sigma_\phi^{\mathrm{DM}}$, even more so than the typical merger.

The systematic offset arises from differences in how quickly the two components are stripped from the infalling merger. The stars, which are deeper in a satellite's potential well, tend to get stripped later in the merging process and are deposited toward the center of the MW. The associated DM, on the other hand, is stripped from the infalling merger sooner and deposited at higher orbital energies. Especially for late-time mergers, when the MW's potential well is deepest and allows for such large differences in orbital energy, this biases the DM speed distribution to higher speeds relative to stars from the same merger, resulting in systematically broader velocity dispersions for the DM. This effect is studied in more detail in \autoref{appendix:boost}.
Due to these systematic discrepancies in velocity dispersions, we implement a boost to the stellar velocities to increase the observed dispersion while maintaining a fixed mean. The resulting boosted velocities, $v^\text{b}$, better trace the DM speed distribution.

For each Traceable merger, we define a one-parameter boosted stellar velocity as follows:
\begin{gather}
    v_i^{\mathrm{b}} = \frac{\Delta\sigma + \sigma_i^\star}{\sigma_i^\star}\left( v_i^\star-\bar{v}_i^\star \right) + \bar{v}_i^\star \, ,\label{eq:boost}\\
    \qquad\text{with} \, \Delta\sigma = \frac{1}{3}\sum_i \left( \sigma^\text{DM}_i - \sigma^\star_i \right)\label{eq:delta_sigma}
\end{gather}
and $i\in\{r,\phi,\theta\}$. This boost keeps the mean value for each component of the stellar velocity fixed, but increases the dispersion by $\Delta\sigma = 34_{-11}^{+10}~\kms{}$, the characteristic offset between DM and stellar velocity dispersions.
The DM and stellar debris from these mergers frequently corotates with the stellar disk, with mean $v_\phi$ displaced from zero toward positive velocities. As a result, simply increasing the dispersions without taking care to maintain a fixed mean would result in poorer agreement between the DM and boosted stars.

The corresponding speed distributions for the boosted stars are shown in \autoref{fig:4}~(dashed orange), and the agreement is evident. While the EMDs between the uncorrected stellar speed distributions and the DM speed distributions are 49, 48, and 40~\kms{} for the left, middle, and right panels, respectively, the boosted stars are only 4, 5, and 5~\kms{} away from the corresponding DM distributions, respectively.

\begin{figure}
    \centering
    \includegraphics{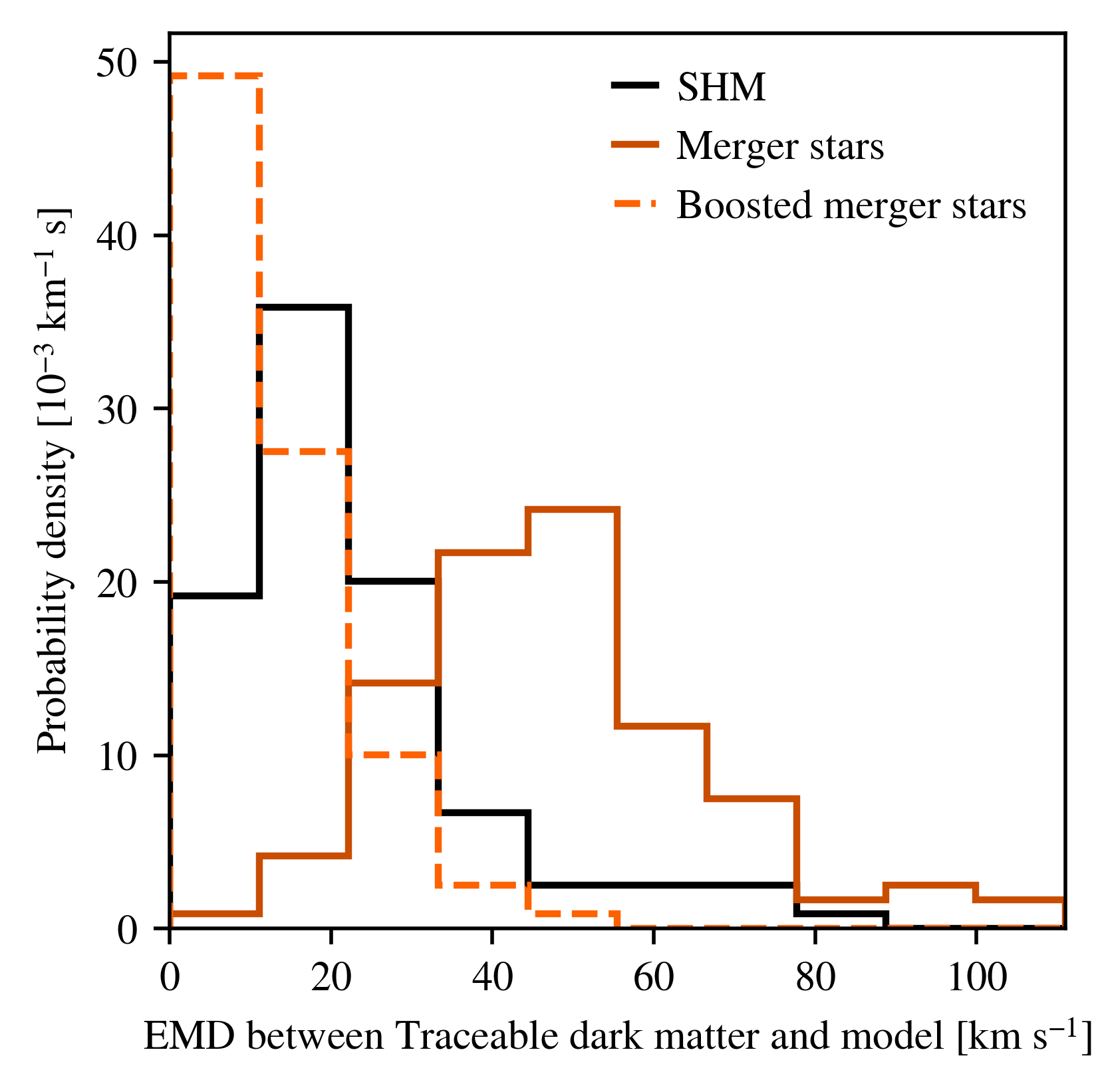}
    \caption{EMDs between the speed distributions of Traceable DM and the stars accreted from the same merger, across all 108 Traceable mergers. The EMD between the DM and the uncorrected stellar distributions~(solid orange) is $47^{+19}_{-15}~\kms{}$. The EMDs, after applying the boost described in \autoref{eq:boost} to the stellar velocities, are shown as Boosted stars~(dashed orange). The boost reduces the EMD to $10^{+11}_{-5}~\kms{}$, indicating that the boosted stars serve as better tracers for the underlying DM speed distribution. The boosted stellar tracer model is also better than the SHM (solid black), which has an EMD of $18^{+16}_{-8}~\kms$ from the DM speed distribution.
    }
    \label{fig:6}
\end{figure}

\autoref{fig:6} shows the result of this boost across the sample of 108 Traceable mergers. The EMD between the DM speed distributions and the uncorrected stars~(solid orange) is $47^{+19}_{-15}~\kms{}$, with a maximum EMD of 111~\kms{}. The EMD between the DM speed distributions and the boosted stars (dashed orange) is shown, and the agreement is much better, as the EMDs are reduced to $10^{+11}_{-5}~\kms{}$ with a maximum of 55~\kms{}. These results show that the boosted stellar distribution is a good model for the Traceable DM speeds. Indeed, it is an improvement upon the SHM (solid black), which exhibits an EMD of $18^{+16}_{-8}~\kms$ to the Traceable DM speed distribution.

\begin{figure*}
    \centering
    \includegraphics{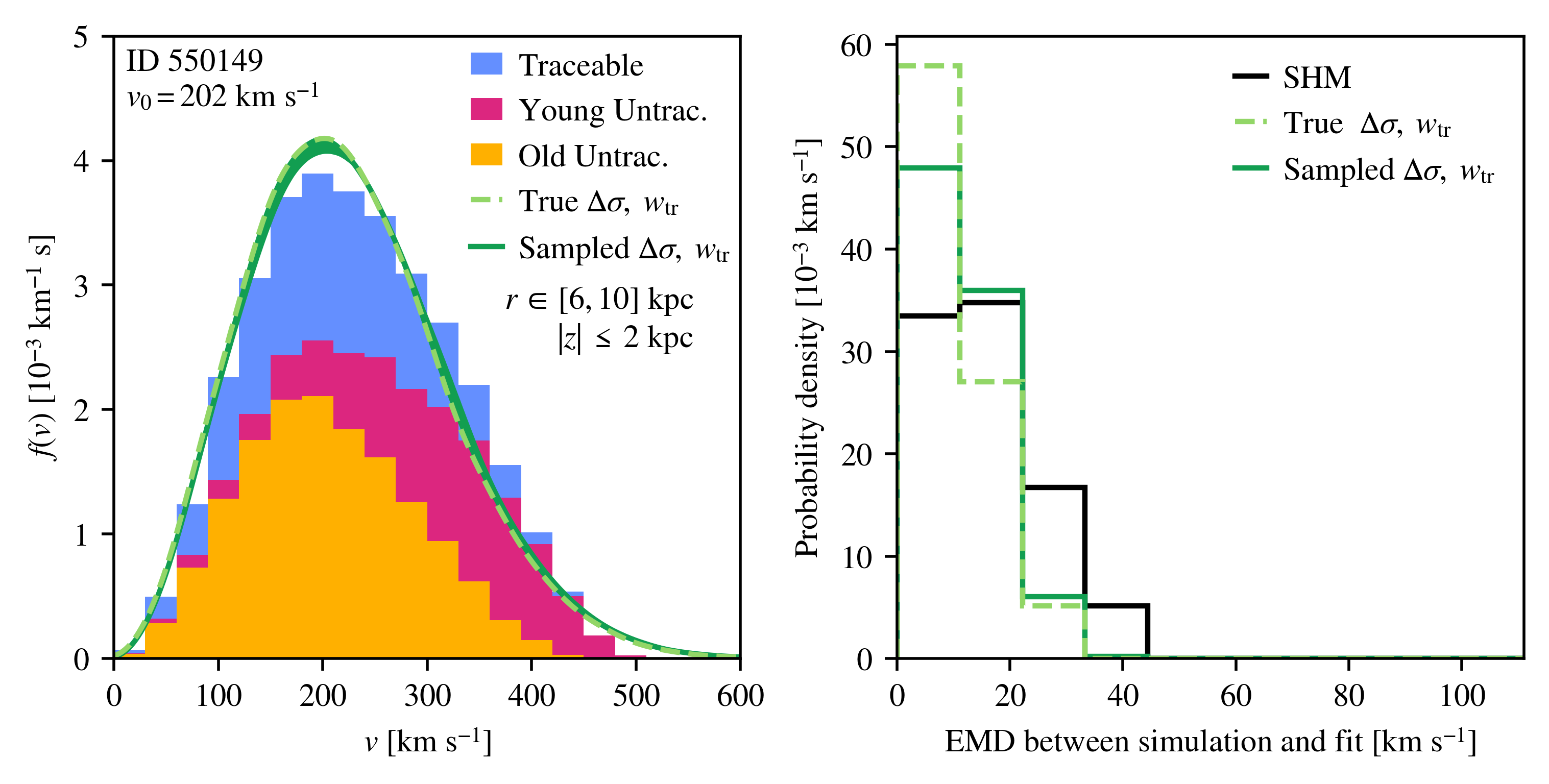}
    \caption{Full reconstruction of the local DM speed distribution. (Left) Speed distributions of the Traceable~(blue), Young Untraceable~(magenta), and Old Untraceable~(yellow) components in the solar annulus for an example MW analog, stacked such that the outer envelope is the total exact DM speed distribution. The reconstructions are comprised of a Maxwell--Boltzmann distribution and an empirical distribution derived from accreted stars according to \autoref{eq:model}. The reconstruction that uses the true values for the Traceable DM fraction (\ftr) and the boost applied to the stellar tracers~($\Delta\sigma$, \autoref{eq:boost}) is shown in light dashed green and has an EMD of 10~\kms{} from the exact speed distribution. The solid dark green band indicates the 16th--84th percentile range across all reconstructions, sampling over the uncertainty induced by varying these parameters. Even with this uncertainty, the sampled distributions closely match the exact DM distribution, with an EMD between them of $8^{+3}_{-2}~\kms{}$. The SHM alone (not shown here) aligns well with the reconstructions for this example. (Right) Distribution of EMDs between the exact speed distributions and the reconstructions across all 70 MW analogs with Traceable mergers, for models using the true values of $\Delta\sigma$ and \ftr{} (dashed light green) and those that sample the uncertainty in these parameters (solid dark green). The true-value models are $10^{+6}_{-4}~\kms{}$ away from the exact speed distributions, while accounting for uncertainty in the parameters modestly inflates these EMDs to $11^{+6}_{-4}~\kms{}$. These results indicate that the model provides a good fit, even when accounting for uncertainty in the parameters. We also compare the exact distributions to the SHM alone~(solid black), which yields an EMD of $14^{+15}_{-6}~\kms{}$. This shows that the SHM provides a reasonable overall approximation, but the inference of the speed distribution improves when including a model of the Traceable component.} 
    \label{fig:7}
\end{figure*}

\subsection{Total Local DM} \label{subsec:total}

With an understanding of the Untraceable DM~(\autoref{subsec:old_untraceable}) and the Traceable DM~(\autoref{subsec:traceable}) components, we are equipped to reconstruct the total DM speed distribution in the ROI. We highlight two parameters for this reconstruction that are difficult to obtain observationally: (i)~the boost factor to correct the distribution of stellar tracers, $\Delta\sigma$ (\autoref{eq:boost}), and (ii)~the fraction of DM in the solar neighborhood traced by these stars, \ftr{}. To illustrate the impact of the uncertainties in these parameters, we perform two versions of the analysis, one in which these values are fixed to the truth for each Traceable merger and one in which they are sampled from the distribution of values extracted from TNG50. 

We neglect uncertainty due to the determination of $v_0$~(\autoref{eq:v0}), as it is well constrained for the MW~\citep{2021EPJC...81..907B}. In the case of analogs with multiple Traceable mergers, we further neglect the uncertainty in the relative weight of each and instead divide the fraction \ftr{} according to the total stellar mass of each merger at infall, $m_\star$, such that each Traceable merger is given a weight $\ftr{}\,m_\star/\Mstar$, for $\Mstar$ the sum of all $m_\star$ values.
In general, the inferred relative weight $m_\star/M_\star$ is $14^{+23}_{-6}$~percentage points from the true value. If the relative weight is instead determined by the stellar mass within the ROI from each merger, the inferred relative weight is $19^{+31}_{-11}\%$~points from the true fraction. This suggests that using $m_\star/M_\star$ at infall provides a reasonable approximation for each merger's contribution to \ftr{}. The discrepancy induced by this approximation is modest relative to the uncertainty arising from the dispersion offset between DM and stars (\autoref{subsec:traceable}) and does not dominate the reconstruction. In principle, the stellar mass--halo mass relation and differences in tidal stripping as the mergers orbit their host will change the true value of this fraction, and we refer the interested reader to \citet{2019ApJ...883...27N} for a more careful study of this relative weighting.

The total speed distribution $f_\mathrm{tot}(v)$ is constructed as a weighted sum of the modeled components, 
\begin{equation}\label{eq:model}
    f_\mathrm{tot}(v) = (1-\ftr)\ \operatorname{SHM}(v | v_0) + \ftr\hspace{-1em}\sum_{\#\ \mathrm{mergers}}\frac{m_\star}{\Mstar} f^\mathrm{b}(v) \, .
\end{equation}
The first term corresponds to the Maxwell--Boltzmann distribution of \autoref{eq:maxwell} with $v_0$ set by \autoref{eq:v0}. The second term is a sum of 
the speed distributions $f^\mathrm{b}(v)$ for each Traceable merger, which are in turn constructed as a Gaussian kernel density estimation of the boosted stellar velocities, $v^\mathrm{b}_\star$, defined in \autoref{eq:boost}.

The left panel of \autoref{fig:7} shows this construction on an example halo, ID~550149, the MW analog presented in the central panels of \autoref{fig:2} and \autoref{fig:4}. The full speed  distribution is broken down into the three constituent components: Old Untraceable DM~(yellow), Young Untraceable DM~(magenta), and Traceable DM~(blue), such that the outer envelope of the histogram is the exact speed distribution, summed across all components. The inferred total speed distribution according to \autoref{eq:model}~(dashed green) is shown, with all parameters (\ftr{}, $v_0$, $m_\star$, and $\Delta\sigma$) fixed to their true values. The resulting distribution has an EMD of 10~\kms{} from the exact distribution. However, these parameters are not easily determined in realistic observational settings. As such, we wish to sample over possible values for them. To this end, we construct probability distributions for $\ftr$ and $\Delta\sigma$ from the simulation data. 

From the left panel of \autoref{fig:1}, it is clear that \ftr{} is sensitive to the number of Traceable mergers; in halos with more Traceable mergers, more DM is contributed by Traceable mergers. Therefore, the probability distribution for \ftr{} used in this reconstruction is equal to the distribution observed in the simulation conditioned on the number of Traceable mergers (i.e., the marginal distribution shown at the top edge of the figure). The halo in \autoref{fig:7} has a single Traceable merger, so \ftr{} is sampled from the $18^{+15}_{-5}\%$ range corresponding to this distribution, constructed from the 37 single-merger MW analogs.

As for $\Delta\sigma$, the sampled distribution is the $43_{-10}^{+11}~\kms$ range obtained for the 26 GSE-like mergers, as the Traceable merger in this halo is itself GSE-like. Were the merger not GSE-like, we would use the full sample of 108 mergers, which has  $\Delta\sigma$ in the range $34_{-11}^{+10}~\kms$. Since there are no additional Traceable mergers in this halo, $m_\star/\Mstar = 1$, and the only contribution to \ftr{} comes from this GSE-like merger.

To account for the uncertainty in these two parameters, we consider each of the $37\times 26 = 962$ combinations of \ftr{} and $\Delta\sigma$ and construct an inferred speed distribution from every choice. This is a maximally conservative approach, since it does not assume any correlation between the two. At each point along the horizontal axis in the left panel of \autoref{fig:7}, the median probability density across these distributions (solid green) is shown, and the 16th--84th percentile range (shaded green) is indicated as a band. The EMD for these reconstructions is $8^{+3}_{-2}~\kms{}$, comparable to the 10~\kms{} found using the true values of the parameters.

The right panel of \autoref{fig:7} shows the results of this reconstruction across the subsample of 70 MW analogs with Traceable mergers.\footnote{The 28 analogs with no Traceable mergers have an EMD of $10^{+5}_{-4}~\kms{}$ from their reconstruction, though this reconstruction consists only of a Maxwell--Boltzmann component, with no stellar tracer component.} The EMD between the exact speed distributions and the reconstructions performed with all parameters fixed to their true values (dashed green) is $10^{+6}_{-4}~\kms{}$. For each halo, we also sample over the uncertainties in \ftr{} and $\Delta\sigma$ as we did for ID~550149. As in the worked example, we set the possible values of \ftr{} equal to the values produced by the halos with the same number of Traceable mergers. Further, when modeling a GSE-like merger, the $\Delta\sigma$ is sampled from the values for all GSE-like mergers; otherwise, it is sampled from the full distribution of $\Delta\sigma$ values. For each MW analog, this gives a range of EMDs corresponding to the distance between the exact speed distribution and the $f_\mathrm{tot}(v)$ model. The choice of $\Delta\sigma$ and $\ftr$ does not strongly impact the recovered EMDs; the distribution of recovered EMDs in each galaxy typically has a width of $\osim2$--$4$~\kms{}  (16th--84th percentile). 
Sampling over all reconstructions\footnote{Note that the number of possible reconstructions varies from analog to analog. We have ensured that each analog contributes the same amount of probability density to the Sampled curve in \autoref{fig:7}.} for all 70 MW analogs with traceable mergers yields the Sampled EMD distribution~(solid green) in \autoref{fig:7}, which has a typical EMD of $11^{+6}_{-4}~\kms{}$, providing a modestly worse---though still comparable---reconstruction than the true parameter values. Together, these two distributions indicate that the reconstruction is still well determined when the exact values of \ftr{} and $\Delta\sigma$ are unknown, suggesting that the method is applicable to observational data. 

For comparison, the right panel of \autoref{fig:7} also compares the exact speed distribution to the SHM alone (solid black). The SHM achieves a median EMD of $14^{+15}_{-6}~\kms{}$, which demonstrates that while it provides a reasonable overall approximation to the exact distributions, the  reconstruction does improve once the Traceable components are explicitly modeled. This is true even in halos with one Traceable merger, when the Young Untraceable fraction comprises a nonnegligible fraction of the DM in the ROI. For halos with especially large Young Untraceable fraction, the reconstructions become $\osim 10~\kms{}$ farther from the exact distribution because the Young Untraceable component is particularly non-Maxwellian. However, the uncertainty in the reconstruction parameters $(\Delta\sigma, \ftr{})$ is larger than the variation induced by the Young Untraceable DM; as such, the Sampled EMDs are largely insensitive to the Young Untraceable fraction.

\section{Discussion} \label{sec:discussion}

\begin{figure}
    \centering
    \includegraphics{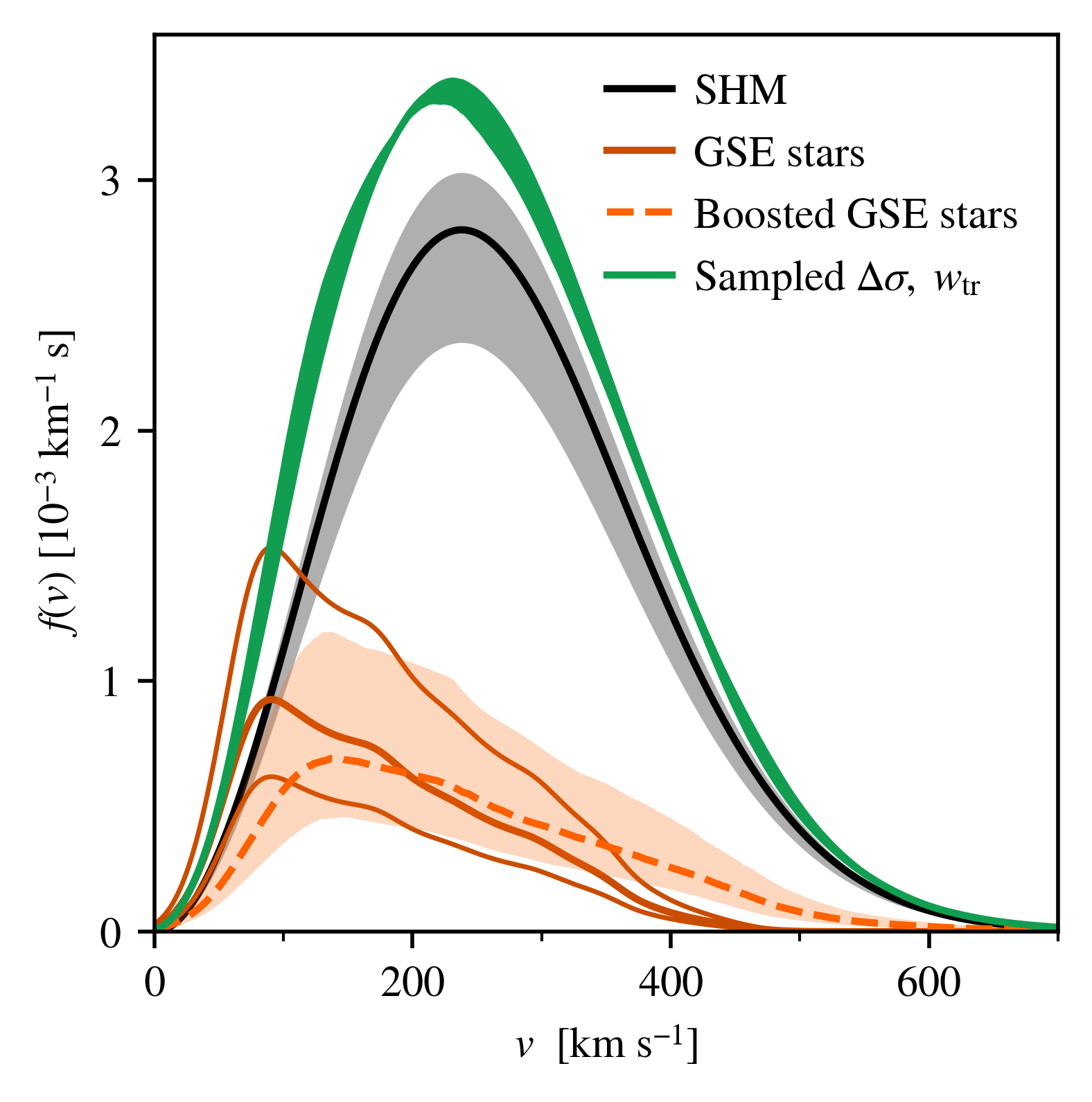}
\caption{Projected local DM speed distribution for the MW in the Galactocentric frame. The SHM with $v_0=238~\kms{}$ is shown~(solid black), as well as high-eccentricity ex situ stars selected from the \citet{2020A&A...636A..75O} catalog as a model for the GSE stellar debris~(solid orange). These GSE stars are boosted using the $\Delta\sigma$ values extracted from GSE-like mergers in TNG50~(dashed orange). Each of these speed distributions has further uncertainty induced by \ftr{}, the relative weighting of the SHM and the stellar tracers, taken from the single Traceable merger halos in the simulation. Summing these two components yields the full reconstructions~(solid green), which lie $6^{+5}_{-3}~\kms{}$ from an SHM-only model. 
    }
    \label{fig:8}
\end{figure}

This section applies the formalism developed in \autoref{sec:methods} and \autoref{sec:results} to our own Galaxy.  \autoref{fig:8} presents the reconstruction of the local DM speed distribution for the MW in the Galactocentric frame. As in \autoref{subsec:old_untraceable}, we describe the Untraceable DM with the SHM~(black), with scale velocity set to the local standard-of-rest speed $v_0=238~\kms{}$ \citep{2004ApJ...616..872R,2016ARA&A..54..529B,2021EPJC...81..907B,2021A&A...647A..59G}. We neglect uncertainty in this parameter, as it is estimated to be only 1.5~\kms{}~\citep{2021EPJC...81..907B}. The shaded gray band indicates uncertainty in the Untraceable DM fraction, described below.

Though many individual mergers are thought to contribute to the MW's ex situ stellar population, by far the largest contribution around the Sun is from the GSE merger \citep{2018Natur.563...85H,2018ApJ...856L..26M, 2020ApJ...901...48N}. Therefore, we take the GSE to be the only Traceable merger in our Galaxy and assume that no other merger contributes more than 10\% of the local DM. To model this component, we select GSE-contributed stars from the catalog of \citet{2020A&A...636A..75O}, which consists of stars from the Gaia DR2 dataset~\citep{2018A&A...616A...1G} flagged by a neural network as likely being ex situ based on their kinematics. 
To isolate the GSE population from this sample, we apply an eccentricity cut of $e>0.7$ and a spatial cut ($r \in [6,10]$~kpc, $|z|\leq 2$~kpc)~\citep{2020ApJ...901...48N}. 
We have tested alternative GSE selections and confirmed that this choice yields a stellar speed distribution consistent with the literature~\citep{2019PhRvD..99b3012E,2019ApJ...874....3N,2019MNRAS.486..378L,2020ApJ...903...25N,2020PhRvD.101b3006O,2024ApJ...974..167Z}. While we take this to be a first approximation to the speed distribution of the GSE debris, a more detailed clustering analysis of the \citet{2020A&A...636A..75O} stars in this region would refine this analysis by reducing contamination from, e.g., high-$\alpha$ disk stars, which may contribute a modest ($\osim 25~\kms$) prograde rotation in our GSE sample. See \autoref{appendix:gaia_GSE} for a more detailed discussion.

The resulting speed distribution for the GSE stars is shown in \autoref{fig:8}~(solid orange), where it is normalized by the weight parameter \ftr{} drawn from the distribution of single Traceable merger MW analogs in TNG50 (see \autoref{subsec:total}). The thinner lines indicate the 16th--84th percentiles of these samples. As described in \autoref{subsec:traceable}, we apply a correction factor to these stellar velocities to give a prediction for the GSE-contributed DM. The values of $\Delta\sigma$ are taken from the GSE-like mergers in TNG50. The boosted stellar distribution~(dashed orange), constructed using this boost, is shown with uncertainty from sampling over both $\Delta\sigma$ and \ftr{}.

The total reconstructed speed distribution, formed by combining the SHM and the boosted stellar component, is shown as a green band that samples over both model parameters, where at each point along the horizontal axis, we have shaded the 16th--84th percentiles of the reconstructions. These sampled distributions have an EMD of $6^{+5}_{-3}~\kms{}$ from the SHM, indicating that the model incorporating GSE stars is similar to the Maxwell--Boltzmann-only model. In contrast, constructing the total speed distribution without applying the boost to the stellar velocities yields an EMD of $17^{+11}_{-6}~\kms{}$, as the unboosted GSE stellar speeds are typically much slower than $v_0$. Even after the boost is applied, the GSE stars are generally slower than $v_0$, leading to a suppression of the probability on the high-speed tail from the $(1-\ftr{})$ weighting of the SHM. Due to this effect, the 95th percentile of DM speed is $467^{+5}_{-5}~\kms{}$ for our model, compared to $470~\kms{}$ for the SHM alone.

We can compare our final results to other predictions in the literature for the local DM speed distribution.  For example, \citet{2019ApJ...883...27N} and \citet{2024ApJ...974..167Z} sought to model the DM arising specifically from the luminous mergers, including the GSE, using ex situ stars as tracers. In the Galactocentric frame, we find that the stellar debris from the GSE is slower than the 238~\kms{} local standard-of-rest speed in the MW, with a mode stellar speed of 90~\kms{}---compared to 150~\kms{} and 170~\kms{} for \citet{2019ApJ...883...27N} and \citet{2024ApJ...974..167Z}, respectively. When we use the boosted stellar model, our mode speed increases to $140_{-12}^{+13}$~\kms{} (see \autoref{fig:8}), comparable to the previously published results.

Because we include the contribution of dark accretion in the total DM distribution, we can compare the final result to the SHM$^{++}$ model~\citep{2019PhRvD..99b3012E}, which adds an analytic model for the GSE to the SHM. \cite{2019PhRvD..99b3012E} estimate that the DM fraction in the solar neighborhood due to the GSE is $\osim10\%$--30\%, consistent with the prediction for \ftr{} (13\%--33\%) that we obtain from TNG50. In the Galactocentric frame, their GSE model has larger velocity dispersions than our reconstruction (see \autoref{fig:14}), but its speed distribution is still peaked below the local standard-of-rest speed. We note that the high dispersion of their GSE model results in a total DM distribution in the heliocentric frame that is slightly faster than predicted for the isotropic SHM. In contrast, our total distribution is peaked below the SHM when boosted to the heliocentric frame.

\section{Conclusions} \label{sec:conclusions}
In this paper, we have studied how to empirically build a model for the local distribution of DM speeds. Using a sample of 98 MW analogs from the TNG50 simulation, we demonstrated how to model the contribution of DM from both dark accretion and luminous mergers in the solar neighborhood, defined as the cylindrical shell $r\in[6,10]$ kpc and $|z|\leq 2$~kpc.
Our results demonstrate that the contribution from dark accretion, as well as  early ($\zacc{} > 3$) luminous mergers, can be modeled as a Maxwellian.  Additionally, the stellar debris from recent and significant  luminous mergers can reliably trace the corresponding DM.

This is the first study showing how to properly treat the contribution of dark accretion to the solar neighborhood.  We accounted for the DM that originated from early-accretion events, as well as late-time accretion from smaller mergers. In general, the former has time to virialize and is, per se, well modeled today by a Maxwell--Boltzmann distribution.  The latter can exhibit more variation in its speed distributions.  However, in most cases, it is subdominant to the early-accretion events, such that the Maxwellian continues to be a good model for the total speed distribution of these two contributions, with typical EMDs of $13_{-6}^{+13}~\kms{}$ from the Maxwellian model. This confirms that this ``Untraceable'' component of the local DM can be modeled with this parameterization.

For the remainder of the DM, which comes from large mergers that deposit significant amounts of debris into the ROI, the speed distribution can be reconstructed using the kinematics of the stars from these mergers. However, the stellar debris from these larger mergers is systematically slower than the corresponding ``Traceable'' DM. This is because the DM is stripped from the infalling mergers first and is deposited at higher orbital energies. Despite this discrepancy, the velocities of the DM and stellar populations are strongly correlated. Applying a boost to its velocity dispersion (\autoref{eq:boost}) enables the stellar debris to serve as a good model for the non-Maxwellian component of the DM speed distribution, with an EMD of $10_{-5}^{+11}~\kms{}$ between the DM and boosted stars. 

We combined these results to reconstruct the total local DM speed distribution by adding the stellar-based model for the Traceable component to the Maxwell--Boltzmann model for the Untraceable background. As shown in \autoref{fig:7}, this reconstruction yields strong agreement with the exact distributions across all 70 MW analogs that have stellar tracers, even when accounting for uncertainty in the relative weighting of the Maxwell--Boltzmann and Traceable components of the model. Sampling over this uncertainty produces models with median EMDs of $11_{-4}^{+6}~\kms{}$ from the exact speed distribution. Importantly, the same agreement holds for the subset of analogs that experienced GSE-like mergers, indicating that our conclusions are consistent with expectations for the MW's accretion history.

Finally, in \autoref{sec:discussion}, we applied this approach to the MW itself. Tracers for the GSE DM are taken to be high-eccentricity ex situ stars from the catalog of \citet{2020A&A...636A..75O}. While the GSE stars themselves are slower than the SHM Maxwellian, with a median speed of $180~\kms{}$ (rather than the $\osim260~\kms{}$ expected from the local standard-of-rest velocity), applying our correction factor to the tracers' velocities yields a median speed of $\osim220~\kms{}$ for the GSE-contributed DM, with the exact value set by the correction factor used. As a result, incorporating the empirical speed distribution of the GSE stars only modestly modifies the prediction set by the SHM alone, at least for the DM speed. It also leads to a modest $\osim20\%$ suppression on the high-speed tail compared to the pure Maxwellian. The resulting models for the MW DM speed distribution are available online.\footnote{\url{https://github.com/Tal-Shpigel/stellar-dm-velocity-distributions}} The results of this work motivate improved reconstructions for the velocity distributions of GSE stars in the solar region, which can be used to continue improving the reconstruction of the local DM speed distribution.

This work focuses only on the DM speed distribution, rather than a directionally sensitive velocity distribution. This is sufficient for predicting the scattering rates in an isotropic detector material, but the directional information provided by the stellar tracers may be leveraged by an anisotropic DM detector. Indeed, \autoref{fig:5} suggests that many mergers deposit debris with nonzero azimuthal velocity, contrary to the typical assumption of isotropy. Implications of this will be explored in further work. 

This paper develops a set prescription for how to build an empirical speed distribution, verifying it with a large sample of MW-like systems from TNG50.  These results motivate further study to stress-test the procedure on increasingly large samples of MW-like systems with improved resolution, as well as variation on the subgrid feedback prescriptions.  The latter is likely an important source of systematic uncertainty that should be accounted for when using empirical speed distributions in direct detection analyses.  In terms of the overall methodology, the conclusions of our work are consistent with the prior work of \cite{2019ApJ...883...27N}, which studied two galaxies using FIRE-2 physics~\citep{2016ApJ...827L..23W}, as well as the results in the recent work by \cite{2026arXiv260325783Z}, which expands the sample size to five FIRE-2 MW-like simulations, studying the five mergers in each galaxy that contribute the most stars to the solar annulus. 

\citet{2026arXiv260325783Z} expands upon the prior work in FIRE-2 by considering 
the contribution of dark accretion.  They find---in agreement with our study---that the Untraceable component is well modeled by a Maxwellian distribution, even though the FIRE-2 halos form later than the halos here and have therefore not had as long for their DM to equilibrate. This may be due to the short dynamical timescales in the inner galaxy. 

We find a modest offset between the ex situ stellar velocity dispersions and the Traceable DM counterparts, on the order of a few tens of kilometers per second. This is in contrast with both \cite{2019ApJ...883...27N} and \cite{2026arXiv260325783Z}, who find a tighter correlation between the two distributions for GSE-like mergers. The difference between the FIRE-2 and TNG50 results can be attributed to multiple effects, both numerical and physical, which are challenging to disentangle at this stage.  

On the numerical side, the Latte suite of FIRE-2 has a higher resolution than TNG50, with over an order of magnitude lower mass per simulation particle. Results from \citet{2025arXiv251204157L} suggest that DM speed distributions around our ROI may be resolution-dependent, and it is yet unclear whether TNG50 has sufficient resolution for convergence. Further, \cite{2026arXiv260325783Z} find that the correlation between DM and stellar speed distributions degrades with lower resolution. Additionally, the work in FIRE-2 uses the \textsc{Rockstar} halo-finding algorithm for defining DM halos rather than \subfind{}, which may affect how well DM can be traced through the simulation history. 

Physically, the local environment of the two simulations differs, as the TNG50 cosmological volume holds two large clusters, while the FIRE-2 halos are simulated as zoom-ins with more isolated initial conditions. This denser environment likely biases the TNG50 halos to form earlier than those in FIRE-2, resulting in more concentrated halos at the present day. This concentration is exacerbated by the baryonic feedback physics, as FIRE-2 has a more aggressive prescription for baryonic feedback, which generally yields less dense halos. Both of these will affect tidal disruption of satellites that are orbiting the host systems and the resulting distribution of orbital energies for the DM and stars. 

In general, it is beneficial to vary over these possible systematics by considering many possible simulations. As the sample size of simulated galaxies increases, the connection between a merger's DM and stars will be probed in more depth and with increasing statistical power. Further, simulations performed at higher resolution will be able to identify smaller mergers than those considered here, allowing more of the MW's known mergers to be traced. Advancing hand-in-hand with these computational efforts, future observational surveys of the stellar halo, such as 4MOST~\citep{2019Msngr.175...23H}, WEAVE~\citep{2024MNRAS.530.2688J}, and the LSST~\citep{2019ApJ...873..111I}, will yield more information about the Galaxy's merger history. These efforts will ensure that a detailed empirical model can be borne out to further improve the precision with which one may infer the local DM speed distribution.

\needspace{2\baselineskip}
\section*{Acknowledgments}
We thank Carlos Blanco, Akaxia Cruz, Andreas Thoyas, and Xiuyuan Zhang for useful conversations. M.L. and D.F. are supported by the Department of Energy~(DOE) under Award No. DE-SC0007968. M.L. is also supported by the Simons Investigator in Physics Award. D.F. is additionally supported by the Joseph H. Taylor Graduate Student Fellowship. L.N. is supported by the Sloan Fellowship and the NSF CAREER award 2337864. This work was performed in part at the Aspen Center for Physics, which is supported by National Science Foundation grant PHY-2210452. The computations in this paper were run on the FASRC cluster supported by the FAS Division of Science Research Computing Group at Harvard University. The IllustrisTNG simulations were undertaken with compute time awarded by the Gauss Centre for Supercomputing~(GCS) under GCS Large-Scale Projects GCS-ILLU and GCS-DWAR on the GCS share of the supercomputer Hazel Hen at the High Performance Computing Center Stuttgart~(HLRS), as well as on the machines of the Max Planck Computing and Data Facility~(MPCDF) in Garching, Germany.

This report was prepared as an account of work sponsored by an agency of the United States Government. Neither the United States Government nor any agency thereof, nor any of their employees, makes any warranty, express or implied, or assumes any legal liability or responsibility for the accuracy, completeness, or usefulness of any information, apparatus, product, or process disclosed, or represents that its use would not infringe privately owned rights. Reference herein to any specific commercial product, process, or service by trade name, trademark, manufacturer, or otherwise does not necessarily constitute or imply its endorsement, recommendation, or favoring by the United States Government or any agency thereof. The views and opinions of authors expressed herein do not necessarily state or reflect those of the United States Government or any agency thereof.

\section*{Data availability}
All data used in this work are publicly available at \href{https://tng-project.org/}{https://tng-project.org/}. The inferred MW DM speed distributions presented in \autoref{fig:8} and \autoref{sec:discussion}, as well as a Jupyter Notebook to use these data, are available at \url{https://github.com/Tal-Shpigel/stellar-dm-velocity-distributions} and preserved on Zenodo at \dataset[doi:10.5281/zenodo.19024864]{\doi{10.5281/zenodo.19024864}}.

\section*{ORCID \MakeLowercase{i}Ds}
\noindent Tal Shpigel~\orcidlinkf{0009-0003-5629-5848}\par
\noindent Dylan Folsom~\orcidlinkf{0000-0002-1544-1381}\par
\noindent Mariangela Lisanti~\orcidlinkf{0000-0002-8495-8659}\par
\noindent Lina Necib~\orcidlinkf{0000-0003-2806-1414}\par
\noindent Mark Vogelsberger~\orcidlinkf{0000-0001-8593-7692}\par
\noindent Lars Hernquist~\orcidlinkf{0000-0001-6950-1629}
\onecolumngrid
\clearpage

\renewcommand{\thefigure}{A1}\begin{figure*}
    \centering
    \includegraphics{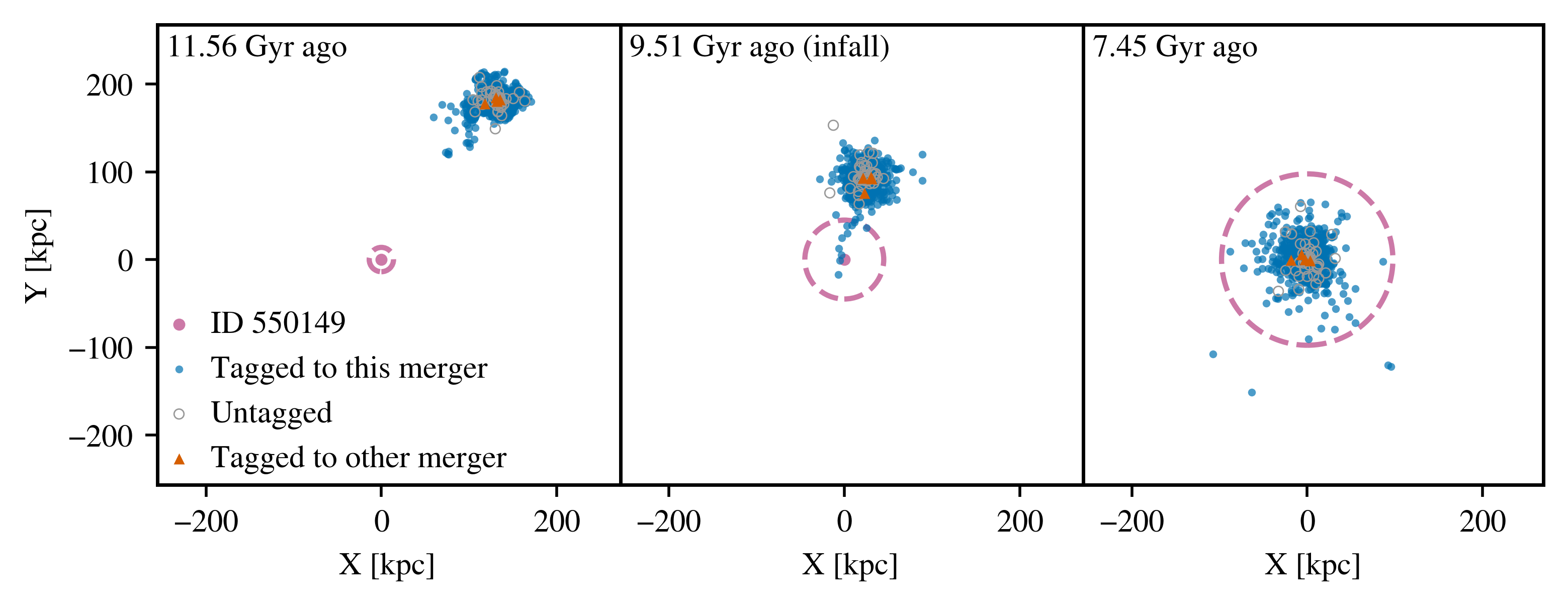}
    \caption{Spatial distribution of DM particles in MW 550149, focusing on the Traceable merger highlighted in the middle panel of \autoref{fig:4}. This merger's DM particles are shown 2 Gyr before infall (left), at infall (middle), and 2 Gyr after infall (right). All panels display the same pool of 58,541 DM particles that \subfind{} identifies as bound to the merger 2 Gyr before infall, subsampled such that they are more visible in this image. We apply the DM tracking algorithm to divide the particles into three categories: tagged to this merger (blue filled circles), untagged (gray unfilled circles), and tagged to another merger (orange triangles). The host MW is indicated by a point (purple) with its virial radius shown as a circle (dashed purple). Of the selected particles, 96\% are tagged to this merger, 4\% are not tagged to any merger, and $<0.1\%$ are tagged to another merger. This example demonstrates that the tracking procedure robustly associates the majority of DM particles with their correct merger, with very few left untagged or mistagged. Note that this figure assesses the completeness of the algorithm rather than the purity, as particles that are not bound to this merger 2~Gyr before infall are not shown.}
    \label{fig:9}
\end{figure*}

\twocolumngrid

\renewcommand{\thefigure}{A2}\begin{figure}
    \centering
    \includegraphics[width=\columnwidth]{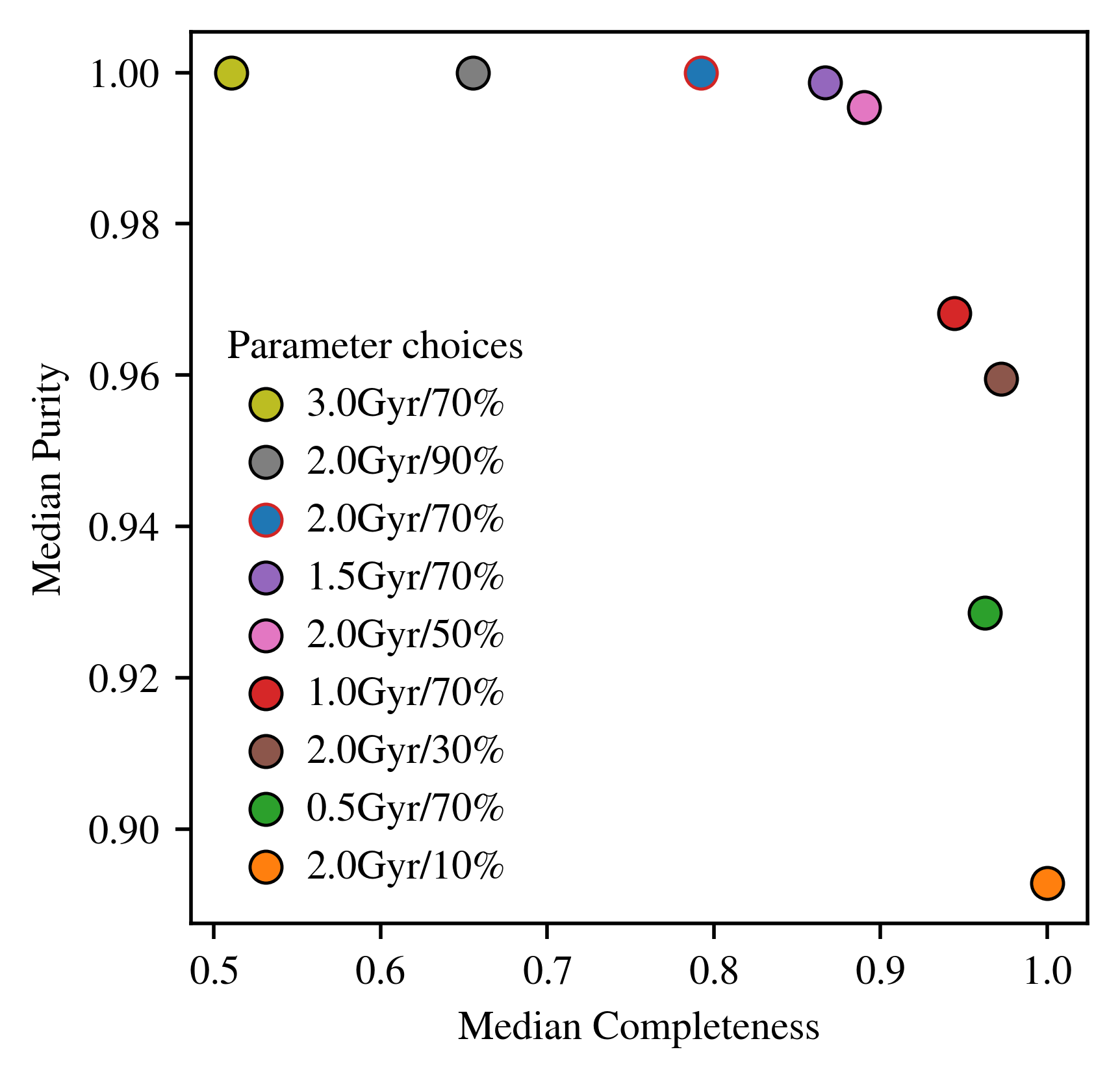}
    \caption{Median purity and completeness across all 98 MW analogs for nine combinations of merger-tracking parameters (see \autoref{subsec:mergers}). Each point shows the median values computed over all mergers with peak mass above $10^9~\Msun{}$. Completeness is defined as the fraction of particles bound to the merger before infall that are tagged to the merger, while purity is the fraction of particles tagged to the merger that are present in the halo before infall. The combinations vary the lookback time window (1--3 Gyr) and the required fraction of snapshots for which a DM particle must be bound to a subhalo~(50\%--90\%); they are labeled by these two parameters. The blue marker with red edge (2.0 Gyr/70\%) denotes the fiducial choice adopted in this work, which maximizes purity while maintaining a completeness above 0.75.}
    \label{fig:10}
\end{figure}

\appendix
\section{Merger Tagging Systematics} \label{appendix:merger_association}

This Appendix presents systematic checks of the DM--merger tagging algorithm introduced in \autoref{subsec:mergers}. Recall that the procedure identifies DM particles that are contributed by a given merger by considering a fixed time interval before the particle is bound to the MW, requiring that the particle remain bound to the subhalo for a sufficient fraction of the snapshots within that time window. Our fiducial choice of parameters is a time window of 2~Gyr, and we require that a particle be bound to its merger for 70\% of the snapshots within this window.

For reference, \autoref{fig:9} illustrates the performance of this fiducial choice for a single merger, showing the positions of a subsample of DM particles at different times relative to the merger's infall time (from left to right: 2 Gyr before infall, at infall, and 2 Gyr after infall), projected onto a coordinate system centered at the location of the MW. We show DM particles that are tagged to the merger~(blue filled circles), particles that are not tagged to any merger~(gray unfilled circles), and particles tagged to a different merger~(orange triangles). In this case, the algorithm correctly tags approximately 96\% of the particles that are bound to the subhalo 2~Gyr before infall, fails to tag about 4\%, and mistags fewer than 0.1\%. This example demonstrates that our fiducial algorithm correctly finds most of the merger debris and, importantly, does so with extremely low contamination. 

To quantify this more systematically, we consider the completeness and the purity of this choice of algorithm parameters. These are assessed  based on the \subfind{} halos 1~Gyr before infall:
\begin{enumerate}[wide=0pt, leftmargin=\parindent]
    \item \textit{Completeness} is the fraction of DM particles bound to the merger halo in any of the three snapshots 1~Gyr before the merger's infall (i.e., the snapshot closest to 1~Gyr before infall, plus or minus one snapshot) that are tagged to the merger by the algorithm.
    \item \textit{Purity} is the fraction of DM particles tagged to a merger that are bound to the merger in any of the three snapshots, 1~Gyr before infall (again, plus or minus one snapshot).
\end{enumerate}

Algorithms with low completeness are more likely to leave the least-bound DM as untagged. These loosely bound DM particles will generally be deposited at the highest orbital energies and have the highest present-day speeds. Therefore, an algorithm with low completeness will likely underestimate a merger's contribution to the high-speed tail. 
On the other hand, algorithms with low purity lead to reconstructions for a merger's DM speed distribution that are contaminated by particles from a different merger, resulting in unpredictable biases to the speed distribution. 
Ideally, the algorithm will have high completeness \emph{and} purity, but optimizing for one often comes at the cost of the other; more selective algorithms will have higher purity, but lower completeness, and vice versa for less restrictive algorithms. Since the low-completeness case introduces a bias that is easier to predict and understand, we prioritize purity, but still aim for high completeness in our choice of tagging algorithm. 

\autoref{fig:10} summarizes how the purity and completeness vary due to the choice of algorithmic parameters. We test nine different combinations of lookback time window (from 0.5 to 3.0 Gyr) and the required fraction of snapshots for which a DM particle must be bound to a subhalo (from 10\% to 90\%); each parameter point is labeled by these values. For each choice of these parameters, we calculate the purity and completeness across all mergers with peak masses above $10^9$~M$_\odot$ and report the medians.

As stated above, there is a trade-off between purity and completeness: relaxing the tagging criteria---by requiring that a particle be bound to a merger for fewer snapshots---raises completeness, since a larger fraction of a merger's particles are recovered. However, this comes at the cost of lower purity, as more particles are incorrectly tagged as belonging to the merger. Conversely, increasing the required bound fraction or extending the lookback time improves purity but excludes some legitimately bound particles, thus lowering completeness. The fiducial choice adopted in this work, 2.0 Gyr/70\%, provides an optimal balance along this trade-off, achieving a median purity of 0.99 and a completeness of 0.79. This combination ensures that nearly all tagged particles are truly bound to the merger, while retaining a sufficient sample size to characterize the merger's kinematics. 

\autoref{fig:11} tests the sensitivity of our reconstruction results to these parameter variations. The left panel shows the distribution of EMDs between the exact DM speed distribution and the sampled reconstruction~(the solid green line in the right panel of \autoref{fig:7}) for each parameter set. While some choices of algorithm parameters result in greater EMDs---indicating worse speed distribution reconstruction---our fiducial choice is among the best-performing algorithms, and the differences between our choice and the other best-performing choices are minimal.

The right panel of \autoref{fig:11} shows the distribution of the boost factor $\Delta\sigma$, as defined in \autoref{eq:delta_sigma}, for each parameter choice. These boosts are consistently biased toward $\osim30~\kms{}$, indicating that the offset is a real physical effect and not the result of our tagging procedure. 

Though these diagnostics are largely insensitive to the choice of parameters used, we caution that \ftr{} does exhibit fairly large variations. The TNG50 halos are rapidly assembling around a redshift of 3, which is 2~Gyr after the start of the simulation. Algorithms with a smaller time window than this threshold can resolve mergers earlier in the simulation history, meaning less DM is classified as Old Untraceable, and more is considered either Traceable or Young Untraceable. This is purely due to a choice in the threshold between ``Old'' and ``Young'' and does not meaningfully impact the physics presented in this work.

These results demonstrate that our reconstruction method is largely insensitive to the specific parameter choice. The fiducial 2.0 Gyr/70\% choice performs comparably to the best alternatives, producing low EMDs and stable boost factors while maintaining high purity and completeness. For these reasons, we adopt this configuration for the analysis presented in the main text.

\onecolumngrid
\clearpage

\renewcommand{\thefigure}{A3}\begin{figure*}
    \centering
    \includegraphics[width=\linewidth]{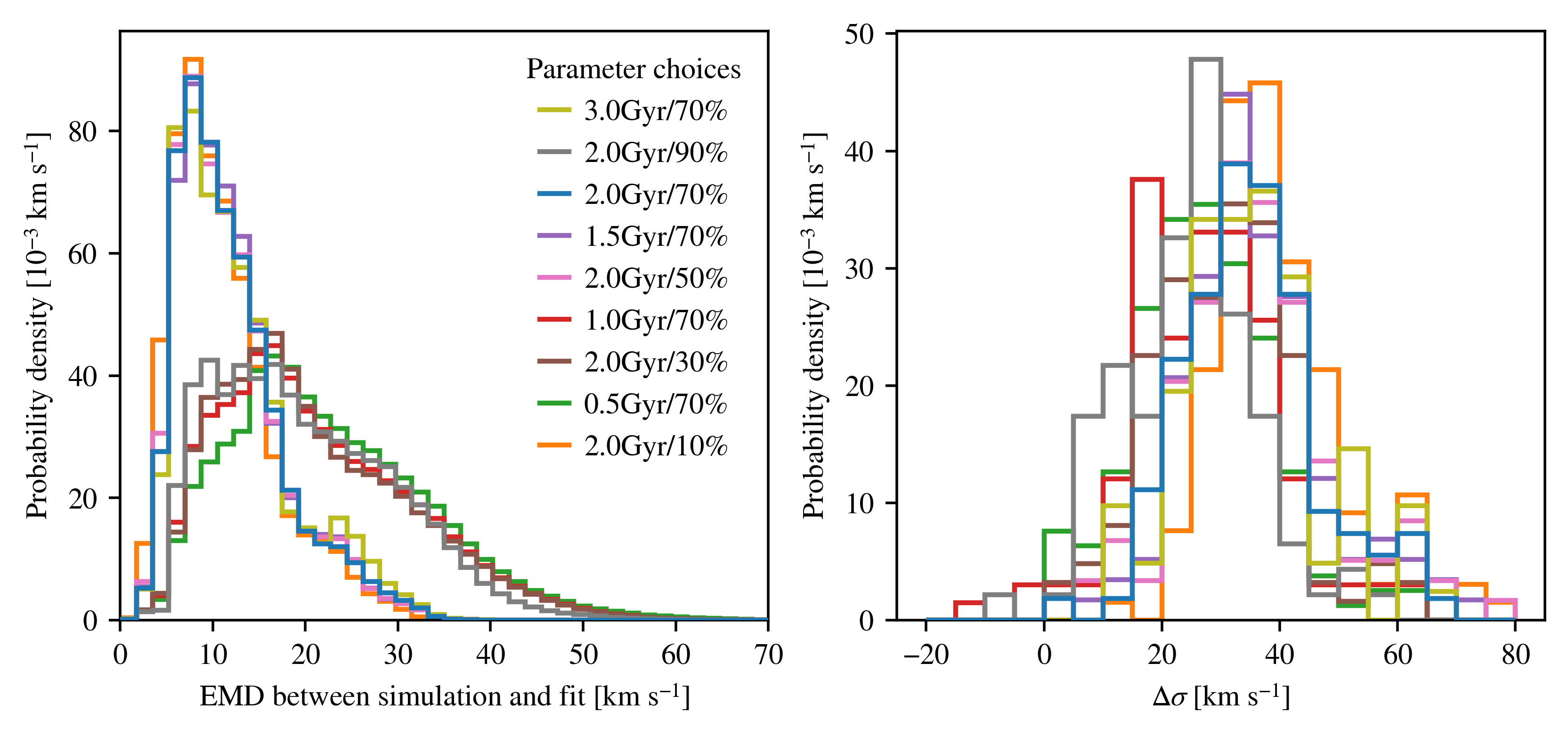}
    \caption{Comparison of reconstruction performance across the nine merger-tracking parameter sets, labeled by the time window (in Gyr) and the fraction of snapshots that a particle must be bound to be considered tagged (as a percentage). (Left) Distribution of EMDs between the exact DM speed distributions and the sampled--$\Delta\sigma,\ \ftr$ reconstructions (see \autoref{fig:7}) for all Traceable mergers across the 98 MW analogs. While low-purity choices of parameters (i.e., those with smaller time windows or less restrictive fractions) tend to have larger EMDs---and therefore less accurate reconstructions---our fiducial choice is among the most accurate, with a few comparable algorithms. 
    (Right) Distribution of the boost factor $\Delta\sigma$ (\autoref{eq:delta_sigma}) across all Traceable mergers for each parameter set. The results here are qualitatively the same regardless of the choice of DM tagging parameters. There is consistently an offset between the DM and stellar velocity dispersions, suggesting that this offset is a physical result and not due to our tagging procedure.}
    \label{fig:11}
\end{figure*}

\twocolumngrid

\renewcommand{\thefigure}{B1}\begin{figure}
    \centering
    \includegraphics{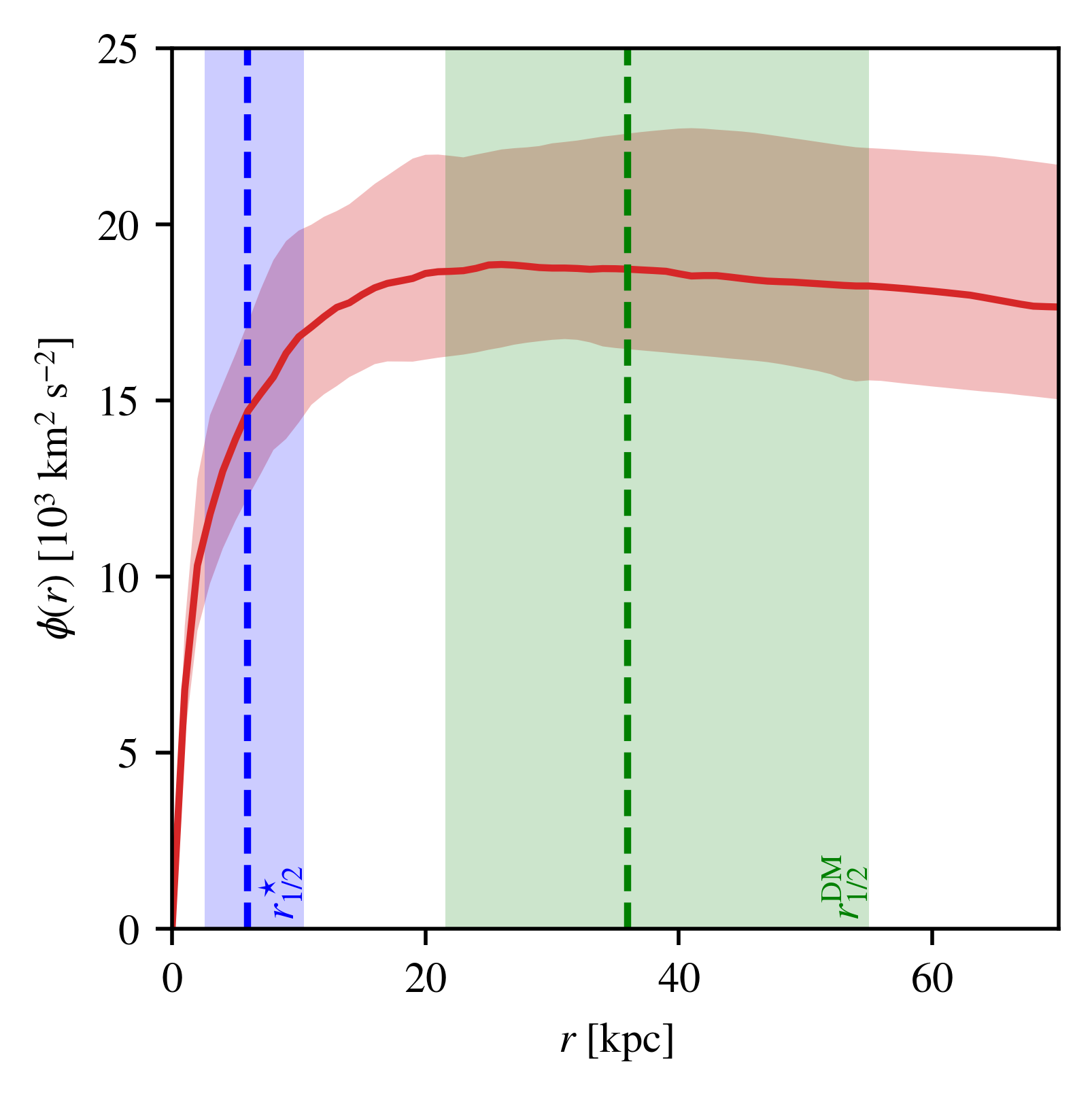}
    \caption{Gravitational potential per unit mass, $\phi(r)$, at redshift zero across our 98 MW analogs (solid red, with shaded regions indicating 16th--84th percentiles across the analogs). The vertical bands indicate the galactocentric spherical radii that enclose half of the DM debris ($r_{1/2}^\mathrm{DM}$, green) and half of the stellar debris ($r_{1/2}^\star$, blue). The stars are deposited at radii of $6^{+4}_{-3}$~kpc, while the DM is further from the galactic center, at $36^{+19}_{-14}$~kpc. The DM debris typically has higher orbital energy and exhibits a larger velocity dispersion than the stellar debris, yielding the offset shown in \autoref{fig:5}.}
    \label{fig:12}
\end{figure}
\renewcommand{\thefigure}{B2}\begin{figure}
    \centering
    \includegraphics[width=\columnwidth]{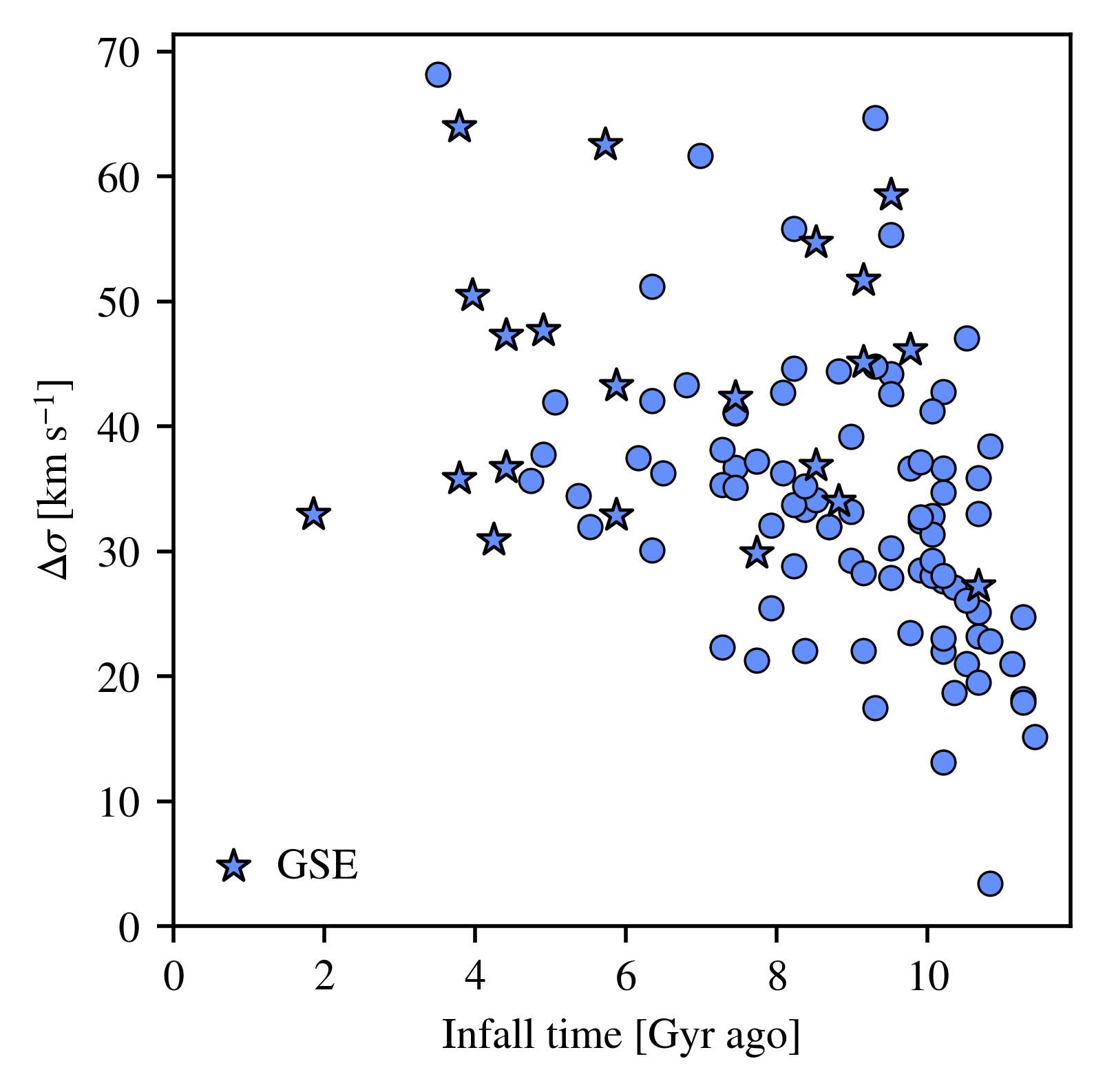}
    \caption{Boost factor $\Delta\sigma$ as a function of the infall time for all Traceable mergers (circles). GSE-like mergers are highlighted separately (stars). Mergers with earlier infall times tend to show smaller dispersion offsets. Those with infall times greater than 11~Gyr have a median offset of $9^{+2}_{-5}~\kms{}$, while those with infall times less than 11~Gyr have a median offset of $15^{+6}_{-7}~\kms{}$. This trend is consistent with the mechanism highlighted in \autoref{fig:12}. When mergers fall in earlier, both their DM and stars are deposited at smaller radii because the MW itself was smaller, reducing the difference in potential energies and thus the dispersion offset.}
    \label{fig:13}
\end{figure}
\renewcommand{\thefigure}{C1}\begin{figure*}
    \centering
    \includegraphics{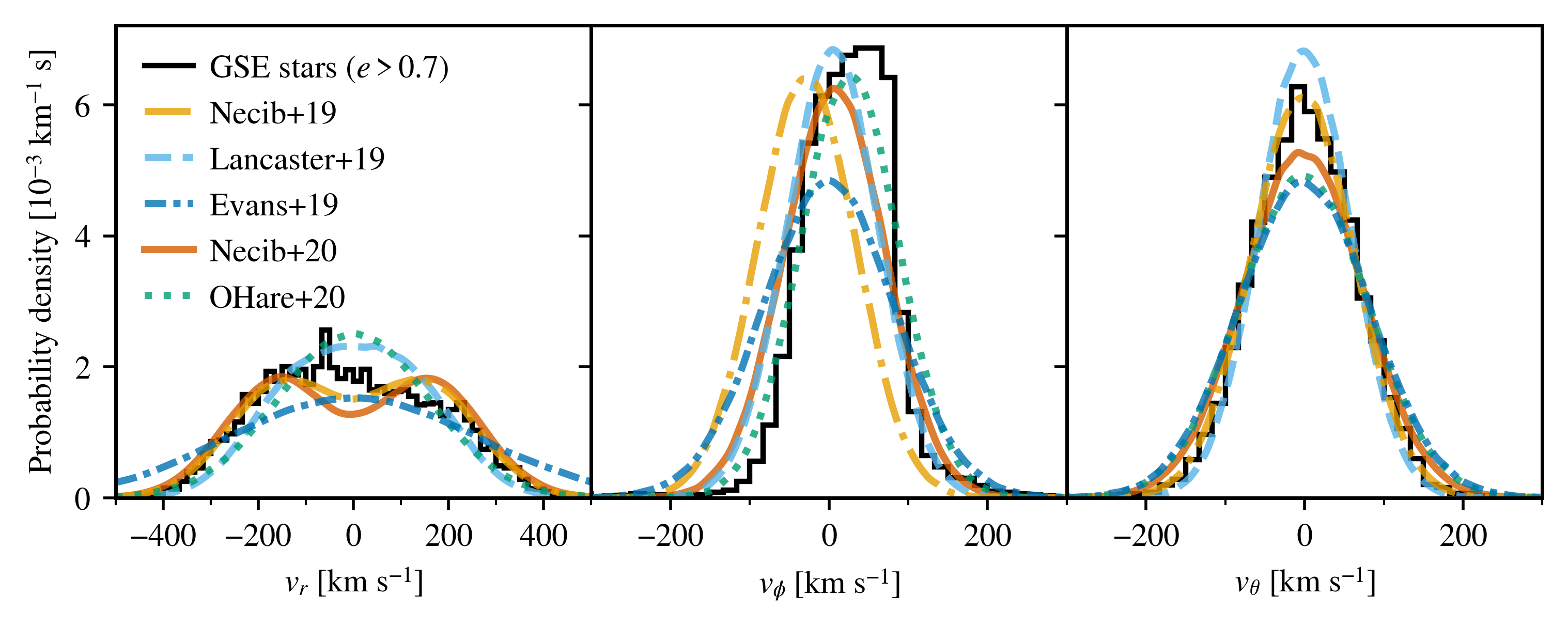}
    \caption{Velocity distributions of Gaia DR2 GSE stars selected with eccentricity $e>0.7$ and restricted to the ROI (black histograms), shown in spherical coordinates: radial ($v_r$, left), azimuthal ($v_\phi$, center), and polar ($v_\theta$, right). For comparison, we show results from previous studies: \citet[][solid yellow]{2019ApJ...874....3N}, \citet[][dashed turquoise]{2019MNRAS.486..378L}, \citet[][dashed-dotted blue]{2019PhRvD..99b3012E}, \citet[][solid orange]{2020ApJ...903...25N}, and \citet[][dotted green]{2020PhRvD.101b3006O}. The distributions of our selected GSE stars are largely consistent with those reported in earlier works.}
    \label{fig:14}
\end{figure*}

\section{DM and Stellar Dispersions}
\label{appendix:boost}
\setcounter{equation}{0}
\renewcommand{\theequation}{B\arabic{equation}}

\autoref{sec:results} showed that the DM associated with a particular merger tends to have a larger velocity dispersion than the corresponding stellar debris, as highlighted in \autoref{fig:5}. This Appendix investigates the origin of this effect. This offset arises because DM is stripped from the merging subhalos first and is left at larger galactocentric radii and higher orbital energies than the stars. The stars, being more tightly bound, remain in the subhalo for longer and are deposited deeper in the MW's potential well, with lower orbital energies. This difference yields DM speed distributions that are biased higher than those of the stars. 

\autoref{fig:12} illustrates this effect. We model the potential well of each host galaxy using its enclosed mass $M(r)$, with 
\begin{equation}
\phi(r) = \frac{GM(r)}{r}
\end{equation}
for each of the 98 MW analogs at redshift zero (solid red, with shaded regions indicating the 16th--84th percentiles across the hosts). For each Traceable merger, we compute the galactocentric spherical radius enclosing half of its stellar debris, $r_{1/2}^\star$~(dashed blue), and half of its DM debris, $r_{1/2}^\mathrm{DM}$~(dashed green), and find that the stars are deposited $6^{+4}_{-3}$~kpc from the galactic center, while the DM is $36^{+19}_{-14}$~kpc from the galactic center. This demonstrates that the stellar debris is deposited at systematically smaller radii than the DM, corresponding to regions deeper in the potential well, with lower orbital energies.

This difference in orbital energies is most distinct for late-infall mergers, when the host potential is particularly deep. Mergers occurring early in a halo's assembly typically exhibit smaller offsets between the DM and stars. This is shown in  \autoref{fig:13}, which plots the boost factor $\Delta\sigma$ for a given merger against its infall time.  The results are shown for all Traceable mergers, and  GSE-like mergers are highlighted with star-shaped markers. Most of the TNG50 halos assemble early and have accreted $\osim60\%$ of their DM by a lookback time of 11~Gyr ago (redshift $\sim 2.5$). The mergers that occur after this time exhibit greater offsets between their DM and stars, with $\Delta\sigma = 15^{+6}_{-7}~\kms{}$. However, very early mergers tend to show smaller offsets, as the hosts' potential wells have not yet reached their full depth. For these mergers, $\Delta\sigma = 9^{+2}_{-5}~\kms{}$. 

This also gives a physical mechanism to explain why we find bigger offsets between a merger's stellar and DM debris compared to \citet{2019ApJ...883...27N} and \citet{2026arXiv260325783Z}. The FIRE-2 halos used in these works assemble more slowly than those in TNG50, lowering $\Delta\sigma$ for more recent mergers and yielding better agreement between the FIRE-2 DM and stars. Furthermore, the baryonic feedback prescription tends to produce shallower potential wells in FIRE-2 than in TNG50 \citep[e.g.,][]{2025arXiv250114868H}, which would further reduce $\Delta\sigma$.

\section{GSE Stellar Sample} \label{appendix:gaia_GSE}

This Appendix describes the data used for the GSE stellar distribution in \autoref{sec:discussion} and \autoref{fig:8}. We provide an account of the source catalog, our subselection of GSE stars, and compare the  results to other selections performed in the literature.

\citet{2020A&A...636A..75O} produced a catalog of accreted stars from the Gaia DR2 dataset \citep{2018A&A...616A...1G}. They train a deep learning classifier on mock Gaia catalogs produced from the FIRE-2 simulation, such that their classifier is able to transfer chemodynamical correlations derived from the simulation to real-world data. This deep learning technique is advantageous, as extracting ex situ stars from a sample containing many in situ MW stars using simple chemodynamical cuts is prone to contamination from the disk and in situ stellar halo \citep{2022MNRAS.510.5119L,2024MNRAS.527.2165C,2024NewAR..9901706D}. The catalog produced by \citet{2020A&A...636A..75O} reproduces known ex situ structures such as the GSE, Helmi streams, and Nyx. This makes it a natural starting point for constructing a sample of GSE stars. 

The neural network classifier outputs a score, $S$, that corresponds to the likelihood that a star is accreted; it ranges from $0$ to $1$ for the least and most likely accreted stars, respectively. We use a likelihood cut of $S>0.75$, which \citet{2020A&A...636A..75O} show most accurately reproduces the ex situ stellar speed distribution. Using the 6D phase-space information of these stars, we compute their eccentricity following the procedure of \citet{2022MNRAS.510.5119L}.\footnote{See \url{https://github.com/jamesmlane/mw-dfs} for an implementation of this method, which we applied to the Gaia~DR2 data.} In particular, we use \textsc{galpy} \citep{2015ApJS..216...29B} to compute actions through the St\"ackel fudge method of \citet{2012MNRAS.426.1324B} and \citet{ 2018PASP..130k4501M}. This approach locally approximates the axisymmetric \textsc{MWPotential2014} potential \citep{2015ApJS..216...29B} as a St\"ackel function, with focal length estimated according to \citet{2012MNRAS.426..128S}. This allows us to compute orbital eccentricities from the Gaia phase-space data.

We consider stars within the solar annulus ROI~$(r \in [6,10]~\mathrm{kpc},\ |z|\leq2~\mathrm{kpc})$ in Galactocentric coordinates. A complete analysis for identifying the GSE stars in this region would require performing a Gaussian mixture analysis, or other clustering method, similar to what was done by \citet{2020ApJ...903...25N} with the same catalog.  While such an analysis is beyond the scope of this work, as a first approximation, we impose an eccentricity cut of $e > 0.7$, in keeping with the recommendation of \citet{2020ApJ...901...48N}. This recommendation is also supported by the findings of \citet{2022MNRAS.510.5119L}, who suggest using an eccentricity cut to achieve a high-completeness selection, which we find suitable given that \citet{2020A&A...636A..75O} already provide a high-purity population of ex situ stars. The resulting distributions of radial, azimuthal, and polar velocities are shown in \autoref{fig:14}, where we compare our selection (black histograms) to others in the literature, namely \citet[][solid yellow]{2019ApJ...874....3N}, \citet[][dashed turquoise]{2019MNRAS.486..378L}, \citet[][dashed-dotted blue]{2019PhRvD..99b3012E}, \citet[][solid orange]{2020ApJ...903...25N}, and \citet[][dotted green]{2020PhRvD.101b3006O}. Our distributions are broadly consistent with these results, though we recover a modest prograde motion to the GSE stars, with $v_\phi = 25^{+50}_{-55}~\kms$. This is possibly due to contamination from the high-$\alpha$ disk, which does contribute stars to the ROI with ex situ likelihood $S > 0.75$ in the \citet{2020A&A...636A..75O} sample. These stars tend to be on circular orbits, so the eccentricity cut removes the majority of them; 62\% of the stars in the ROI are removed by the eccentricity selection, and the majority of these are prograde.

It is important to note that the recommendation of \citet{2020ApJ...901...48N} is not solely an eccentricity cut, but includes cuts to excise the high-$\alpha$ disk, in situ halo, and Sagittarius dwarf~\citep{1994Natur.370..194I}. We neglect these additional cuts because (i) the \citet{2020A&A...636A..75O} sample does not contain the metallicity information required to completely remove the high-$\alpha$ disk and in situ halo components and  (ii) the Sagittarius dwarf is not expected to have deposited many stars within our ROI, as its debris is primarily at Galactocentric radii $r \gtrsim 15$~kpc \citep[e.g.,][]{2020ApJ...900..103J}. We also checked more stringent eccentricity cuts and found that the resulting velocity distributions are similar, though with narrower azimuthal velocity dispersions (i.e., lower $\sigma_\phi$). Because the choice of $e > 0.7$ yields distributions most similar to those recovered elsewhere in the literature, we adopt it as our fiducial case.

The velocity anisotropy of the adopted GSE sample provides an additional diagnostic of its orbital structure. We find a value of $\beta=0.86$, consistent with the highly radial nature of the GSE stars. This result is in good agreement with other studies of the GSE \citep{2018ApJ...862L...1D, 2018ApJ...856L..26M, 2019MNRAS.486..378L, 2019ApJ...874....3N, 2021MNRAS.502.5686I, 2021ApJ...923...92N}.

\vfill\null
\bibliographystyle{aasjournal}
\bibliography{main}

@article{2010MNRAS.401.1796A,
  title     = {The Birth and Growth of Neutralino Haloes},
  author    = {Angulo, R. E. and White, S. D. M.},
  year      = {2010},
  month     = jan,
  journal   = {MNRAS},
  volume    = {401},
  pages     = {1796--1803},
  publisher = {OUP},
  issn      = {0035-8711},
  doi       = {10.1111/j.1365-2966.2009.15742.x},
  url       = {https://ui.adsabs.harvard.edu/abs/2010MNRAS.401.1796A},
}

@article{2021EPJC...81..907B,
  title     = {Recommended Conventions for Reporting Results from Direct Dark Matter Searches},
  author    = {Baxter, D. and Bloch, I. M. and Bodnia, E. and Chen, X. and Conrad, J. and Di Gangi, P. and Dobson, J. E. Y. and Durnford, D. and Haselschwardt, S. J. and Kaboth, A. and Lang, R. F. and Lin, Q. and Lippincott, W. H. and Liu, J. and Manalaysay, A. and McCabe, C. and Mor{\aa}, K. D. and Naim, D. and Neilson, R. and Olcina, I. and Piro, M. -C. and Selvi, M. and {von Krosigk}, B. and Westerdale, S. and Yang, Y. and Zhou, N.},
  year      = {2021},
  month     = oct,
  journal   = {EPJC},
  volume    = {81},
  pages     = {907},
  publisher = {Springer},
  issn      = {1434-6044},
  doi       = {10.1140/epjc/s10052-021-09655-y},
  url       = {https://ui.adsabs.harvard.edu/abs/2021EPJC...81..907B},
}

@article{2018MNRAS.478..611B,
  title     = {Co-Formation of the Disc and the Stellar Halo},
  author    = {Belokurov, V. and Erkal, D. and Evans, N. W. and Koposov, S. E. and Deason, A. J.},
  year      = {2018},
  month     = jul,
  journal   = {MNRAS},
  volume    = {478},
  pages     = {611--619},
  publisher = {OUP},
  issn      = {0035-8711},
  doi       = {10.1093/mnras/sty982},
  url       = {https://ui.adsabs.harvard.edu/abs/2018MNRAS.478..611B},
}

@article{2022ApJ...928L...5B,
  title     = {The {{Local Group Mass}} in the {{Light}} of {{Gaia}}},
  author    = {Benisty, David and Vasiliev, Eugene and Evans, N. Wyn and Davis, Anne-Christine and Hartl, Odelia V. and Strigari, Louis E.},
  year      = {2022},
  month     = mar,
  journal   = {ApJ},
  volume    = {928},
  pages     = {L5},
  publisher = {IOP},
  issn      = {0004-637X},
  doi       = {10.3847/2041-8213/ac5c42},
  url       = {https://ui.adsabs.harvard.edu/abs/2022ApJ...928L...5B},
}

@article{2012MNRAS.426.1324B,
  title = {Actions for Axisymmetric Potentials},
  author = {Binney, James},
  year = {2012},
  month = oct,
  journal = {MNRAS},
  volume = {426},
  pages = {1324--1327},
  publisher = {OUP},
  issn = {0035-8711},
  doi = {10.1111/j.1365-2966.2012.21757.x},
  url = {https://ui.adsabs.harvard.edu/abs/2012MNRAS.426.1324B}
}

@article{2016ARA&A..54..529B,
  title      = {The {{Galaxy}} in {{Context}}: {{Structural}}, {{Kinematic}}, and {{Integrated Properties}}},
  shorttitle = {The {{Galaxy}} in {{Context}}},
  author     = {{Bland-Hawthorn}, Joss and Gerhard, Ortwin},
  year       = {2016},
  month      = sep,
  journal    = {ARA\&A},
  volume     = {54},
  pages      = {529--596},
  issn       = {0066-4146},
  doi        = {10.1146/annurev-astro-081915-023441},
  url        = {https://ui.adsabs.harvard.edu/abs/2016ARA&A..54..529B},
}

@article{2020MNRAS.498.4943B,
  title      = {Are the {{Milky Way}} and {{Andromeda}} Unusual? {{A}} Comparison with {{Milky Way}} and {{Andromeda}} Analogues},
  shorttitle = {Are the {{Milky Way}} and {{Andromeda}} Unusual?},
  author     = {Boardman, N. and Zasowski, G. and Newman, J. A. and Andrews, B. and Fielder, C. and Bershady, M. and Brinkmann, J. and Drory, N. and Krishnarao, D. and Lane, R. R. and Mackereth, T. and Masters, K. and Stringfellow, G. S.},
  year       = {2020},
  month      = nov,
  journal    = {MNRAS},
  volume     = {498},
  pages      = {4943--4954},
  publisher  = {OUP},
  issn       = {0035-8711},
  doi        = {10.1093/mnras/staa2731},
  url        = {https://ui.adsabs.harvard.edu/abs/2020MNRAS.498.4943B},
}

@article{2020ApJ...897L..18B,
  title     = {Timing the {{Early Assembly}} of the {{Milky Way}} with the {{H3 Survey}}},
  author    = {Bonaca, Ana and Conroy, Charlie and Cargile, Phillip A. and Naidu, Rohan P. and Johnson, Benjamin D. and Zaritsky, Dennis and Ting, Yuan-Sen and Caldwell, Nelson and Han, Jiwon Jesse and {van Dokkum}, Pieter},
  year      = {2020},
  month     = jul,
  journal   = {ApJ},
  volume    = {897},
  pages     = {L18},
  publisher = {IOP},
  issn      = {0004-637X},
  doi       = {10.3847/2041-8213/ab9caa},
  url       = {https://ui.adsabs.harvard.edu/abs/2020ApJ...897L..18B},
}

@article{2015ApJS..216...29B,
  title = {Galpy: {{A}} Python {{Library}} for {{Galactic Dynamics}}},
  shorttitle = {Galpy},
  author = {Bovy, Jo},
  year = {2015},
  month = feb,
  journal = {ApJS},
  volume = {216},
  pages = {29},
  publisher = {IOP},
  issn = {0067-0049},
  doi = {10.1088/0067-0049/216/2/29},
  url = {https://ui.adsabs.harvard.edu/abs/2015ApJS..216...29B}
}

@article{2013ApJ...779..115B,
  title     = {A {{Direct Dynamical Measurement}} of the {{Milky Way}}'s {{Disk Surface Density Profile}}, {{Disk Scale Length}}, and {{Dark Matter Profile}} at 4 Kpc {$<$}{\textasciitilde} {{R}} {$<$}{\textasciitilde} 9 Kpc},
  author    = {Bovy, Jo and Rix, Hans-Walter},
  year      = {2013},
  month     = dec,
  journal   = {ApJ},
  volume    = {779},
  pages     = {115},
  publisher = {IOP},
  issn      = {0004-637X},
  doi       = {10.1088/0004-637X/779/2/115},
  url       = {https://ui.adsabs.harvard.edu/abs/2013ApJ...779..115B},
}

@article{2017IJMPA..3230016B,
  title   = {Implications of Hydrodynamical Simulations for the Interpretation of Direct Dark Matter Searches},
  author  = {Bozorgnia, Nassim and Bertone, Gianfranco},
  year    = {2017},
  month   = jul,
  journal = {IJMPA},
  volume  = {32},
  pages   = {1730016},
  issn    = {0217-751X},
  doi     = {10.1142/S0217751X17300162},
  url     = {https://ui.adsabs.harvard.edu/abs/2017IJMPA..3230016B},
}

@article{2016JCAP...05..024B,
  title      = {Simulated {{Milky Way}} Analogues: Implications for Dark Matter Direct Searches},
  shorttitle = {Simulated {{Milky Way}} Analogues},
  author     = {Bozorgnia, Nassim and Calore, Francesca and Schaller, Matthieu and Lovell, Mark and Bertone, Gianfranco and Frenk, Carlos S. and Crain, Robert A. and Navarro, Julio F. and Schaye, Joop and Theuns, Tom},
  year       = {2016},
  month      = may,
  journal    = {JCAP},
  volume     = {2016},
  pages      = {024},
  publisher  = {IOP},
  issn       = {1475-7516},
  doi        = {10.1088/1475-7516/2016/05/024},
  url        = {https://ui.adsabs.harvard.edu/abs/2016JCAP...05..024B},
}

@article{2019JCAP...06..045B,
  title   = {On the Correlation between the Local Dark Matter and Stellar Velocities},
  author  = {Bozorgnia, Nassim and Fattahi, Azadeh and Cerde{\~n}o, David G. and Frenk, Carlos S. and G{\'o}mez, Facundo A. and Grand, Robert J. J. and Marinacci, Federico and Pakmor, R{\"u}diger},
  year    = {2019},
  month   = jun,
  journal = {JCAP},
  volume  = {2019},
  pages   = {045},
  issn    = {1475-7516},
  doi     = {10.1088/1475-7516/2019/06/045},
  url     = {https://ui.adsabs.harvard.edu/abs/2019JCAP...06..045B},
}

@article{2020JCAP...07..036B,
  title     = {The Dark Matter Component of the {{Gaia}} Radially Anisotropic Substructure},
  author    = {Bozorgnia, Nassim and Fattahi, Azadeh and Frenk, Carlos S. and Cheek, Andrew and Cerde{\~n}o, David G. and G{\'o}mez, Facundo A. and Grand, Robert J. J. and Marinacci, Federico},
  year      = {2020},
  month     = jul,
  journal   = {JCAP},
  volume    = {2020},
  pages     = {036},
  publisher = {IOP},
  issn      = {1475-7516},
  doi       = {10.1088/1475-7516/2020/07/036},
  url       = {https://ui.adsabs.harvard.edu/abs/2020JCAP...07..036B},
}

@article{2024ApJ...971...79B,
  title      = {Milky {{Way-est}}: {{Cosmological Zoom-in Simulations}} with {{Large Magellanic Cloud}} and {{Gaia}}--{{Sausage}}--{{Enceladus Analogs}}},
  shorttitle = {Milky {{Way-est}}},
  author     = {Buch, Deveshi and Nadler, Ethan O. and Wechsler, Risa H. and Mao, Yao-Yuan},
  year       = {2024},
  month      = aug,
  journal    = {ApJ},
  volume     = {971},
  pages      = {79},
  publisher  = {IOP},
  issn       = {0004-637X},
  doi        = {10.3847/1538-4357/ad554c},
  url        = {https://ui.adsabs.harvard.edu/abs/2024ApJ...971...79B},
}

@article{2016MNRAS.462..663B,
  title      = {{{NIHAO}} Project {{II}}: Halo Shape, Phase-Space Density and Velocity Distribution of Dark Matter in Galaxy Formation Simulations},
  shorttitle = {{{NIHAO}} Project {{II}}},
  author     = {Butsky, Iryna and Macci{\`o}, Andrea V. and Dutton, Aaron A. and Wang, Liang and Obreja, Aura and Stinson, Greg S. and Penzo, Camilla and Kang, Xi and Keller, Ben W. and Wadsley, James},
  year       = {2016},
  month      = oct,
  journal    = {MNRAS},
  volume     = {462},
  pages      = {663--680},
  publisher  = {OUP},
  issn       = {0035-8711},
  doi        = {10.1093/mnras/stw1688},
  url        = {https://ui.adsabs.harvard.edu/abs/2016MNRAS.462..663B},
}

@article{2024MNRAS.527.2165C,
  title      = {Can We Really Pick and Choose? {{Benchmarking}} Various Selections of {{Gaia Enceladus}}/{{Sausage}} Stars in Observations with Simulations},
  shorttitle = {Can We Really Pick and Choose?},
  author     = {Carrillo, Andreia and Deason, Alis J. and Fattahi, Azadeh and Callingham, Thomas M. and Grand, Robert J. J.},
  year       = {2024},
  month      = jan,
  journal    = {MNRAS},
  volume     = {527},
  pages      = {2165--2184},
  publisher  = {OUP},
  issn       = {0035-8711},
  doi        = {10.1093/mnras/stad3274},
  url        = {https://ui.adsabs.harvard.edu/abs/2024MNRAS.527.2165C},
}

@article{2020MNRAS.494.4291C,
  title     = {The Milky Way Total Mass Profile as Inferred from {{Gaia DR2}}},
  author    = {Cautun, Marius and {Ben{\'i}tez-Llambay}, Alejandro and Deason, Alis J. and Frenk, Carlos S. and Fattahi, Azadeh and G{\'o}mez, Facundo A. and Grand, Robert J. J. and Oman, Kyle A. and Navarro, Julio F. and Simpson, Christine M.},
  year      = {2020},
  month     = may,
  journal   = {MNRAS},
  volume    = {494},
  pages     = {4291--4313},
  publisher = {OUP},
  issn      = {0035-8711},
  doi       = {10.1093/mnras/staa1017},
  url       = {https://ui.adsabs.harvard.edu/abs/2020MNRAS.494.4291C},
}

@article{1985ApJ...292..371D,
  title   = {The Evolution of Large-Scale Structure in a Universe Dominated by Cold Dark Matter},
  author  = {Davis, M. and Efstathiou, G. and Frenk, C. S. and White, S. D. M.},
  year    = {1985},
  month   = may,
  journal = {ApJ},
  volume  = {292},
  pages   = {371--394},
  issn    = {0004-637X},
  doi     = {10.1086/163168},
  url     = {https://ui.adsabs.harvard.edu/abs/1985ApJ...292..371D},
}

@article{2024NewAR..9901706D,
  title     = {Galactic {{Archaeology}} with {{Gaia}}},
  author    = {Deason, Alis J. and Belokurov, Vasily},
  year      = {2024},
  month     = dec,
  journal   = {NewAR},
  volume    = {99},
  pages     = {101706},
  publisher = {Elsevier},
  issn      = {1387-6473},
  doi       = {10.1016/j.newar.2024.101706},
  url       = {https://ui.adsabs.harvard.edu/abs/2024NewAR..9901706D},
}

@article{2018ApJ...862L...1D,
  title      = {Apocenter {{Pile-up}}: {{Origin}} of the {{Stellar Halo Density Break}}},
  shorttitle = {Apocenter {{Pile-up}}},
  author     = {Deason, Alis J. and Belokurov, Vasily and Koposov, Sergey E. and Lancaster, Lachlan},
  year       = {2018},
  month      = jul,
  journal    = {ApJ},
  volume     = {862},
  pages      = {L1},
  publisher  = {IOP},
  issn       = {0004-637X},
  doi        = {10.3847/2041-8213/aad0ee},
  url        = {https://ui.adsabs.harvard.edu/abs/2018ApJ...862L...1D},
}

@article{2014arXiv1404.4130D,
  title      = {Halo-Independent Comparison of Direct Dark Matter Detection Data: A Review},
  shorttitle = {Halo-Independent Comparison of Direct Dark Matter Detection Data},
  author     = {Del Nobile, Eugenio},
  year       = {2014},
  month      = apr,
  journal    = {AdHEP},
  volume     = {2014},
  pages      = {1--14},
  doi        = {10.1155/2014/604914},
  url        = {https://ui.adsabs.harvard.edu/abs/2014arXiv1404.4130D},
}

@article{2008Natur.454..735D,
  title   = {Clumps and Streams in the Local Dark Matter Distribution},
  author  = {Diemand, J. and Kuhlen, M. and Madau, P. and Zemp, M. and Moore, B. and Potter, D. and Stadel, J.},
  year    = {2008},
  month   = aug,
  journal = {Natur},
  volume  = {454},
  pages   = {735--738},
  issn    = {0028-0836},
  doi     = {10.1038/nature07153},
  url     = {https://ui.adsabs.harvard.edu/abs/2008Natur.454..735D},
}

@article{2009MNRAS.399..497D,
  title     = {Substructures in Hydrodynamical Cluster Simulations},
  author    = {Dolag, K. and Borgani, S. and Murante, G. and Springel, V.},
  year      = {2009},
  month     = oct,
  journal   = {MNRAS},
  volume    = {399},
  pages     = {497--514},
  publisher = {OUP},
  issn      = {0035-8711},
  doi       = {10.1111/j.1365-2966.2009.15034.x},
  url       = {https://ui.adsabs.harvard.edu/abs/2009MNRAS.399..497D},
}

@article{1986PhRvD..33.3495D,
  title     = {Detecting Cold Dark-Matter Candidates},
  author    = {Drukier, Andrzej K. and Freese, Katherine and Spergel, David N.},
  year      = {1986},
  month     = jun,
  journal   = {PhRvD},
  volume    = {33},
  pages     = {3495--3508},
  publisher = {APS},
  issn      = {1550-79980556-2821},
  doi       = {10.1103/PhysRevD.33.3495},
  url       = {https://ui.adsabs.harvard.edu/abs/1986PhRvD..33.3495D},
}

@article{2019PhRvD..99b3012E,
  title     = {Refinement of the Standard Halo Model for Dark Matter Searches in Light of the {{Gaia Sausage}}},
  author    = {Evans, N. Wyn and O'Hare, Ciaran A. J. and McCabe, Christopher},
  year      = {2019},
  month     = jan,
  journal   = {PhRvD},
  volume    = {99},
  pages     = {023012},
  publisher = {APS},
  issn      = {1550-79980556-2821},
  doi       = {10.1103/PhysRevD.99.023012},
  url       = {https://ui.adsabs.harvard.edu/abs/2019PhRvD..99b3012E},
}

@article{2019MNRAS.484.4471F,
  title   = {The Origin of Galactic Metal-Rich Stellar Halo Components with Highly Eccentric Orbits},
  author  = {Fattahi, Azadeh and Belokurov, Vasily and Deason, Alis J. and Frenk, Carlos S. and G{\'o}mez, Facundo A. and Grand, Robert J. J. and Marinacci, Federico and Pakmor, R{\"u}diger and Springel, Volker},
  year    = {2019},
  month   = apr,
  journal = {MNRAS},
  volume  = {484},
  pages   = {4471--4483},
  issn    = {0035-8711},
  doi     = {10.1093/mnras/stz159},
  url     = {https://ui.adsabs.harvard.edu/abs/2019MNRAS.484.4471F},
}

@article{2021MNRAS.508.1489F,
  title      = {Selecting Accreted Populations: Metallicity, Elemental Abundances, and Ages of the {{Gaia-Sausage-Enceladus}} and {{Sequoia}} Populations},
  shorttitle = {Selecting Accreted Populations},
  author     = {Feuillet, Diane K. and Sahlholdt, Christian L. and Feltzing, Sofia and Casagrande, Luca},
  year       = {2021},
  month      = nov,
  journal    = {MNRAS},
  volume     = {508},
  pages      = {1489--1508},
  publisher  = {OUP},
  issn       = {0035-8711},
  doi        = {10.1093/mnras/stab2614},
  url        = {https://ui.adsabs.harvard.edu/abs/2021MNRAS.508.1489F},
}

@article{2006MNRAS.372.1149F,
  title   = {On the Mass-to-Light Ratio of the Local {{Galactic}} Disc and the Optical Luminosity of the {{Galaxy}}},
  author  = {Flynn, Chris and Holmberg, Johan and Portinari, Laura and Fuchs, Burkhard and Jahrei{\ss}, Hartmut},
  year    = {2006},
  month   = nov,
  journal = {MNRAS},
  volume  = {372},
  pages   = {1149--1160},
  issn    = {0035-8711},
  doi     = {10.1111/j.1365-2966.2006.10911.x},
  url     = {https://ui.adsabs.harvard.edu/abs/2006MNRAS.372.1149F},
}

@article{2025arXiv250507924F,
  title = {Dark {{Matter Velocity Distributions}} for {{Direct Detection}}: {{Astrophysical Uncertainties Are Smaller Than They Appear}}},
  shorttitle = {Dark {{Matter Velocity Distributions}} for {{Direct Detection}}},
  author = {Folsom, Dylan and Blanco, Carlos and Lisanti, Mariangela and Necib, Lina and Vogelsberger, Mark and Hernquist, Lars},
  year = 2025,
  month = nov,
  journal = {PhRvL},
  volume = {135},
  pages = {211004},
  publisher = {APS},
  issn = {0031-9007},
  doi = {10.1103/wmpq-mw4h},
  url = {https://ui.adsabs.harvard.edu/abs/2025PhRvL.135u1004F}
}

@article{2025ApJ...983..119F,
  title      = {Cosmological {{Simulations}} of {{Stellar Halos}} with {{Gaia Sausage}}--{{Enceladus Analogs}}: {{Two Sausages}}, {{One Bun}}?},
  shorttitle = {Cosmological {{Simulations}} of {{Stellar Halos}} with {{Gaia Sausage}}--{{Enceladus Analogs}}},
  author     = {Folsom, Dylan and Lisanti, Mariangela and Necib, Lina and Horta, Danny and Vogelsberger, Mark and Hernquist, Lars},
  year       = {2025},
  month      = apr,
  journal    = {ApJ},
  volume     = {983},
  pages      = {119},
  publisher  = {IOP},
  issn       = {0004-637X},
  doi        = {10.3847/1538-4357/adbe31},
  url        = {https://ui.adsabs.harvard.edu/abs/2025ApJ...983..119F},
}

@article{2013RvMP...85.1561F,
  title      = {Colloquium: {{Annual}} Modulation of Dark Matter},
  shorttitle = {Colloquium},
  author     = {Freese, Katherine and Lisanti, Mariangela and Savage, Christopher},
  year       = {2013},
  month      = oct,
  journal    = {RvMP},
  volume     = {85},
  pages      = {1561--1581},
  publisher  = {APS},
  issn       = {0034-6861},
  doi        = {10.1103/RevModPhys.85.1561},
  url        = {https://ui.adsabs.harvard.edu/abs/2013RvMP...85.1561F},
}

@article{2018A&A...616A...1G,
  title   = {Gaia {{Data Release}} 2. {{Summary}} of the Contents and Survey Properties},
  author  = {{Gaia Collaboration} and Brown, A. G. A. and Vallenari, A. and Prusti, T. and {de Bruijne}, J. H. J. and Babusiaux, C. and {Bailer-Jones}, C. A. L. and Biermann, M. and Evans, D. W. and Eyer, L. and Jansen, F. and Jordi, C. and Klioner, S. A. and Lammers, U. and Lindegren, L. and Luri, X. and Mignard, F. and Panem, C. and Pourbaix, D. and Randich, S. and Sartoretti, P. and Siddiqui, H. I. and Soubiran, C. and {van Leeuwen}, F. and Walton, N. A. and Arenou, F. and Bastian, U. and Cropper, M. and Drimmel, R. and Katz, D. and Lattanzi, M. G. and Bakker, J. and Cacciari, C. and Casta{\~n}eda, J. and Chaoul, L. and Cheek, N. and De Angeli, F. and Fabricius, C. and Guerra, R. and Holl, B. and Masana, E. and Messineo, R. and Mowlavi, N. and Nienartowicz, K. and Panuzzo, P. and Portell, J. and Riello, M. and Seabroke, G. M. and Tanga, P. and Th{\'e}venin, F. and {Gracia-Abril}, G. and Comoretto, G. and {Garcia-Reinaldos}, M. and Teyssier, D. and Altmann, M. and Andrae, R. and Audard, M. and {Bellas-Velidis}, I. and Benson, K. and Berthier, J. and Blomme, R. and Burgess, P. and Busso, G. and Carry, B. and Cellino, A. and Clementini, G. and Clotet, M. and Creevey, O. and Davidson, M. and De Ridder, J. and Delchambre, L. and Dell'Oro, A. and Ducourant, C. and {Fern{\'a}ndez-Hern{\'a}ndez}, J. and Fouesneau, M. and Fr{\'e}mat, Y. and Galluccio, L. and {Garc{\'i}a-Torres}, M. and {Gonz{\'a}lez-N{\'u}{\~n}ez}, J. and {Gonz{\'a}lez-Vidal}, J. J. and Gosset, E. and Guy, L. P. and Halbwachs, J. -L. and Hambly, N. C. and Harrison, D. L. and Hern{\'a}ndez, J. and Hestroffer, D. and Hodgkin, S. T. and Hutton, A. and Jasniewicz, G. and {Jean-Antoine-Piccolo}, A. and Jordan, S. and Korn, A. J. and {Krone-Martins}, A. and Lanzafame, A. C. and Lebzelter, T. and L{\"o}ffler, W. and Manteiga, M. and Marrese, P. M. and {Mart{\'i}n-Fleitas}, J. M. and Moitinho, A. and Mora, A. and Muinonen, K. and Osinde, J. and Pancino, E. and Pauwels, T. and Petit, J. -M. and {Recio-Blanco}, A. and Richards, P. J. and Rimoldini, L. and Robin, A. C. and Sarro, L. M. and Siopis, C. and Smith, M. and Sozzetti, A. and S{\"u}veges, M. and Torra, J. and {van Reeven}, W. and Abbas, U. and Abreu Aramburu, A. and Accart, S. and Aerts, C. and Altavilla, G. and {\'A}lvarez, M. A. and Alvarez, R. and Alves, J. and Anderson, R. I. and Andrei, A. H. and Anglada Varela, E. and Antiche, E. and Antoja, T. and Arcay, B. and Astraatmadja, T. L. and Bach, N. and Baker, S. G. and {Balaguer-N{\'u}{\~n}ez}, L. and Balm, P. and Barache, C. and Barata, C. and Barbato, D. and Barblan, F. and Barklem, P. S. and Barrado, D. and Barros, M. and Barstow, M. A. and Bartholom{\'e} Mu{\~n}oz, S. and Bassilana, J. -L. and Becciani, U. and Bellazzini, M. and Berihuete, A. and Bertone, S. and Bianchi, L. and Bienaym{\'e}, O. and {Blanco-Cuaresma}, S. and Boch, T. and Boeche, C. and Bombrun, A. and Borrachero, R. and Bossini, D. and Bouquillon, S. and Bourda, G. and Bragaglia, A. and Bramante, L. and Breddels, M. A. and Bressan, A. and Brouillet, N. and Br{\"u}semeister, T. and Brugaletta, E. and Bucciarelli, B. and Burlacu, A. and Busonero, D. and Butkevich, A. G. and Buzzi, R. and Caffau, E. and Cancelliere, R. and Cannizzaro, G. and {Cantat-Gaudin}, T. and Carballo, R. and Carlucci, T. and Carrasco, J. M. and Casamiquela, L. and Castellani, M. and {Castro-Ginard}, A. and Charlot, P. and Chemin, L. and Chiavassa, A. and Cocozza, G. and Costigan, G. and Cowell, S. and Crifo, F. and Crosta, M. and Crowley, C. and Cuypers, J. and Dafonte, C. and Damerdji, Y. and Dapergolas, A. and David, P. and David, M. and {de Laverny}, P. and De Luise, F. and De March, R. and {de Martino}, D. and {de Souza}, R. and {de Torres}, A. and Debosscher, J. and {del Pozo}, E. and Delbo, M. and Delgado, A. and Delgado, H. E. and Di Matteo, P. and Diakite, S. and Diener, C. and Distefano, E. and Dolding, C. and Drazinos, P. and Dur{\'a}n, J. and Edvardsson, B. and Enke, H. and Eriksson, K. and Esquej, P. and Eynard Bontemps, G. and Fabre, C. and Fabrizio, M. and Faigler, S. and Falc{\~a}o, A. J. and Farr{\`a}s Casas, M. and Federici, L. and Fedorets, G. and Fernique, P. and Figueras, F. and Filippi, F. and Findeisen, K. and Fonti, A. and Fraile, E. and Fraser, M. and Fr{\'e}zouls, B. and Gai, M. and Galleti, S. and Garabato, D. and {Garc{\'i}a-Sedano}, F. and Garofalo, A. and Garralda, N. and Gavel, A. and Gavras, P. and Gerssen, J. and Geyer, R. and Giacobbe, P. and Gilmore, G. and Girona, S. and Giuffrida, G. and Glass, F. and Gomes, M. and Granvik, M. and Gueguen, A. and Guerrier, A. and Guiraud, J. and {Guti{\'e}rrez-S{\'a}nchez}, R. and Haigron, R. and Hatzidimitriou, D. and Hauser, M. and Haywood, M. and Heiter, U. and Helmi, A. and Heu, J. and Hilger, T. and Hobbs, D. and Hofmann, W. and Holland, G. and Huckle, H. E. and Hypki, A. and Icardi, V. and Jan{\ss}en, K. and {Jevardat de Fombelle}, G. and Jonker, P. G. and Juh{\'a}sz, {\'A}. L. and Julbe, F. and Karampelas, A. and Kewley, A. and Klar, J. and Kochoska, A. and Kohley, R. and Kolenberg, K. and Kontizas, M. and Kontizas, E. and Koposov, S. E. and Kordopatis, G. and {Kostrzewa-Rutkowska}, Z. and Koubsky, P. and Lambert, S. and Lanza, A. F. and Lasne, Y. and Lavigne, J. -B. and Le Fustec, Y. and {Le Poncin-Lafitte}, C. and Lebreton, Y. and Leccia, S. and Leclerc, N. and {Lecoeur-Taibi}, I. and Lenhardt, H. and Leroux, F. and Liao, S. and Licata, E. and Lindstr{\o}m, H. E. P. and Lister, T. A. and Livanou, E. and Lobel, A. and L{\'o}pez, M. and Managau, S. and Mann, R. G. and Mantelet, G. and Marchal, O. and Marchant, J. M. and Marconi, M. and Marinoni, S. and Marschalk{\'o}, G. and Marshall, D. J. and Martino, M. and Marton, G. and Mary, N. and Massari, D. and Matijevi{\v c}, G. and Mazeh, T. and McMillan, P. J. and Messina, S. and Michalik, D. and Millar, N. R. and Molina, D. and Molinaro, R. and Moln{\'a}r, L. and Montegriffo, P. and Mor, R. and Morbidelli, R. and Morel, T. and Morris, D. and Mulone, A. F. and Muraveva, T. and Musella, I. and Nelemans, G. and Nicastro, L. and Noval, L. and O'Mullane, W. and Ord{\'e}novic, C. and {Ord{\'o}{\~n}ez-Blanco}, D. and Osborne, P. and Pagani, C. and Pagano, I. and Pailler, F. and Palacin, H. and Palaversa, L. and Panahi, A. and Pawlak, M. and Piersimoni, A. M. and Pineau, F. -X. and Plachy, E. and Plum, G. and Poggio, E. and Poujoulet, E. and Pr{\v s}a, A. and Pulone, L. and Racero, E. and Ragaini, S. and Rambaux, N. and {Ramos-Lerate}, M. and Regibo, S. and Reyl{\'e}, C. and Riclet, F. and Ripepi, V. and Riva, A. and Rivard, A. and Rixon, G. and Roegiers, T. and Roelens, M. and {Romero-G{\'o}mez}, M. and Rowell, N. and Royer, F. and {Ruiz-Dern}, L. and Sadowski, G. and Sagrist{\`a} Sell{\'e}s, T. and Sahlmann, J. and Salgado, J. and Salguero, E. and Sanna, N. and {Santana-Ros}, T. and Sarasso, M. and Savietto, H. and Schultheis, M. and Sciacca, E. and Segol, M. and Segovia, J. C. and S{\'e}gransan, D. and Shih, I. -C. and Siltala, L. and Silva, A. F. and Smart, R. L. and Smith, K. W. and Solano, E. and Solitro, F. and Sordo, R. and Soria Nieto, S. and Souchay, J. and Spagna, A. and Spoto, F. and Stampa, U. and Steele, I. A. and Steidelm{\"u}ller, H. and Stephenson, C. A. and Stoev, H. and Suess, F. F. and Surdej, J. and Szabados, L. and {Szegedi-Elek}, E. and Tapiador, D. and Taris, F. and Tauran, G. and Taylor, M. B. and Teixeira, R. and Terrett, D. and Teyssandier, P. and Thuillot, W. and Titarenko, A. and Torra Clotet, F. and Turon, C. and Ulla, A. and Utrilla, E. and Uzzi, S. and Vaillant, M. and Valentini, G. and Valette, V. and {van Elteren}, A. and Van Hemelryck, E. and {van Leeuwen}, M. and Vaschetto, M. and Vecchiato, A. and Veljanoski, J. and Viala, Y. and Vicente, D. and Vogt, S. and {von Essen}, C. and Voss, H. and Votruba, V. and Voutsinas, S. and Walmsley, G. and Weiler, M. and Wertz, O. and Wevers, T. and Wyrzykowski, {\L}. and Yoldas, A. and {\v Z}erjal, M. and Ziaeepour, H. and Zorec, J. and Zschocke, S. and Zucker, S. and Zurbach, C. and Zwitter, T.},
  year    = {2018},
  month   = aug,
  journal = {A\&A},
  volume  = {616},
  pages   = {A1},
  issn    = {0004-6361},
  doi     = {10.1051/0004-6361/201833051},
  url     = {https://ui.adsabs.harvard.edu/abs/2018A&A...616A...1G},
}

@article{2019NatAs...3..932G,
  title   = {Uncovering the Birth of the {{Milky Way}} through Accurate Stellar Ages with {{Gaia}}},
  author  = {Gallart, Carme and Bernard, Edouard J. and Brook, Chris B. and {Ruiz-Lara}, Tom{\'a}s and Cassisi, Santi and Hill, Vanessa and Monelli, Matteo},
  year    = {2019},
  month   = jul,
  journal = {NatAs},
  volume  = {3},
  pages   = {932--939},
  issn    = {2397-3366},
  doi     = {10.1038/s41550-019-0829-5},
  url     = {https://ui.adsabs.harvard.edu/abs/2019NatAs...3..932G},
}

@article{2017MNRAS.467..179G,
  title      = {The {{Auriga Project}}: The Properties and Formation Mechanisms of Disc Galaxies across Cosmic Time},
  shorttitle = {The {{Auriga Project}}},
  author     = {Grand, Robert J. J. and G{\'o}mez, Facundo A. and Marinacci, Federico and Pakmor, R{\"u}diger and Springel, Volker and Campbell, David J. R. and Frenk, Carlos S. and Jenkins, Adrian and White, Simon D. M.},
  year       = {2017},
  month      = may,
  journal    = {MNRAS},
  volume     = {467},
  pages      = {179--207},
  publisher  = {OUP},
  issn       = {0035-8711},
  doi        = {10.1093/mnras/stx071},
  url        = {https://ui.adsabs.harvard.edu/abs/2017MNRAS.467..179G},
}

@article{2024A&A...692A.242G,
  title     = {Improving Constraints on the Extended Mass Distribution in the {{Galactic}} Center with Stellar Orbits},
  author    = {{GRAVITY Collaboration} and Abd El Dayem, K. and Abuter, R. and Aimar, N. and Amaro Seoane, P. and Amorim, A. and Beck, J. and Berger, J. P. and Bonnet, H. and Bourdarot, G. and Brandner, W. and Cardoso, V. and Capuzzo Dolcetta, R. and Cl{\'e}net, Y. and Davies, R. and {de Zeeuw}, P. T. and Drescher, A. and Eckart, A. and Eisenhauer, F. and Feuchtgruber, H. and Finger, G. and F{\"o}rster Schreiber, N. M. and Foschi, A. and Gao, F. and Garcia, P. and Gendron, E. and Genzel, R. and Gillessen, S. and Hartl, M. and Haubois, X. and Haussmann, F. and Hei{\ss}el, G. and Henning, T. and Hippler, S. and Horrobin, M. and Jochum, L. and Jocou, L. and Kaufer, A. and Kervella, P. and Lacour, S. and Lapeyr{\`e}re, V. and Le Bouquin, J. -B. and L{\'e}na, P. and Lutz, D. and Mang, F. and More, N. and Ott, T. and Paumard, T. and Perraut, K. and Perrin, G. and Pfuhl, O. and Rabien, S. and Ribeiro, D. C. and Sadun Bordoni, M. and Scheithauer, S. and Shangguan, J. and Shimizu, T. and Stadler, J. and Straub, O. and Straubmeier, C. and Sturm, E. and Tacconi, L. J. and Urso, I. and Vincent, F. and {von Fellenberg}, S. D. and Widmann, F. and Wieprecht, E. and Woillez, J. and Zhang, F.},
  year      = {2024},
  month     = dec,
  journal   = {A\&A},
  volume    = {692},
  pages     = {A242},
  publisher = {EDP},
  issn      = {0004-6361},
  doi       = {10.1051/0004-6361/202452274},
  url       = {https://ui.adsabs.harvard.edu/abs/2024A&A...692A.242G},
}

@article{2021A&A...647A..59G,
  title   = {Improved {{GRAVITY}} Astrometric Accuracy from Modeling Optical Aberrations},
  author  = {{GRAVITY Collaboration} and Abuter, R. and Amorim, A. and Baub{\"o}ck, M. and Berger, J. P. and Bonnet, H. and Brandner, W. and Cl{\'e}net, Y. and Davies, R. and {de Zeeuw}, P. T. and Dexter, J. and Dallilar, Y. and Drescher, A. and Eckart, A. and Eisenhauer, F. and F{\"o}rster Schreiber, N. M. and Garcia, P. and Gao, F. and Gendron, E. and Genzel, R. and Gillessen, S. and Habibi, M. and Haubois, X. and Hei{\ss}el, G. and Henning, T. and Hippler, S. and Horrobin, M. and {Jim{\'e}nez-Rosales}, A. and Jochum, L. and Jocou, L. and Kaufer, A. and Kervella, P. and Lacour, S. and Lapeyr{\`e}re, V. and Le Bouquin, J. -B. and L{\'e}na, P. and Lutz, D. and Nowak, M. and Ott, T. and Paumard, T. and Perraut, K. and Perrin, G. and Pfuhl, O. and Rabien, S. and {Rodr{\'i}guez-Coira}, G. and Shangguan, J. and Shimizu, T. and Scheithauer, S. and Stadler, J. and Straub, O. and Straubmeier, C. and Sturm, E. and Tacconi, L. J. and Vincent, F. and {von Fellenberg}, S. and Waisberg, I. and Widmann, F. and Wieprecht, E. and Wiezorrek, E. and Woillez, J. and Yazici, S. and Young, A. and Zins, G.},
  year    = {2021},
  month   = mar,
  journal = {A\&A},
  volume  = {647},
  pages   = {A59},
  issn    = {0004-6361},
  doi     = {10.1051/0004-6361/202040208},
  url     = {https://ui.adsabs.harvard.edu/abs/2021A&A...647A..59G},
}

@article{2017JPhG...44h4001G,
  title     = {Astrophysical Uncertainties on the Local Dark Matter Distribution and Direct Detection Experiments},
  author    = {Green, Anne M.},
  year      = {2017},
  month     = aug,
  journal   = {JPhG},
  volume    = {44},
  pages     = {084001},
  publisher = {IOP},
  issn      = {0954-3899},
  doi       = {10.1088/1361-6471/aa7819},
  url       = {https://ui.adsabs.harvard.edu/abs/2017JPhG...44h4001G},
}

@article{2010JCAP...10..034G,
  title     = {Dependence of Direct Detection Signals on the {{WIMP}} Velocity Distribution},
  author    = {Green, Anne M.},
  year      = {2010},
  month     = oct,
  journal   = {JCAP},
  volume    = {2010},
  pages     = {034},
  publisher = {IOP},
  issn      = {1475-7516},
  doi       = {10.1088/1475-7516/2010/10/034},
  url       = {https://ui.adsabs.harvard.edu/abs/2010JCAP...10..034G},
}

@article{2011ApJ...742...76G,
  title      = {Forming {{Realistic Late-type Spirals}} in a {{$\Lambda$CDM Universe}}: {{The Eris Simulation}}},
  shorttitle = {Forming {{Realistic Late-type Spirals}} in a {{$\Lambda$CDM Universe}}},
  author     = {Guedes, Javiera and Callegari, Simone and Madau, Piero and Mayer, Lucio},
  year       = {2011},
  month      = dec,
  journal    = {ApJ},
  volume     = {742},
  pages      = {76},
  publisher  = {IOP},
  issn       = {0004-637X},
  doi        = {10.1088/0004-637X/742/2/76},
  url        = {https://ui.adsabs.harvard.edu/abs/2011ApJ...742...76G},
}

@article{2007ApJ...662..322H,
  title      = {The {{Milky Way}}, an {{Exceptionally Quiet Galaxy}}: {{Implications}} for the {{Formation}} of {{Spiral Galaxies}}},
  shorttitle = {The {{Milky Way}}, an {{Exceptionally Quiet Galaxy}}},
  author     = {Hammer, F. and Puech, M. and Chemin, L. and Flores, H. and Lehnert, M. D.},
  year       = {2007},
  month      = jun,
  journal    = {ApJ},
  volume     = {662},
  pages      = {322--334},
  publisher  = {IOP},
  issn       = {0004-637X},
  doi        = {10.1086/516727},
  url        = {https://ui.adsabs.harvard.edu/abs/2007ApJ...662..322H},
}

@article{2006JCAP...01..014H,
  title     = {A Universal Velocity Distribution of Relaxed Collisionless Structures},
  author    = {Hansen, Steen H. and Moore, Ben and Zemp, Marcel and Stadel, Joachim},
  year      = {2006},
  month     = jan,
  journal   = {JCAP},
  volume    = {2006},
  pages     = {014},
  publisher = {IOP},
  issn      = {1475-7516},
  doi       = {10.1088/1475-7516/2006/01/014},
  url       = {https://ui.adsabs.harvard.edu/abs/2006JCAP...01..014H},
}

@article{2019Msngr.175...23H,
  title      = {{{4MOST Consortium Survey}} 1: {{The Milky Way Halo Low-Resolution Survey}}},
  shorttitle = {{{4MOST Consortium Survey}} 1},
  author     = {Helmi, A. and Irwin, M. and Deason, A. and Balbinot, E. and Belokurov, V. and {Bland-Hawthorn}, J. and Christlieb, N. and Cioni, M. -R. L. and Feltzing, S. and Grebel, E. K. and Kordopatis, G. and Starkenburg, E. and Walton, N. and Worley, C. C.},
  year       = {2019},
  month      = mar,
  journal    = {Msngr},
  volume     = {175},
  pages      = {23--25},
  issn       = {0722-6691},
  doi        = {10.18727/0722-6691/5120},
  url        = {https://ui.adsabs.harvard.edu/abs/2019Msngr.175...23H},
}

@article{2020ARA&A..58..205H,
  title   = {Streams, {{Substructures}}, and the {{Early History}} of the {{Milky Way}}},
  author  = {Helmi, Amina},
  year    = {2020},
  month   = aug,
  journal = {ARA\&A},
  volume  = {58},
  pages   = {205--256},
  issn    = {0066-4146},
  doi     = {10.1146/annurev-astro-032620-021917},
  url     = {https://ui.adsabs.harvard.edu/abs/2020ARA&A..58..205H},
}

@article{2018Natur.563...85H,
  title   = {The Merger That Led to the Formation of the {{Milky Way}}'s Inner Stellar Halo and Thick Disk},
  author  = {Helmi, Amina and Babusiaux, Carine and Koppelman, Helmer H. and Massari, Davide and Veljanoski, Jovan and Brown, Anthony G. A.},
  year    = {2018},
  month   = oct,
  journal = {Natur},
  volume  = {563},
  pages   = {85--88},
  issn    = {0028-0836},
  doi     = {10.1038/s41586-018-0625-x},
  url     = {https://ui.adsabs.harvard.edu/abs/2018Natur.563...85H},
}

@article{2018PhRvL.120d1102H,
  title     = {Empirical {{Determination}} of {{Dark Matter Velocities Using Metal-Poor Stars}}},
  author    = {{Herzog-Arbeitman}, Jonah and Lisanti, Mariangela and Madau, Piero and Necib, Lina},
  year      = {2018},
  month     = jan,
  journal   = {PhRvL},
  volume    = {120},
  pages     = {041102},
  publisher = {APS},
  issn      = {0031-9007},
  doi       = {10.1103/PhysRevLett.120.041102},
  url       = {https://ui.adsabs.harvard.edu/abs/2018PhRvL.120d1102H},
}

@article{2018MNRAS.480..800H,
  title      = {{{FIRE-2}} Simulations: Physics versus Numerics in Galaxy Formation},
  shorttitle = {{{FIRE-2}} Simulations},
  author     = {Hopkins, Philip F. and Wetzel, Andrew and Kere{\v s}, Du{\v s}an and {Faucher-Gigu{\`e}re}, Claude-Andr{\'e} and Quataert, Eliot and {Boylan-Kolchin}, Michael and Murray, Norman and Hayward, Christopher C. and {Garrison-Kimmel}, Shea and Hummels, Cameron and Feldmann, Robert and Torrey, Paul and Ma, Xiangcheng and {Angl{\'e}s-Alc{\'a}zar}, Daniel and Su, Kung-Yi and Orr, Matthew and Schmitz, Denise and Escala, Ivanna and Sanderson, Robyn and Grudi{\'c}, Michael Y. and Hafen, Zachary and Kim, Ji-Hoon and Fitts, Alex and Bullock, James S. and Wheeler, Coral and Chan, T. K. and Elbert, Oliver D. and Narayanan, Desika},
  year       = {2018},
  month      = oct,
  journal    = {MNRAS},
  volume     = {480},
  pages      = {800--863},
  publisher  = {OUP},
  issn       = {0035-8711},
  doi        = {10.1093/mnras/sty1690},
  url        = {https://ui.adsabs.harvard.edu/abs/2018MNRAS.480..800H},
}

@article{2020JHEP...07..081H,
  title     = {Impact of Uncertainties in the Halo Velocity Profile on Direct Detection of Sub-{{GeV}} Dark Matter},
  author    = {Hryczuk, Andrzej and Karukes, Ekaterina and Roszkowski, Leszek and Talia, Matthew},
  year      = {2020},
  month     = jul,
  journal   = {JHEP},
  volume    = {2020},
  pages     = {81},
  publisher = {Springer},
  issn      = {1029-8479},
  doi       = {10.1007/JHEP07(2020)081},
  url       = {https://ui.adsabs.harvard.edu/abs/2020JHEP...07..081H},
}

@misc{2025arXiv250114868H,
  title = {preprint},
  author = {Hussein, Abdelaziz and Necib, Lina and Kaplinghat, Manoj and Kim, Stacy Y. and Wetzel, Andrew and Read, Justin I. and Rey, Martin P. and Agertz, Oscar},
  year = {2025},
  month = jan,
  publisher = {arXiv},
  doi = {10.48550/arXiv.2501.14868}
}

@article{1994Natur.370..194I,
  title   = {A Dwarf Satellite Galaxy in {{Sagittarius}}},
  author  = {Ibata, R. A. and Gilmore, G. and Irwin, M. J.},
  year    = {1994},
  month   = jul,
  journal = {Natur},
  volume  = {370},
  pages   = {194--196},
  issn    = {0028-0836},
  doi     = {10.1038/370194a0},
  url     = {https://ui.adsabs.harvard.edu/abs/1994Natur.370..194I},
}

@article{2021MNRAS.502.5686I,
  title      = {Chemo-Kinematics of the {{Gaia RR Lyrae}}: The Halo and the Disc},
  shorttitle = {Chemo-Kinematics of the {{Gaia RR Lyrae}}},
  author     = {Iorio, Giuliano and Belokurov, Vasily},
  year       = {2021},
  month      = apr,
  journal    = {MNRAS},
  volume     = {502},
  pages      = {5686--5710},
  publisher  = {OUP},
  issn       = {0035-8711},
  doi        = {10.1093/mnras/stab005},
  url        = {https://ui.adsabs.harvard.edu/abs/2021MNRAS.502.5686I},
}

@article{2019ApJ...873..111I,
  title      = {{{LSST}}: {{From Science Drivers}} to {{Reference Design}} and {{Anticipated Data Products}}},
  shorttitle = {{{LSST}}},
  author     = {Ivezi{\'c}, {\v Z}eljko and Kahn, Steven M. and Tyson, J. Anthony and Abel, Bob and Acosta, Emily and Allsman, Robyn and Alonso, David and AlSayyad, Yusra and Anderson, Scott F. and Andrew, John and Angel, James Roger P. and Angeli, George Z. and Ansari, Reza and Antilogus, Pierre and Araujo, Constanza and Armstrong, Robert and Arndt, Kirk T. and Astier, Pierre and Aubourg, {\'E}ric and Auza, Nicole and Axelrod, Tim S. and Bard, Deborah J. and Barr, Jeff D. and Barrau, Aurelian and Bartlett, James G. and Bauer, Amanda E. and Bauman, Brian J. and Baumont, Sylvain and Bechtol, Ellen and Bechtol, Keith and Becker, Andrew C. and Becla, Jacek and Beldica, Cristina and Bellavia, Steve and Bianco, Federica B. and Biswas, Rahul and Blanc, Guillaume and Blazek, Jonathan and Blandford, Roger D. and Bloom, Josh S. and Bogart, Joanne and Bond, Tim W. and Booth, Michael T. and Borgland, Anders W. and Borne, Kirk and Bosch, James F. and Boutigny, Dominique and Brackett, Craig A. and Bradshaw, Andrew and Brandt, William Nielsen and Brown, Michael E. and Bullock, James S. and Burchat, Patricia and Burke, David L. and Cagnoli, Gianpietro and Calabrese, Daniel and Callahan, Shawn and Callen, Alice L. and Carlin, Jeffrey L. and Carlson, Erin L. and Chandrasekharan, Srinivasan and {Charles-Emerson}, Glenaver and Chesley, Steve and Cheu, Elliott C. and Chiang, Hsin-Fang and Chiang, James and Chirino, Carol and Chow, Derek and Ciardi, David R. and Claver, Charles F. and {Cohen-Tanugi}, Johann and Cockrum, Joseph J. and Coles, Rebecca and Connolly, Andrew J. and Cook, Kem H. and Cooray, Asantha and Covey, Kevin R. and Cribbs, Chris and Cui, Wei and Cutri, Roc and Daly, Philip N. and Daniel, Scott F. and Daruich, Felipe and Daubard, Guillaume and Daues, Greg and Dawson, William and Delgado, Francisco and Dellapenna, Alfred and {de Peyster}, Robert and {de Val-Borro}, Miguel and Digel, Seth W. and Doherty, Peter and Dubois, Richard and {Dubois-Felsmann}, Gregory P. and Durech, Josef and Economou, Frossie and Eifler, Tim and Eracleous, Michael and Emmons, Benjamin L. and Fausti Neto, Angelo and Ferguson, Henry and Figueroa, Enrique and {Fisher-Levine}, Merlin and Focke, Warren and Foss, Michael D. and Frank, James and Freemon, Michael D. and Gangler, Emmanuel and Gawiser, Eric and Geary, John C. and Gee, Perry and Geha, Marla and Gessner, Charles J. B. and Gibson, Robert R. and Gilmore, D. Kirk and Glanzman, Thomas and Glick, William and Goldina, Tatiana and Goldstein, Daniel A. and Goodenow, Iain and Graham, Melissa L. and Gressler, William J. and Gris, Philippe and Guy, Leanne P. and Guyonnet, Augustin and Haller, Gunther and Harris, Ron and Hascall, Patrick A. and Haupt, Justine and Hernandez, Fabio and Herrmann, Sven and Hileman, Edward and Hoblitt, Joshua and Hodgson, John A. and Hogan, Craig and Howard, James D. and Huang, Dajun and Huffer, Michael E. and Ingraham, Patrick and Innes, Walter R. and Jacoby, Suzanne H. and Jain, Bhuvnesh and Jammes, Fabrice and Jee, M. James and Jenness, Tim and Jernigan, Garrett and Jevremovi{\'c}, Darko and Johns, Kenneth and Johnson, Anthony S. and Johnson, Margaret W. G. and Jones, R. Lynne and {Juramy-Gilles}, Claire and Juri{\'c}, Mario and Kalirai, Jason S. and Kallivayalil, Nitya J. and Kalmbach, Bryce and Kantor, Jeffrey P. and Karst, Pierre and Kasliwal, Mansi M. and Kelly, Heather and Kessler, Richard and Kinnison, Veronica and Kirkby, David and Knox, Lloyd and Kotov, Ivan V. and Krabbendam, Victor L. and Krughoff, K. Simon and Kub{\'a}nek, Petr and Kuczewski, John and Kulkarni, Shri and Ku, John and Kurita, Nadine R. and Lage, Craig S. and Lambert, Ron and Lange, Travis and Langton, J. Brian and Le Guillou, Laurent and Levine, Deborah and Liang, Ming and Lim, Kian-Tat and Lintott, Chris J. and Long, Kevin E. and Lopez, Margaux and Lotz, Paul J. and Lupton, Robert H. and Lust, Nate B. and MacArthur, Lauren A. and Mahabal, Ashish and Mandelbaum, Rachel and Markiewicz, Thomas W. and Marsh, Darren S. and Marshall, Philip J. and Marshall, Stuart and May, Morgan and McKercher, Robert and McQueen, Michelle and Meyers, Joshua and Migliore, Myriam and Miller, Michelle and Mills, David J. and Miraval, Connor and Moeyens, Joachim and Moolekamp, Fred E. and Monet, David G. and Moniez, Marc and Monkewitz, Serge and Montgomery, Christopher and Morrison, Christopher B. and Mueller, Fritz and Muller, Gary P. and Mu{\~n}oz Arancibia, Freddy and Neill, Douglas R. and Newbry, Scott P. and Nief, Jean-Yves and Nomerotski, Andrei and Nordby, Martin and O'Connor, Paul and Oliver, John and Olivier, Scot S. and Olsen, Knut and O'Mullane, William and Ortiz, Sandra and Osier, Shawn and Owen, Russell E. and Pain, Reynald and Palecek, Paul E. and Parejko, John K. and Parsons, James B. and Pease, Nathan M. and Peterson, J. Matt and Peterson, John R. and Petravick, Donald L. and Libby Petrick, M. E. and Petry, Cathy E. and Pierfederici, Francesco and Pietrowicz, Stephen and Pike, Rob and Pinto, Philip A. and Plante, Raymond and Plate, Stephen and Plutchak, Joel P. and Price, Paul A. and Prouza, Michael and Radeka, Veljko and Rajagopal, Jayadev and Rasmussen, Andrew P. and Regnault, Nicolas and Reil, Kevin A. and Reiss, David J. and Reuter, Michael A. and Ridgway, Stephen T. and Riot, Vincent J. and Ritz, Steve and Robinson, Sean and Roby, William and Roodman, Aaron and Rosing, Wayne and Roucelle, Cecille and Rumore, Matthew R. and Russo, Stefano and Saha, Abhijit and Sassolas, Benoit and Schalk, Terry L. and Schellart, Pim and Schindler, Rafe H. and Schmidt, Samuel and Schneider, Donald P. and Schneider, Michael D. and Schoening, William and Schumacher, German and Schwamb, Megan E. and Sebag, Jacques and Selvy, Brian and Sembroski, Glenn H. and Seppala, Lynn G. and Serio, Andrew and Serrano, Eduardo and Shaw, Richard A. and Shipsey, Ian and Sick, Jonathan and Silvestri, Nicole and Slater, Colin T. and Smith, J. Allyn and Smith, R. Chris and Sobhani, Shahram and Soldahl, Christine and {Storrie-Lombardi}, Lisa and Stover, Edward and Strauss, Michael A. and Street, Rachel A. and Stubbs, Christopher W. and Sullivan, Ian S. and Sweeney, Donald and Swinbank, John D. and Szalay, Alexander and Takacs, Peter and Tether, Stephen A. and Thaler, Jon J. and Thayer, John Gregg and Thomas, Sandrine and Thornton, Adam J. and Thukral, Vaikunth and Tice, Jeffrey and Trilling, David E. and Turri, Max and Van Berg, Richard and Vanden Berk, Daniel and Vetter, Kurt and Virieux, Francoise and Vucina, Tomislav and Wahl, William and Walkowicz, Lucianne and Walsh, Brian and Walter, Christopher W. and Wang, Daniel L. and Wang, Shin-Yawn and Warner, Michael and Wiecha, Oliver and Willman, Beth and Winters, Scott E. and Wittman, David and Wolff, Sidney C. and {Wood-Vasey}, W. Michael and Wu, Xiuqin and Xin, Bo and Yoachim, Peter and Zhan, Hu},
  year       = {2019},
  month      = mar,
  journal    = {ApJ},
  volume     = {873},
  pages      = {111},
  publisher  = {IOP},
  issn       = {0004-637X},
  doi        = {10.3847/1538-4357/ab042c},
  url        = {https://ui.adsabs.harvard.edu/abs/2019ApJ...873..111I},
}

@article{2024MNRAS.530.2688J,
  title      = {The Wide-Field, Multiplexed, Spectroscopic Facility {{WEAVE}}: {{Survey}} Design, Overview, and Simulated Implementation},
  shorttitle = {The Wide-Field, Multiplexed, Spectroscopic Facility {{WEAVE}}},
  author     = {Jin, Shoko and Trager, Scott C. and Dalton, Gavin B. and Aguerri, J. Alfonso L. and Drew, J. E. and {Falc{\'o}n-Barroso}, Jes{\'u}s and G{\"a}nsicke, Boris T. and Hill, Vanessa and Iovino, Angela and Pieri, Matthew M. and Poggianti, Bianca M. and Smith, D. J. B. and Vallenari, Antonella and Abrams, Don Carlos and Aguado, David S. and Antoja, Teresa and {Arag{\'o}n-Salamanca}, Alfonso and Ascasibar, Yago and Babusiaux, Carine and Balcells, Marc and Barrena, R. and Battaglia, Giuseppina and Belokurov, Vasily and Bensby, Thomas and Bonifacio, Piercarlo and Bragaglia, Angela and Carrasco, Esperanza and Carrera, Ricardo and Cornwell, Daniel J. and {Dom{\'i}nguez-Palmero}, Lilian and Duncan, Kenneth J. and Famaey, Benoit and Fari{\~n}a, Cecilia and Gonzalez, Oscar A. and Guest, Steve and Hatch, Nina A. and Hess, Kelley M. and Hoskin, Matthew J. and Irwin, Mike and Knapen, Johan H. and Koposov, Sergey E. and Kuchner, Ulrike and Laigle, Clotilde and Lewis, Jim and Longhetti, Marcella and Lucatello, Sara and {M{\'e}ndez-Abreu}, Jairo and Mercurio, Amata and Molaeinezhad, Alireza and Mongui{\'o}, Maria and Morrison, Sean and Murphy, David N. A. and {Peralta de Arriba}, Luis and P{\'e}rez, Isabel and {P{\'e}rez-R{\`a}fols}, Ignasi and Pic{\'o}, Sergio and Raddi, Roberto and {Romero-G{\'o}mez}, Merc{\`e} and Royer, Fr{\'e}d{\'e}ric and Siebert, Arnaud and Seabroke, George M. and Som, Debopam and Terrett, David and Thomas, Guillaume and Wesson, Roger and Worley, C. Clare and Alfaro, Emilio J. and Allende Prieto, Carlos and {Alonso-Santiago}, Javier and Amos, Nicholas J. and Ashley, Richard P. and {Balaguer-N{\'u}{\~n}ez}, Lola and Balbinot, Eduardo and Bellazzini, Michele and Benn, Chris R. and Berlanas, Sara R. and Bernard, Edouard J. and Best, Philip and Bettoni, Daniela and Bianco, Andrea and Bishop, Georgia and Blomqvist, Michael and Boeche, Corrado and Bolzonella, Micol and Bonoli, Silvia and Bosma, Albert and Britavskiy, Nikolay and Busarello, Gianni and Caffau, Elisabetta and {Cantat-Gaudin}, Tristan and {Castro-Ginard}, Alfred and Couto, Guilherme and {Carbajo-Hijarrubia}, Juan and Carter, David and Casamiquela, Laia and Conrado, Ana M. and {Corcho-Caballero}, Pablo and Costantin, Luca and Deason, Alis and {de Burgos}, Abel and De Grandi, Sabrina and Di Matteo, Paola and {Dom{\'i}nguez-G{\'o}mez}, Jes{\'u}s and Dorda, Ricardo and Drake, Alyssa and Dutta, Rajeshwari and Erkal, Denis and Feltzing, Sofia and {Ferr{\'e}-Mateu}, Anna and Feuillet, Diane and Figueras, Francesca and Fossati, Matteo and Franciosini, Elena and Frasca, Antonio and Fumagalli, Michele and Gallazzi, Anna and {Garc{\'i}a-Benito}, Rub{\'e}n and Gentile Fusillo, Nicola and Gebran, Marwan and Gilbert, James and Gledhill, T. M. and Gonz{\'a}lez Delgado, Rosa M. and Greimel, Robert and Guarcello, Mario Giuseppe and Guerra, Jose and Gullieuszik, Marco and Haines, Christopher P. and Hardcastle, Martin J. and Harris, Amy and Haywood, Misha and Helmi, Amina and Hernandez, Nauzet and Herrero, Artemio and Hughes, Sarah and Ir{\v s}i{\v c}, Vid and Jablonka, Pascale and Jarvis, Matt J. and Jordi, Carme and Kondapally, Rohit and Kordopatis, Georges and Krogager, Jens-Kristian and La Barbera, Francesco and Lam, Man I. and Larsen, S{\o}ren S. and Lemasle, Bertrand and Lewis, Ian J. and Lhom{\'e}, Emilie and Lind, Karin and Lodi, Marcello and Longobardi, Alessia and Lonoce, Ilaria and Magrini, Laura and Ma{\'i}z Apell{\'a}niz, Jes{\'u}s and Marchal, Olivier and Marco, Amparo and Martin, Nicolas F. and Matsuno, Tadafumi and Maurogordato, Sophie and Merluzzi, Paola and {Miralda-Escud{\'e}}, Jordi and Molinari, Emilio and Monari, Giacomo and Morelli, Lorenzo and Mottram, Christopher J. and Naylor, Tim and Negueruela, Ignacio and O{\~n}orbe, Jose and Pancino, Elena and Peirani, S{\'e}bastien and Peletier, Reynier F. and Pozzetti, Lucia and Rainer, Monica and Ramos, Pau and Read, Shaun C. and Rossi, Elena Maria and R{\"o}ttgering, Huub J. A. and {Rubi{\~n}o-Mart{\'i}n}, Jose Alberto and Sabater, Jose and San Juan, Jos{\'e} and Sanna, Nicoletta and Schallig, Ellen and Schiavon, Ricardo P. and Schultheis, Mathias and Serra, Paolo and Shimwell, Timothy W. and {Sim{\'o}n-D{\'i}az}, Sergio and Smith, Russell J. and Sordo, Rosanna and Sorini, Daniele and Soubiran, Caroline and Starkenburg, Else and Steele, Iain A. and Stott, John and Stuik, Remko and Tolstoy, Eline and Tortora, Crescenzo and Tsantaki, Maria and {Van der Swaelmen}, Mathieu and {van Weeren}, Reinout J. and Vergani, Daniela and Verheijen, Marc A. W. and Verro, Kristiina and Vink, Jorick S. and Vioque, Miguel and Walcher, C. Jakob and Walton, Nicholas A. and Wegg, Christopher and Weijmans, Anne-Marie and Williams, Wendy L. and Wilson, Andrew J. and Wright, Nicholas J. and {Xylakis-Dornbusch}, Theodora and Youakim, Kris and Zibetti, Stefano and Zurita, Cristina},
  year       = {2024},
  month      = may,
  journal    = {MNRAS},
  volume     = {530},
  pages      = {2688--2730},
  publisher  = {OUP},
  issn       = {0035-8711},
  doi        = {10.1093/mnras/stad557},
  url        = {https://ui.adsabs.harvard.edu/abs/2024MNRAS.530.2688J},
}

@article{2020ApJ...900..103J,
  title = {A {{Diffuse Metal-poor Component}} of the {{Sagittarius Stream Revealed}} by the {{H3 Survey}}},
  author = {Johnson, Benjamin D. and Conroy, Charlie and Naidu, Rohan P. and Bonaca, Ana and Zaritsky, Dennis and Ting, Yuan-Sen and Cargile, Phillip A. and Han, Jiwon Jesse and Speagle, Joshua S.},
  year = {2020},
  month = sep,
  journal = {ApJ},
  volume = {900},
  pages = {103},
  publisher = {IOP},
  issn = {0004-637X},
  doi = {10.3847/1538-4357/abab08},
  url = {https://ui.adsabs.harvard.edu/abs/2020ApJ...900..103J}
}

@article{1996PhR...267..195J,
  title     = {Supersymmetric Dark Matter},
  author    = {Jungman, G. and Kamionkowski, M. and Griest, K.},
  year      = {1996},
  month     = mar,
  journal   = {PhR},
  volume    = {267},
  pages     = {195--373},
  publisher = {Elsevier},
  issn      = {0370-1573},
  doi       = {10.1016/0370-1573(95)00058-5},
  url       = {https://ui.adsabs.harvard.edu/abs/1996PhR...267..195J},
}

@article{2018MNRAS.475.4043K,
  title      = {The Need for Speed: Escape Velocity and Dynamical Mass Measurements of the {{Andromeda}} Galaxy},
  shorttitle = {The Need for Speed},
  author     = {Kafle, Prajwal R. and Sharma, Sanjib and Lewis, Geraint F. and Robotham, Aaron S. G. and Driver, Simon P.},
  year       = {2018},
  month      = apr,
  journal    = {MNRAS},
  volume     = {475},
  pages      = {4043--4054},
  publisher  = {OUP},
  issn       = {0035-8711},
  doi        = {10.1093/mnras/sty082},
  url        = {https://ui.adsabs.harvard.edu/abs/2018MNRAS.475.4043K},
}

@article{2005AJ....129..178K,
  title     = {The {{Local Group}} and {{Other Neighboring Galaxy Groups}}},
  author    = {Karachentsev, I. D.},
  year      = {2005},
  month     = jan,
  journal   = {AJ},
  volume    = {129},
  pages     = {178--188},
  publisher = {IOP},
  issn      = {0004-6256},
  doi       = {10.1086/426368},
  url       = {https://ui.adsabs.harvard.edu/abs/2005AJ....129..178K},
}

@article{2016JCAP...08..071K,
  title     = {The Impact of Baryons on the Direct Detection of Dark Matter},
  author    = {Kelso, Chris and Savage, Christopher and Valluri, Monica and Freese, Katherine and Stinson, Gregory S. and Bailin, Jeremy},
  year      = {2016},
  month     = aug,
  journal   = {JCAP},
  volume    = {2016},
  pages     = {071},
  publisher = {IOP},
  issn      = {1475-7516},
  doi       = {10.1088/1475-7516/2016/08/071},
  url       = {https://ui.adsabs.harvard.edu/abs/2016JCAP...08..071K},
}

@article{2011ApJ...740..102K,
  title      = {Dark {{Matter Halos}} in the {{Standard Cosmological Model}}: {{Results}} from the {{Bolshoi Simulation}}},
  shorttitle = {Dark {{Matter Halos}} in the {{Standard Cosmological Model}}},
  author     = {Klypin, Anatoly A. and {Trujillo-Gomez}, Sebastian and Primack, Joel},
  year       = {2011},
  month      = oct,
  journal    = {ApJ},
  volume     = {740},
  pages      = {102},
  publisher  = {IOP},
  issn       = {0004-637X},
  doi        = {10.1088/0004-637X/740/2/102},
  url        = {https://ui.adsabs.harvard.edu/abs/2011ApJ...740..102K},
}

@article{2010JCAP...02..030K,
  title     = {Dark Matter Direct Detection with Non-{{Maxwellian}} Velocity Structure},
  author    = {Kuhlen, Michael and Weiner, Neal and Diemand, J{\"u}rg and Madau, Piero and Moore, Ben and Potter, Doug and Stadel, Joachim and Zemp, Marcel},
  year      = {2010},
  month     = feb,
  journal   = {JCAP},
  volume    = {2010},
  pages     = {030},
  publisher = {IOP},
  issn      = {1475-7516},
  doi       = {10.1088/1475-7516/2010/02/030},
  url       = {https://ui.adsabs.harvard.edu/abs/2010JCAP...02..030K},
}

@article{2019MNRAS.486..378L,
  title     = {The Halo's Ancient Metal-Rich Progenitor Revealed with {{BHB}} Stars},
  author    = {Lancaster, Lachlan and Koposov, Sergey E. and Belokurov, Vasily and Evans, N. Wyn and Deason, Alis J.},
  year      = {2019},
  month     = jun,
  journal   = {MNRAS},
  volume    = {486},
  pages     = {378--389},
  publisher = {OUP},
  issn      = {0035-8711},
  doi       = {10.1093/mnras/stz853},
  url       = {https://ui.adsabs.harvard.edu/abs/2019MNRAS.486..378L},
}

@article{2022MNRAS.510.5119L,
  title = {The Kinematic Properties of {{Milky Way}} Stellar Halo Populations},
  author = {Lane, James M. M. and Bovy, Jo and Mackereth, J. Ted},
  year = {2022},
  month = mar,
  journal = {MNRAS},
  volume = {510},
  pages = {5119--5141},
  publisher = {OUP},
  issn = {0035-8711},
  doi = {10.1093/mnras/stab3755},
  url = {https://ui.adsabs.harvard.edu/abs/2022MNRAS.510.5119L}
}

@article{2023MNRAS.524.2606L,
  title      = {Gusts in the Headwind: Uncertainties in Direct Dark Matter Detection},
  shorttitle = {Gusts in the Headwind},
  author     = {Lawrence, Grace E. and Duffy, Alan R. and Blake, Chris A. and Hopkins, Philip F.},
  year       = {2023},
  month      = sep,
  journal    = {MNRAS},
  volume     = {524},
  pages      = {2606--2623},
  publisher  = {OUP},
  issn       = {0035-8711},
  doi        = {10.1093/mnras/stac2447},
  url        = {https://ui.adsabs.harvard.edu/abs/2023MNRAS.524.2606L},
}

@article{2023ApJ...956...15L,
  title      = {Carbon {{Stars}} as {{Standard Candles}}: {{An Empirical Test}} for the {{Reddening}}, {{Metallicity}}, and {{Age Sensitivity}} of the {{J-region Asymptotic Giant Branch}} ({{JAGB}}) {{Method}}},
  shorttitle = {Carbon {{Stars}} as {{Standard Candles}}},
  author     = {Lee, Abigail J.},
  year       = {2023},
  month      = oct,
  journal    = {ApJ},
  volume     = {956},
  pages      = {15},
  publisher  = {IOP},
  issn       = {0004-637X},
  doi        = {10.3847/1538-4357/acee69},
  url        = {https://ui.adsabs.harvard.edu/abs/2023ApJ...956...15L},
}

@article{2021ApJ...920...84L,
  title     = {A {{Sub-2}}\% {{Distance}} to {{M31}} from {{Photometrically Homogeneous Near-infrared Cepheid Period-Luminosity Relations Measured}} with the {{Hubble Space Telescope}}},
  author    = {Li, Siyang and Riess, Adam G. and Busch, Michael P. and Casertano, Stefano and Macri, Lucas M. and Yuan, Wenlong},
  year      = {2021},
  month     = oct,
  journal   = {ApJ},
  volume    = {920},
  pages     = {84},
  publisher = {IOP},
  issn      = {0004-637X},
  doi       = {10.3847/1538-4357/ac1597},
  url       = {https://ui.adsabs.harvard.edu/abs/2021ApJ...920...84L},
}

@article{2016ApJ...831...71L,
  title      = {Sizing {{Up}} the {{Milky Way}}: {{A Bayesian Mixture Model Meta-analysis}} of {{Photometric Scale Length Measurements}}},
  shorttitle = {Sizing {{Up}} the {{Milky Way}}},
  author     = {Licquia, Timothy C. and Newman, Jeffrey A.},
  year       = {2016},
  month      = nov,
  journal    = {ApJ},
  volume     = {831},
  pages      = {71},
  issn       = {0004-637X},
  doi        = {10.3847/0004-637X/831/1/71},
  url        = {https://ui.adsabs.harvard.edu/abs/2016ApJ...831...71L},
}

@article{2015ApJ...806...96L,
  title   = {Improved {{Estimates}} of the {{Milky Way}}'s {{Stellar Mass}} and {{Star Formation Rate}} from {{Hierarchical Bayesian Meta-Analysis}}},
  author  = {Licquia, Timothy C. and Newman, Jeffrey A.},
  year    = {2015},
  month   = jun,
  journal = {ApJ},
  volume  = {806},
  pages   = {96},
  issn    = {0004-637X},
  doi     = {10.1088/0004-637X/806/1/96},
  url     = {https://ui.adsabs.harvard.edu/abs/2015ApJ...806...96L},
}

@article{2016ApJ...833..220L,
  title     = {Does the {{Milky Way Obey Spiral Galaxy Scaling Relations}}?},
  author    = {Licquia, Timothy C. and Newman, Jeffrey A. and Bershady, Matthew A.},
  year      = {2016},
  month     = dec,
  journal   = {ApJ},
  volume    = {833},
  pages     = {220},
  publisher = {IOP},
  issn      = {0004-637X},
  doi       = {10.3847/1538-4357/833/2/220},
  url       = {https://ui.adsabs.harvard.edu/abs/2016ApJ...833..220L},
}

@article{2010JCAP...02..012L,
  title     = {Dark Matter Direct Detection Signals Inferred from a Cosmological {{N-body}} Simulation with Baryons},
  author    = {Ling, F. -S. and Nezri, E. and Athanassoula, E. and Teyssier, R.},
  year      = {2010},
  month     = feb,
  journal   = {JCAP},
  volume    = {2010},
  pages     = {012},
  publisher = {IOP},
  issn      = {1475-7516},
  doi       = {10.1088/1475-7516/2010/02/012},
  url       = {https://ui.adsabs.harvard.edu/abs/2010JCAP...02..012L},
}

@misc{2018arXiv181204114L,
  title     = {preprint},
  author    = {Lisanti, Mariangela and Necib, Lina},
  year      = {2018},
  month     = dec,
  publisher = {arXiv},
  doi       = {10.48550/arXiv.1812.04114},
}

@article{2012PDU.....1..155L,
  title   = {Dark Matter Debris Flows in the {{Milky Way}}},
  author  = {Lisanti, Mariangela and Spergel, David N.},
  year    = {2012},
  month   = nov,
  journal = {PDU},
  volume  = {1},
  pages   = {155--161},
  issn    = {2212-6864},
  doi     = {10.1016/j.dark.2012.10.007},
  url     = {https://ui.adsabs.harvard.edu/abs/2012PDU.....1..155L},
}

@article{2015ApJ...807...14L,
  title   = {Signatures of {{Kinematic Substructure}} in the {{Galactic Stellar Halo}}},
  author  = {Lisanti, Mariangela and Spergel, David N. and Madau, Piero},
  year    = {2015},
  month   = jul,
  journal = {ApJ},
  volume  = {807},
  pages   = {14},
  issn    = {0004-637X},
  doi     = {10.1088/0004-637X/807/1/14},
  url     = {https://ui.adsabs.harvard.edu/abs/2015ApJ...807...14L},
}

@article{2018PASP..130k4501M,
  title = {Fast {{Estimation}} of {{Orbital Parameters}} in {{Milky Way-like Potentials}}},
  author = {Mackereth, J. Ted and Bovy, Jo},
  year = {2018},
  month = nov,
  journal = {PASP},
  volume = {130},
  pages = {114501},
  publisher = {IOP},
  issn = {0004-6280},
  doi = {10.1088/1538-3873/aadcdd},
  url = {https://ui.adsabs.harvard.edu/abs/2018PASP..130k4501M}
}

@article{2022MNRAS.512.5823M,
  title      = {{{SIBELIUS-DARK}}: A Galaxy Catalogue of the Local Volume from a Constrained Realization Simulation},
  shorttitle = {{{SIBELIUS-DARK}}},
  author     = {McAlpine, Stuart and Helly, John C. and Schaller, Matthieu and Sawala, Till and Lavaux, Guilhem and Jasche, Jens and Frenk, Carlos S. and Jenkins, Adrian and Lucey, John R. and Johansson, Peter H.},
  year       = {2022},
  month      = jun,
  journal    = {MNRAS},
  volume     = {512},
  pages      = {5823--5847},
  publisher  = {OUP},
  issn       = {0035-8711},
  doi        = {10.1093/mnras/stac295},
  url        = {https://ui.adsabs.harvard.edu/abs/2022MNRAS.512.5823M},
}

@article{2017MNRAS.465...76M,
  title   = {The Mass Distribution and Gravitational Potential of the {{Milky Way}}},
  author  = {McMillan, Paul J.},
  year    = {2017},
  month   = feb,
  journal = {MNRAS},
  volume  = {465},
  pages   = {76--94},
  issn    = {0035-8711},
  doi     = {10.1093/mnras/stw2759},
  url     = {https://ui.adsabs.harvard.edu/abs/2017MNRAS.465...76M},
}

@article{2011MNRAS.414.2446M,
  title     = {Mass Models of the {{Milky Way}}},
  author    = {McMillan, Paul J.},
  year      = {2011},
  month     = jul,
  journal   = {MNRAS},
  volume    = {414},
  pages     = {2446--2457},
  publisher = {OUP},
  issn      = {0035-8711},
  doi       = {10.1111/j.1365-2966.2011.18564.x},
  url       = {https://ui.adsabs.harvard.edu/abs/2011MNRAS.414.2446M},
}

@article{2021NatAs...5..640M,
  title   = {Chronologically Dating the Early Assembly of the {{Milky Way}}},
  author  = {Montalb{\'a}n, Josefina and Mackereth, J. Ted and Miglio, Andrea and Vincenzo, Fiorenzo and Chiappini, Cristina and Buldgen, Gael and Mosser, Beno{\^i}t and Noels, Arlette and Scuflaire, Richard and Vrard, Mathieu and Willett, Emma and Davies, Guy R. and Hall, Oliver J. and Nielsen, Martin Bo and Khan, Saniya and Rendle, Ben M. and {van Rossem}, Walter E. and Ferguson, Jason W. and Chaplin, William J.},
  year    = {2021},
  month   = may,
  journal = {NatAs},
  volume  = {5},
  pages   = {640--647},
  issn    = {2397-3366},
  doi     = {10.1038/s41550-021-01347-7},
  url     = {https://ui.adsabs.harvard.edu/abs/2021NatAs...5..640M},
}

@article{2018ApJ...856L..26M,
  title     = {The {{Milky Way Halo}} in {{Action Space}}},
  author    = {Myeong, G. C. and Evans, N. W. and Belokurov, V. and Sanders, J. L. and Koposov, S. E.},
  year      = {2018},
  month     = apr,
  journal   = {ApJ},
  volume    = {856},
  pages     = {L26},
  publisher = {IOP},
  issn      = {0004-637X},
  doi       = {10.3847/2041-8213/aab613},
  url       = {https://ui.adsabs.harvard.edu/abs/2018ApJ...856L..26M},
}

@article{2020ApJ...901...48N,
  title = {Evidence from the {{H3 Survey That}} the {{Stellar Halo Is Entirely Comprised}} of {{Substructure}}},
  author = {Naidu, Rohan P. and Conroy, Charlie and Bonaca, Ana and Johnson, Benjamin D. and Ting, Yuan-Sen and Caldwell, Nelson and Zaritsky, Dennis and Cargile, Phillip A.},
  year = {2020},
  month = sep,
  journal = {ApJ},
  volume = {901},
  pages = {48},
  publisher = {IOP},
  issn = {0004-637X},
  doi = {10.3847/1538-4357/abaef4},
  url = {https://ui.adsabs.harvard.edu/abs/2020ApJ...901...48N}
}

@article{2021ApJ...923...92N,
  title   = {Reconstructing the {{Last Major Merger}} of the {{Milky Way}} with the {{H3 Survey}}},
  author  = {Naidu, Rohan P. and Conroy, Charlie and Bonaca, Ana and Zaritsky, Dennis and Weinberger, Rainer and Ting, Yuan-Sen and Caldwell, Nelson and Tacchella, Sandro and Han, Jiwon Jesse and Speagle, Joshua S. and Cargile, Phillip A.},
  year    = {2021},
  month   = dec,
  journal = {ApJ},
  volume  = {923},
  pages   = {92},
  issn    = {0004-637X},
  doi     = {10.3847/1538-4357/ac2d2d},
  url     = {https://ui.adsabs.harvard.edu/abs/2021ApJ...923...92N},
}

@article{2019ApJ...874....3N,
  title     = {Inferred {{Evidence}} for {{Dark Matter Kinematic Substructure}} with {{SDSS-Gaia}}},
  author    = {Necib, Lina and Lisanti, Mariangela and Belokurov, Vasily},
  year      = {2019},
  month     = mar,
  journal   = {ApJ},
  volume    = {874},
  pages     = {3},
  publisher = {IOP},
  issn      = {0004-637X},
  doi       = {10.3847/1538-4357/ab095b},
  url       = {https://ui.adsabs.harvard.edu/abs/2019ApJ...874....3N},
}

@article{2019ApJ...883...27N,
  title      = {Under the {{FIRElight}}: {{Stellar Tracers}} of the {{Local Dark Matter Velocity Distribution}} in the {{Milky Way}}},
  shorttitle = {Under the {{FIRElight}}},
  author     = {Necib, Lina and Lisanti, Mariangela and {Garrison-Kimmel}, Shea and Wetzel, Andrew and Sanderson, Robyn and Hopkins, Philip F. and {Faucher-Gigu{\`e}re}, Claude-Andr{\'e} and Kere{\v s}, Du{\v s}an},
  year       = {2019},
  month      = sep,
  journal    = {ApJ},
  volume     = {883},
  pages      = {27},
  publisher  = {IOP},
  issn       = {0004-637X},
  doi        = {10.3847/1538-4357/ab3afc},
  url        = {https://ui.adsabs.harvard.edu/abs/2019ApJ...883...27N},
}

@article{2020ApJ...903...25N,
  title     = {Chasing {{Accreted Structures}} within {{Gaia DR2 Using Deep Learning}}},
  author    = {Necib, Lina and Ostdiek, Bryan and Lisanti, Mariangela and Cohen, Timothy and Freytsis, Marat and {Garrison-Kimmel}, Shea},
  year      = {2020},
  month     = nov,
  journal   = {ApJ},
  volume    = {903},
  pages     = {25},
  publisher = {IOP},
  issn      = {0004-637X},
  doi       = {10.3847/1538-4357/abb814},
  url       = {https://ui.adsabs.harvard.edu/abs/2020ApJ...903...25N},
}

@article{2019MNRAS.490.3234N,
  title      = {First Results from the {{TNG50}} Simulation: Galactic Outflows Driven by Supernovae and Black Hole Feedback},
  shorttitle = {First Results from the {{TNG50}} Simulation},
  author     = {Nelson, Dylan and Pillepich, Annalisa and Springel, Volker and Pakmor, R{\"u}diger and Weinberger, Rainer and Genel, Shy and Torrey, Paul and Vogelsberger, Mark and Marinacci, Federico and Hernquist, Lars},
  year       = {2019},
  month      = dec,
  journal    = {MNRAS},
  volume     = {490},
  pages      = {3234--3261},
  publisher  = {OUP},
  issn       = {0035-8711},
  doi        = {10.1093/mnras/stz2306},
  url        = {https://ui.adsabs.harvard.edu/abs/2019MNRAS.490.3234N},
}

@article{2019ComAC...6....2N,
  title      = {The {{IllustrisTNG}} Simulations: Public Data Release},
  shorttitle = {The {{IllustrisTNG}} Simulations},
  author     = {Nelson, Dylan and Springel, Volker and Pillepich, Annalisa and {Rodriguez-Gomez}, Vicente and Torrey, Paul and Genel, Shy and Vogelsberger, Mark and Pakmor, Ruediger and Marinacci, Federico and Weinberger, Rainer and Kelley, Luke and Lovell, Mark and Diemer, Benedikt and Hernquist, Lars},
  year       = {2019},
  month      = may,
  journal    = {ComAC},
  volume     = {6},
  pages      = {2},
  doi        = {10.1186/s40668-019-0028-x},
  url        = {https://ui.adsabs.harvard.edu/abs/2019ComAC...6....2N},
}

@article{2023JCAP...05..012N,
  title      = {Cosmological Simulations of the Same Spiral Galaxy: Connecting the Dark Matter Distribution of the Host Halo with the Subgrid Baryonic Physics},
  shorttitle = {Cosmological Simulations of the Same Spiral Galaxy},
  author     = {{Nu{\~n}ez-Casti{\~n}eyra}, A. and Nezri, E. and Mollitor, P. and Devriendt, J. and Teyssier, R.},
  year       = {2023},
  month      = may,
  journal    = {JCAP},
  volume     = {2023},
  pages      = {012},
  publisher  = {IOP},
  issn       = {1475-7516},
  doi        = {10.1088/1475-7516/2023/05/012},
  url        = {https://ui.adsabs.harvard.edu/abs/2023JCAP...05..012N},
}

@article{2020PhRvD.101b3006O,
  title     = {Velocity Substructure from {{Gaia}} and Direct Searches for Dark Matter},
  author    = {O'Hare, Ciaran A. J. and Evans, N. Wyn and McCabe, Christopher and Myeong, GyuChul and Belokurov, Vasily},
  year      = {2020},
  month     = jan,
  journal   = {PhRvD},
  volume    = {101},
  pages     = {023006},
  publisher = {APS},
  issn      = {1550-79980556-2821},
  doi       = {10.1103/PhysRevD.101.023006},
  url       = {https://ui.adsabs.harvard.edu/abs/2020PhRvD.101b3006O},
}

@article{2020A&A...636A..75O,
  title   = {Cataloging Accreted Stars within {{Gaia DR2}} Using Deep Learning},
  author  = {Ostdiek, B. and Necib, L. and Cohen, T. and Freytsis, M. and Lisanti, M. and {Garrison-Kimmmel}, S. and Wetzel, A. and Sanderson, R. E. and Hopkins, P. F.},
  year    = {2020},
  month   = apr,
  journal = {A\&A},
  volume  = {636},
  pages   = {A75},
  issn    = {0004-6361},
  doi     = {10.1051/0004-6361/201936866},
  url     = {https://ui.adsabs.harvard.edu/abs/2020A&A...636A..75O},
}

@article{2017MNRAS.468.3428P,
  title     = {Orbits of Massive Satellite Galaxies - {{II}}. {{Bayesian}} Estimates of the {{Milky Way}} and {{Andromeda}} Masses Using High-Precision Astrometry and Cosmological Simulations},
  author    = {Patel, Ekta and Besla, Gurtina and Mandel, Kaisey},
  year      = {2017},
  month     = jul,
  journal   = {MNRAS},
  volume    = {468},
  pages     = {3428--3449},
  publisher = {OUP},
  issn      = {0035-8711},
  doi       = {10.1093/mnras/stx698},
  url       = {https://ui.adsabs.harvard.edu/abs/2017MNRAS.468.3428P},
}

@article{2014ApJ...784..161P,
  title     = {The {{Distribution}} of {{Dark Matter}} in the {{Milky Way}}'s {{Disk}}},
  author    = {Pillepich, Annalisa and Kuhlen, Michael and Guedes, Javiera and Madau, Piero},
  year      = {2014},
  month     = apr,
  journal   = {ApJ},
  volume    = {784},
  pages     = {161},
  publisher = {IOP},
  issn      = {0004-637X},
  doi       = {10.1088/0004-637X/784/2/161},
  url       = {https://ui.adsabs.harvard.edu/abs/2014ApJ...784..161P},
}

@article{2019MNRAS.490.3196P,
  title      = {First Results from the {{TNG50}} Simulation: The Evolution of Stellar and Gaseous Discs across Cosmic Time},
  shorttitle = {First Results from the {{TNG50}} Simulation},
  author     = {Pillepich, Annalisa and Nelson, Dylan and Springel, Volker and Pakmor, R{\"u}diger and Torrey, Paul and Weinberger, Rainer and Vogelsberger, Mark and Marinacci, Federico and Genel, Shy and {van der Wel}, Arjen and Hernquist, Lars},
  year       = {2019},
  month      = dec,
  journal    = {MNRAS},
  volume     = {490},
  pages      = {3196--3233},
  publisher  = {OUP},
  issn       = {0035-8711},
  doi        = {10.1093/mnras/stz2338},
  url        = {https://ui.adsabs.harvard.edu/abs/2019MNRAS.490.3196P},
}

@article{2016A&A...594A..13P,
  title   = {Planck 2015 Results. {{XIII}}. {{Cosmological}} Parameters},
  author  = {{Planck Collaboration} and Ade, P. A. R. and Aghanim, N. and Arnaud, M. and Ashdown, M. and Aumont, J. and Baccigalupi, C. and Banday, A. J. and Barreiro, R. B. and Bartlett, J. G. and Bartolo, N. and Battaner, E. and Battye, R. and Benabed, K. and Beno{\^i}t, A. and {Benoit-L{\'e}vy}, A. and Bernard, J. -P. and Bersanelli, M. and Bielewicz, P. and Bock, J. J. and Bonaldi, A. and Bonavera, L. and Bond, J. R. and Borrill, J. and Bouchet, F. R. and Boulanger, F. and Bucher, M. and Burigana, C. and Butler, R. C. and Calabrese, E. and Cardoso, J. -F. and Catalano, A. and Challinor, A. and Chamballu, A. and Chary, R. -R. and Chiang, H. C. and Chluba, J. and Christensen, P. R. and Church, S. and Clements, D. L. and Colombi, S. and Colombo, L. P. L. and Combet, C. and Coulais, A. and Crill, B. P. and Curto, A. and Cuttaia, F. and Danese, L. and Davies, R. D. and Davis, R. J. and {de Bernardis}, P. and {de Rosa}, A. and {de Zotti}, G. and Delabrouille, J. and D{\'e}sert, F. -X. and Di Valentino, E. and Dickinson, C. and Diego, J. M. and Dolag, K. and Dole, H. and Donzelli, S. and Dor{\'e}, O. and Douspis, M. and Ducout, A. and Dunkley, J. and Dupac, X. and Efstathiou, G. and Elsner, F. and En{\ss}lin, T. A. and Eriksen, H. K. and Farhang, M. and Fergusson, J. and Finelli, F. and Forni, O. and Frailis, M. and Fraisse, A. A. and Franceschi, E. and Frejsel, A. and Galeotta, S. and Galli, S. and Ganga, K. and Gauthier, C. and Gerbino, M. and Ghosh, T. and Giard, M. and {Giraud-H{\'e}raud}, Y. and Giusarma, E. and Gjerl{\o}w, E. and {Gonz{\'a}lez-Nuevo}, J. and G{\'o}rski, K. M. and Gratton, S. and Gregorio, A. and Gruppuso, A. and Gudmundsson, J. E. and Hamann, J. and Hansen, F. K. and Hanson, D. and Harrison, D. L. and Helou, G. and {Henrot-Versill{\'e}}, S. and {Hern{\'a}ndez-Monteagudo}, C. and Herranz, D. and Hildebrandt, S. R. and Hivon, E. and Hobson, M. and Holmes, W. A. and Hornstrup, A. and Hovest, W. and Huang, Z. and Huffenberger, K. M. and Hurier, G. and Jaffe, A. H. and Jaffe, T. R. and Jones, W. C. and Juvela, M. and Keih{\"a}nen, E. and Keskitalo, R. and Kisner, T. S. and Kneissl, R. and Knoche, J. and Knox, L. and Kunz, M. and {Kurki-Suonio}, H. and Lagache, G. and L{\"a}hteenm{\"a}ki, A. and Lamarre, J. -M. and Lasenby, A. and Lattanzi, M. and Lawrence, C. R. and Leahy, J. P. and Leonardi, R. and Lesgourgues, J. and Levrier, F. and Lewis, A. and Liguori, M. and Lilje, P. B. and {Linden-V{\o}rnle}, M. and {L{\'o}pez-Caniego}, M. and Lubin, P. M. and {Mac{\'i}as-P{\'e}rez}, J. F. and Maggio, G. and Maino, D. and Mandolesi, N. and Mangilli, A. and Marchini, A. and Maris, M. and Martin, P. G. and Martinelli, M. and {Mart{\'i}nez-Gonz{\'a}lez}, E. and Masi, S. and Matarrese, S. and McGehee, P. and Meinhold, P. R. and Melchiorri, A. and Melin, J. -B. and Mendes, L. and Mennella, A. and Migliaccio, M. and Millea, M. and Mitra, S. and {Miville-Desch{\^e}nes}, M. -A. and Moneti, A. and Montier, L. and Morgante, G. and Mortlock, D. and Moss, A. and Munshi, D. and Murphy, J. A. and Naselsky, P. and Nati, F. and Natoli, P. and Netterfield, C. B. and {N{\o}rgaard-Nielsen}, H. U. and Noviello, F. and Novikov, D. and Novikov, I. and Oxborrow, C. A. and Paci, F. and Pagano, L. and Pajot, F. and Paladini, R. and Paoletti, D. and Partridge, B. and Pasian, F. and Patanchon, G. and Pearson, T. J. and Perdereau, O. and Perotto, L. and Perrotta, F. and Pettorino, V. and Piacentini, F. and Piat, M. and Pierpaoli, E. and Pietrobon, D. and Plaszczynski, S. and Pointecouteau, E. and Polenta, G. and Popa, L. and Pratt, G. W. and Pr{\'e}zeau, G. and Prunet, S. and Puget, J. -L. and Rachen, J. P. and Reach, W. T. and Rebolo, R. and Reinecke, M. and Remazeilles, M. and Renault, C. and Renzi, A. and Ristorcelli, I. and Rocha, G. and Rosset, C. and Rossetti, M. and Roudier, G. and {Rouill{\'e} d'Orfeuil}, B. and {Rowan-Robinson}, M. and {Rubi{\~n}o-Mart{\'i}n}, J. A. and Rusholme, B. and Said, N. and Salvatelli, V. and Salvati, L. and Sandri, M. and Santos, D. and Savelainen, M. and Savini, G. and Scott, D. and Seiffert, M. D. and Serra, P. and Shellard, E. P. S. and Spencer, L. D. and Spinelli, M. and Stolyarov, V. and Stompor, R. and Sudiwala, R. and Sunyaev, R. and Sutton, D. and {Suur-Uski}, A. -S. and Sygnet, J. -F. and Tauber, J. A. and Terenzi, L. and Toffolatti, L. and Tomasi, M. and Tristram, M. and Trombetti, T. and Tucci, M. and Tuovinen, J. and T{\"u}rler, M. and Umana, G. and Valenziano, L. and Valiviita, J. and Van Tent, F. and Vielva, P. and Villa, F. and Wade, L. A. and Wandelt, B. D. and Wehus, I. K. and White, M. and White, S. D. M. and Wilkinson, A. and Yvon, D. and Zacchei, A. and Zonca, A.},
  year    = {2016},
  month   = sep,
  journal = {A\&A},
  volume  = {594},
  pages   = {A13},
  issn    = {0004-6361},
  doi     = {10.1051/0004-6361/201525830},
  url     = {https://ui.adsabs.harvard.edu/abs/2016A&A...594A..13P},
}

@article{2020JCAP...11..016P,
  title     = {Informing Dark Matter Direct Detection Limits with the {{ARTEMIS}} Simulations},
  author    = {{Poole-McKenzie}, Robert and Font, Andreea S. and Boxer, Billy and McCarthy, Ian G. and Burdin, Sergey and Stafford, Sam G. and Brown, Shaun T.},
  year      = {2020},
  month     = nov,
  journal   = {JCAP},
  volume    = {2020},
  pages     = {016},
  publisher = {IOP},
  issn      = {1475-7516},
  doi       = {10.1088/1475-7516/2020/11/016},
  url       = {https://ui.adsabs.harvard.edu/abs/2020JCAP...11..016P},
}

@article{2012ApJ...745..142R,
  title     = {On the {{Assembly}} of the {{Milky Way Dwarf Satellites}} and {{Their Common Mass Scale}}},
  author    = {Rashkov, Valery and Madau, Piero and Kuhlen, Michael and Diemand, J{\"u}rg},
  year      = {2012},
  month     = feb,
  journal   = {ApJ},
  volume    = {745},
  pages     = {142},
  publisher = {IOP},
  issn      = {0004-637X},
  doi       = {10.1088/0004-637X/745/2/142},
  url       = {https://ui.adsabs.harvard.edu/abs/2012ApJ...745..142R},
}

@article{2004ApJ...616..872R,
  title     = {The {{Proper Motion}} of {{Sagittarius A}}*. {{II}}. {{The Mass}} of {{Sagittarius A}}*},
  author    = {Reid, M. J. and Brunthaler, A.},
  year      = {2004},
  month     = dec,
  journal   = {ApJ},
  volume    = {616},
  pages     = {872--884},
  publisher = {IOP},
  issn      = {0004-637X},
  doi       = {10.1086/424960},
  url       = {https://ui.adsabs.harvard.edu/abs/2004ApJ...616..872R},
}

@article{2023MNRAS.521..995R,
  title      = {{{VINTERGATAN-GM}}: {{The}} Cosmological Imprints of Early Mergers on {{Milky-Way-mass}} Galaxies},
  shorttitle = {{{VINTERGATAN-GM}}},
  author     = {Rey, Martin P. and Agertz, Oscar and Starkenburg, Tjitske K. and Renaud, Florent and Joshi, Gandhali D. and Pontzen, Andrew and Martin, Nicolas F. and Feuillet, Diane K. and Read, Justin I.},
  year       = {2023},
  month      = may,
  journal    = {MNRAS},
  volume     = {521},
  pages      = {995--1012},
  issn       = {0035-8711},
  doi        = {10.1093/mnras/stad513},
  url        = {https://ui.adsabs.harvard.edu/abs/2023MNRAS.521..995R},
}

@article{2012MNRAS.426..128S,
  title = {Angle-Action Estimation in a General Axisymmetric Potential},
  author = {Sanders, Jason},
  year = {2012},
  month = oct,
  journal = {MNRAS},
  volume = {426},
  pages = {128--139},
  publisher = {OUP},
  issn = {0035-8711},
  doi = {10.1111/j.1365-2966.2012.21698.x},
  url = {https://ui.adsabs.harvard.edu/abs/2012MNRAS.426..128S}
}

@article{2015MNRAS.448.2941S,
  title      = {Bent by Baryons: The Low-Mass Galaxy-Halo Relation},
  shorttitle = {Bent by Baryons},
  author     = {Sawala, Till and Frenk, Carlos S. and Fattahi, Azadeh and Navarro, Julio F. and Bower, Richard G. and Crain, Robert A. and Dalla Vecchia, Claudio and Furlong, Michelle and Jenkins, Adrian and McCarthy, Ian G. and Qu, Yan and Schaller, Matthieu and Schaye, Joop and Theuns, Tom},
  year       = {2015},
  month      = apr,
  journal    = {MNRAS},
  volume     = {448},
  pages      = {2941--2947},
  publisher  = {OUP},
  issn       = {0035-8711},
  doi        = {10.1093/mnras/stu2753},
  url        = {https://ui.adsabs.harvard.edu/abs/2015MNRAS.448.2941S},
}

@article{2023MNRAS.521.4863S,
  title      = {The {{Local Group}}'s Mass: Probably No More than the Sum of Its Parts},
  shorttitle = {The {{Local Group}}'s Mass},
  author     = {Sawala, Till and Teeriaho, Meri and Johansson, Peter H.},
  year       = {2023},
  month      = jun,
  journal    = {MNRAS},
  volume     = {521},
  pages      = {4863--4877},
  publisher  = {OUP},
  issn       = {0035-8711},
  doi        = {10.1093/mnras/stad883},
  url        = {https://ui.adsabs.harvard.edu/abs/2023MNRAS.521.4863S},
}

@article{2016ApJ...831...93S,
  title     = {Assessing {{Astrophysical Uncertainties}} in {{Direct Detection}} with {{Galaxy Simulations}}},
  author    = {Sloane, Jonathan D. and Buckley, Matthew R. and Brooks, Alyson M. and Governato, Fabio},
  year      = {2016},
  month     = nov,
  journal   = {ApJ},
  volume    = {831},
  pages     = {93},
  publisher = {IOP},
  issn      = {0004-637X},
  doi       = {10.3847/0004-637X/831/1/93},
  url       = {https://ui.adsabs.harvard.edu/abs/2016ApJ...831...93S},
}

@article{2008MNRAS.391.1685S,
  title      = {The {{Aquarius Project}}: The Subhaloes of Galactic Haloes},
  shorttitle = {The {{Aquarius Project}}},
  author     = {Springel, V. and Wang, J. and Vogelsberger, M. and Ludlow, A. and Jenkins, A. and Helmi, A. and Navarro, J. F. and Frenk, C. S. and White, S. D. M.},
  year       = {2008},
  month      = dec,
  journal    = {MNRAS},
  volume     = {391},
  pages      = {1685--1711},
  publisher  = {OUP},
  issn       = {0035-8711},
  doi        = {10.1111/j.1365-2966.2008.14066.x},
  url        = {https://ui.adsabs.harvard.edu/abs/2008MNRAS.391.1685S},
}

@article{2010ARA&A..48..391S,
  title   = {Smoothed {{Particle Hydrodynamics}} in {{Astrophysics}}},
  author  = {Springel, Volker},
  year    = {2010},
  month   = sep,
  journal = {ARA\&A},
  volume  = {48},
  pages   = {391--430},
  issn    = {0066-4146},
  doi     = {10.1146/annurev-astro-081309-130914},
  url     = {https://ui.adsabs.harvard.edu/abs/2010ARA&A..48..391S},
}

@article{2001MNRAS.328..726S,
  author    = {Springel, Volker and White, Simon D. M. and Tormen, Giuseppe and Kauffmann, Guinevere},
  year      = {2001},
  month     = dec,
  journal   = {MNRAS},
  volume    = {328},
  pages     = {726--750},
  publisher = {OUP},
  issn      = {0035-8711},
  doi       = {10.1046/j.1365-8711.2001.04912.x},
  url       = {https://ui.adsabs.harvard.edu/abs/2001MNRAS.328..726S},
  title     = {Populating a Cluster of Galaxies -- {{I}}. {{Results}} at {$z=0$}},
}

@article{2024JCAP...08..022S,
  title      = {Sliding into {{DM}}: Determining the Local Dark Matter Density and Speed Distribution Using Only the Local Circular Speed of the Galaxy},
  shorttitle = {Sliding into {{DM}}},
  author     = {Staudt, Patrick G. and Bullock, James S. and {Boylan-Kolchin}, Michael and Kirkby, David and Wetzel, Andrew and Ou, Xiaowei},
  year       = {2024},
  month      = aug,
  journal    = {JCAP},
  volume     = {2024},
  number     = {08},
  pages      = {022},
  publisher  = {IOP},
  issn       = {1475-7516},
  doi        = {10.1088/1475-7516/2024/08/022},
  url        = {https://ui.adsabs.harvard.edu/abs/2024JCAP...08..022S},
}

@article{2010MNRAS.406..922T,
  title     = {Dark Matter Response to Galaxy Formation},
  author    = {Tissera, Patricia B. and White, Simon D. M. and Pedrosa, Susana and Scannapieco, Cecilia},
  year      = {2010},
  month     = aug,
  journal   = {MNRAS},
  volume    = {406},
  pages     = {922--935},
  publisher = {OUP},
  issn      = {0035-8711},
  doi       = {10.1111/j.1365-2966.2010.16777.x},
  url       = {https://ui.adsabs.harvard.edu/abs/2010MNRAS.406..922T},
}

@article{2025MNRAS.540.3493T,
  title     = {The Emergence of Galactic Thin and Thick Discs across Cosmic History},
  author    = {Tsukui, Takafumi and Wisnioski, Emily and {Bland-Hawthorn}, Joss and Freeman, Ken},
  year      = {2025},
  month     = jul,
  journal   = {MNRAS},
  volume    = {540},
  pages     = {3493--3522},
  publisher = {OUP},
  issn      = {0035-8711},
  doi       = {10.1093/mnras/staf604},
  url       = {https://ui.adsabs.harvard.edu/abs/2025MNRAS.540.3493T},
}

@article{2023PhRvD.107j3003V,
  title     = {Weighing the {{Milky Way}} and {{Andromeda}} Galaxies with Artificial Intelligence},
  author    = {{Villanueva-Domingo}, Pablo and {Villaescusa-Navarro}, Francisco and Genel, Shy and {Angl{\'e}s-Alc{\'a}zar}, Daniel and Hernquist, Lars and Marinacci, Federico and Spergel, David N. and Vogelsberger, Mark and Narayanan, Desika},
  year      = {2023},
  month     = may,
  journal   = {PhRvD},
  volume    = {107},
  pages     = {103003},
  publisher = {APS},
  issn      = {1550-79980556-2821},
  doi       = {10.1103/PhysRevD.107.103003},
  url       = {https://ui.adsabs.harvard.edu/abs/2023PhRvD.107j3003V},
}

@article{2009MNRAS.395..797V,
  title     = {Phase-Space Structure in the Local Dark Matter Distribution and Its Signature in Direct Detection Experiments},
  author    = {Vogelsberger, Mark and Helmi, Amina and Springel, Volker and White, Simon D. M. and Wang, Jie and Frenk, Carlos S. and Jenkins, Adrian and Ludlow, Aaron and Navarro, Julio F.},
  year      = {2009},
  month     = may,
  journal   = {MNRAS},
  volume    = {395},
  pages     = {797--811},
  publisher = {OUP},
  issn      = {0035-8711},
  doi       = {10.1111/j.1365-2966.2009.14630.x},
  url       = {https://ui.adsabs.harvard.edu/abs/2009MNRAS.395..797V},
}

@article{2008MNRAS.385..236V,
  title     = {The Fine-Grained Phase-Space Structure of Cold Dark Matter Haloes},
  author    = {Vogelsberger, Mark and White, Simon D. M. and Helmi, Amina and Springel, Volker},
  year      = {2008},
  month     = mar,
  journal   = {MNRAS},
  volume    = {385},
  pages     = {236--254},
  publisher = {OUP},
  issn      = {0035-8711},
  doi       = {10.1111/j.1365-2966.2007.12746.x},
  url       = {https://ui.adsabs.harvard.edu/abs/2008MNRAS.385..236V},
}

@article{2013MNRAS.430.1722V,
  title     = {Direct Detection of Self-Interacting Dark Matter},
  author    = {Vogelsberger, Mark and Zavala, Jesus},
  year      = {2013},
  month     = apr,
  journal   = {MNRAS},
  volume    = {430},
  pages     = {1722--1735},
  publisher = {OUP},
  issn      = {0035-8711},
  doi       = {10.1093/mnras/sts712},
  url       = {https://ui.adsabs.harvard.edu/abs/2013MNRAS.430.1722V},
}

@article{2011MNRAS.413.1373W,
  title   = {Assembly History and Structure of Galactic Cold Dark Matter Haloes},
  author  = {Wang, J. and Navarro, J. F. and Frenk, C. S. and White, S. D. M. and Springel, V. and Jenkins, A. and Helmi, A. and Ludlow, A. and Vogelsberger, M.},
  year    = {2011},
  month   = may,
  journal = {MNRAS},
  volume  = {413},
  pages   = {1373--1382},
  issn    = {0035-8711},
  doi     = {10.1111/j.1365-2966.2011.18220.x},
  url     = {https://ui.adsabs.harvard.edu/abs/2011MNRAS.413.1373W},
}

@article{2016ApJ...827L..23W,
  title      = {Reconciling {{Dwarf Galaxies}} with {{$\Lambda$CDM Cosmology}}: {{Simulating}} a {{Realistic Population}} of {{Satellites}} around a {{Milky Way-mass Galaxy}}},
  shorttitle = {Reconciling {{Dwarf Galaxies}} with {{$\Lambda$CDM Cosmology}}},
  author     = {Wetzel, Andrew R. and Hopkins, Philip F. and Kim, Ji-hoon and {Faucher-Gigu{\`e}re}, Claude-Andr{\'e} and Kere{\v s}, Du{\v s}an and Quataert, Eliot},
  year       = {2016},
  month      = aug,
  journal    = {ApJ},
  volume     = {827},
  pages      = {L23},
  publisher  = {IOP},
  issn       = {0004-637X},
  doi        = {10.3847/2041-8205/827/2/L23},
  url        = {https://ui.adsabs.harvard.edu/abs/2016ApJ...827L..23W},
}

@article{1978MNRAS.183..341W,
  title      = {Core Condensation in Heavy Halos: A Two-Stage Theory for Galaxy Formation and Clustering.},
  shorttitle = {Core Condensation in Heavy Halos},
  author     = {White, S. D. M. and Rees, M. J.},
  year       = {1978},
  month      = may,
  journal    = {MNRAS},
  volume     = {183},
  pages      = {341--358},
  publisher  = {OUP},
  issn       = {0035-8711},
  doi        = {10.1093/mnras/183.3.341},
  url        = {https://ui.adsabs.harvard.edu/abs/1978MNRAS.183..341W},
}

@article{2005MNRAS.361L...1W,
  title     = {Radial Velocity Moments of Dark Matter Haloes},
  author    = {Wojtak, Rados{\l}aw and {\L}okas, Ewa L. and Gottl{\"o}ber, Stefan and Mamon, Gary A.},
  year      = {2005},
  month     = jul,
  journal   = {MNRAS},
  volume    = {361},
  pages     = {L1-L5},
  publisher = {OUP},
  issn      = {0035-8711},
  doi       = {10.1111/j.1745-3933.2005.00054.x},
  url       = {https://ui.adsabs.harvard.edu/abs/2005MNRAS.361L...1W},
}

@misc{2026arXiv260325783Z,
    title = {preprint},
    author = {{Zhang}, Xiuyuan and {Thoyas}, Andreas and {Necib}, Lina and {Wetzel}, Andrew and {Arora}, Arpit},
    year = 2026,
    month = mar,
    publisher = {arXiv},
    doi = {10.48550/arXiv.2603.25783},
}

@article{2024ApJ...974..167Z,
  title     = {Deciphering the {{Kinematic Substructure}} of {{Local Dark Matter}} with {{LAMOST K Giants}}},
  author    = {Zhu, Hai and Guo, Rui and Shen, Juntai and Liu, Jianglai and Liu, Chao and Xue, Xiang-Xiang and Zhang, Lan and Mao, Shude},
  year      = {2024},
  month     = oct,
  journal   = {ApJ},
  volume    = {974},
  pages     = {167},
  publisher = {IOP},
  issn      = {0004-637X},
  doi       = {10.3847/1538-4357/ad6b17},
  url       = {https://ui.adsabs.harvard.edu/abs/2024ApJ...974..167Z},
}

@article{2025arXiv251204157L,
  title = {The {{DREAMS Project}}: {{Disentangling}} the {{Impact}} of {{Halo-to-Halo Variance}} and {{Baryonic Feedback}} on {{Milky Way Dark Matter Speed Distributions}}},
  shorttitle = {The {{DREAMS Project}}},
  year = 2026,
  month = may,
  journal = {ApJ},
  volume = {1002},
  pages = {168},
  publisher = {IOP},
  issn = {0004-637X},
  doi = {10.3847/1538-4357/ae5c9b},
  url = {https://ui.adsabs.harvard.edu/abs/2026ApJ..1002..168L},
  author = {Lilie, Ethan and Rose, Jonah C. and Lisanti, Mariangela and Garcia, Alex M. and Torrey, Paul and Kollmann, Kassidy E. and Li, Jiaxuan and Mostow, Olivia and Wang, Bonny Y. and O'Neil, Stephanie and Shen, Xuejian and Brooks, Alyson M. and Farahi, Arya and Kallivayalil, Nitya and Necib, Lina and Pace, Andrew B. and Vogelsberger, Mark}
}

\end{document}